\documentclass[aps,prx,showpacs,twocolumn,amsmath,amssymb,superscriptaddress]{revtex4-2}

\usepackage{tabularx}
\usepackage{bm}
\usepackage{graphicx}

\usepackage{hyperref}
\hypersetup{colorlinks=true,urlcolor= blue,citecolor=blue,linkcolor= blue,bookmarks=true,bookmarksopen=false}

\usepackage{color}

\usepackage{amsmath,mathtools}
\usepackage{multirow}
\usepackage{dcolumn}
\usepackage{amssymb,amscd,xypic,bm,wasysym}
\usepackage{float}
\usepackage{cleveref}
\usepackage[caption=false,position=top,captionskip=0pt,farskip=0pt]{subfig}
\usepackage{soul}

\begin{document}
	
\title{First-principles theory of spin magnetic multipole moments in antiferromagnets}
	
\author{Hua Chen}
\email{huachen@colostate.edu}
\affiliation{Department of Physics, Colorado State University, Fort Collins, CO 80523, USA}
\affiliation{School of Advanced Materials Discovery, Colorado State University, Fort Collins, CO 80523, USA}
\author{Guang-Yu Guo}
\affiliation{Department of Physics, National Taiwan University, Taipei 10617, Taiwan}
\affiliation{Physics Division, National Center for Theoretical Sciences, Taipei 10617, Taiwan}
\author{Di Xiao}
\affiliation{Department of Material Science and Engineering, University of Washington, Seattle, Washington 98195, USA}
\affiliation{Department of Physics, University of Washington, Seattle, Washington 98195, USA}

\begin{abstract}
Antiferromagnets with vanishing net magnetization are naturally expected to host higher-order magnetic multipole moments. Understanding and utilizing the multipole degrees of freedom are imperative for novel conceptual designs and applications unique to antiferromagnets. However, a universal, quantitative definition of magnetic multipole moments of antiferromagnetic materials is currently lacking. In this work we provide a unified description of arbitrary-order spin magnetic multipole moments (SM$^3$) of antiferromagnets by introducing a nonlocal spin density in macroscopic Maxwell equations. The formalism makes it transparent how SM$^3$ calculated for translationally invariant bulk systems corresponds to experimental observables when translation symmetry is broken. Through the nonlocal spin density calculated from first principles, we propose a robust scheme to extract arbitrary-order SM$^3$ through symmetry-constrained fitting at long wavelengths. Using this approach, we have calculated SM$^3$ of a few representative antiferromagnets, including $\alpha$-$\rm Fe_2O_3$, Mn$_3$Sn, and Mn$_3$NiN. Moreover, we clarify the role of spin-orbit coupling (SOC) in SM$^3$, especially in the weak SOC limit where clean predictions can be made based on symmetry principles. Our work paves the way for systematically investigating multipolar order parameters of unconventional magnetic materials.
\end{abstract}

\maketitle

\section{Introduction}

In recent years antiferromagnets (AFM) have emerged as a promising class of materials for technological applications traditionally relying on ferromagnets, particularly in areas that desire faster dynamics and robustness against external electromagnetic perturbations. To counter the usual difficulty of probing and manipulating the antiferromagnetic order due to the vanishing net dipolar magnetic moment, new physical principles and techniques have been discovered and developed, such as the anomalous Hall effect in certain antiferromagnets, current-induced torques, magnetic spin-Hall effect, and all-AFM tunnel junctions \cite{Solovyev1997, Tomizawa2009, Ohgushi2000, Shindou2001,chen_2014, Kubler_2014, Nakatsuji_2015, Nayak2016, Zhou2019, Gurung2019, Boldrin2019, Zhao2019, Liu2018, Smejkal2020, chen_2020, chen_2022}. However, most of these developments are still rooted in the conventional wisdom from studies on ferromagnets, a main reason being that vector or dipole order parameters are both conceptually simple and easy to control using dipolar fields.

Multipole expansion has been a standard tool for understanding nonuniform distribution of physical quantities across science and technology. In particular, Cartesian (tensor) multipole moments have been routinely used for systematically coarse-graining microscopic charge, current, and spin densities, and are conveniently conjugate to gradients of electromagnetic fields \cite{Jackson:490457}. Historically, Cartesian magnetic multipole moments have been used to characterize the ordering and dynamics of AFM, by serving as alternative order parameters to the microscopic local magnetic moments \cite{Andreev1978,  Andreev_1980, DZYALOSHINSKII1992579, Andreev1996, Astrov1996}. As a natural generalization of vector order parameters, multipole moments have been widely adopted as a convenient language as magnetism and spintronics research starts to focus on complex spin systems with beyond-dipole order. \cite{kuramoto_2009,santini_2009,sakai_2011,onimaru_2011,onimaru_2016,kubo_2004,mannix_2005,kuwahara_2007,matsumura_2009,watanabe_2018,hayami_2018,gao_2018,gao_2018_2,shitade_2018,shitade_2019,Dubovik_1990,Gorbatsevich_1994,ederer_2007,Spaldin_2008,arima_2005,VanAken_2007,Hayami_2014,Batista_2008,thole_2016,watanabe_2017_1,Suzuki2017,Tahir2023}. However, as already built into its classical definition, multipole moments have inherent issues associated with origin dependence except for the lowest nonvanishing order. In extended condensed matter systems, multipole moments have another apparent indeterminacy associated with the choice of the unit cell \cite{Resta_1994,king-smith_1993,xiao_2005,thonhauser_2005,shi_2007}. These conceptual conundrums have prevented multipole moments from being systematically quantified in real materials. Rather, the use of multipole moments is mostly limited to symmetry analysis of magnetic structure and response properties in complex spin systems \cite{Suzuki2017, PhysRevB.99.174407, PhysRevResearch.2.012045}. 

Pioneering efforts on some low-order electric and magnetic multipole moments, such as polarization, orbital magnetization, and toroidization, etc.\cite{Resta_1994,king-smith_1993,xiao_2005,thonhauser_2005,shi_2007,gao_2018,gao_2018_2,shitade_2018,shitade_2019,lapa_2019,daido_2020} have led to two classes of definitions for multipole moments. One is through an adiabatic pumping picture, where the multipole (e.g. charge polarization \cite{Resta_1994,king-smith_1993}) is defined as the time integral of the corresponding adiabatic current, or directly as volume average of gauge connections \cite{Resta1998}. However, a path to generalizing this definition to higher-order multipoles and metallic systems, as well as their direct experimental manifestation, is still unclear, despite the significant effort on the general subject of higher-order topology \cite{Benalcazar61, Benalcazar2017, Schindler2018, Kang2019, Watanabe2020, Trifunovic2020, Ren2021}. The other definition uses a thermodynamic approach, in which multipoles are conjugate variables of external electromagnetic field gradients \cite{shi_2007, gao_2018, gao_2018_2, shitade_2018, shitade_2019, lapa_2019, daido_2020, Oike2025a, Sato2026}. Such a definition is not only conceptually clean but also gives rich thermodynamic relations among different response functions. However, it is not clear whether or how the multipole moments defined in this manner can be measured directly using existing experimental methodologies, especially those with local probes. A well-known example is orbital magnetization, whose meaning as a bulk quantity as well as its local counterpart is still under debate \cite{Chen2012,Moseni2024,Bianco2013,Seleznev2023}. Another serious issue with the thermodynamic definition is that odd-order charge multipoles identically vanish \cite{daido_2020}. It is therefore imperative to formulate the multipole moments in a quantitative, experimentally verifiable manner.

Recently, altermagnets, AFM featuring spin-orbit-coupled electronic states without the need of relativistic spin-orbit coupling (SOC), have become a very active topic in magnetism and spintronic research \cite{Smejkal2022a, Smejkal2022, Liu2022, Guo2025, Shim2025, Tamang2025, Song2025, Jungwirth2026}. Ignoring SOC as a legitimate first approximation in many AFM allows one to constrain physical properties using generally higher symmetry described by spin groups \cite{Brinkman1966, Litvin1974, Litvin1977, Liu2022, Chen2024, Jiang2024, Xiao2024, Watanabe2024, Etxebarria2025, Liu2025, Liu2026} than their magnetic space group symmetry. Since the dipole moments of ferromagnets are insensitive to SOC, one may naively expect that higher-order magnetic multipole moments are weakly dependent on SOC as well. In this work we examine whether this is indeed the case. Moreover, we will show that in cases that SOC is negligible for spin magnetic multipole moments (SM$^3$), there exist close connections between SM$^3$ and local properties near domain walls. 

\begin{figure}[h]
    \centering
         \includegraphics[width=0.3\textwidth]{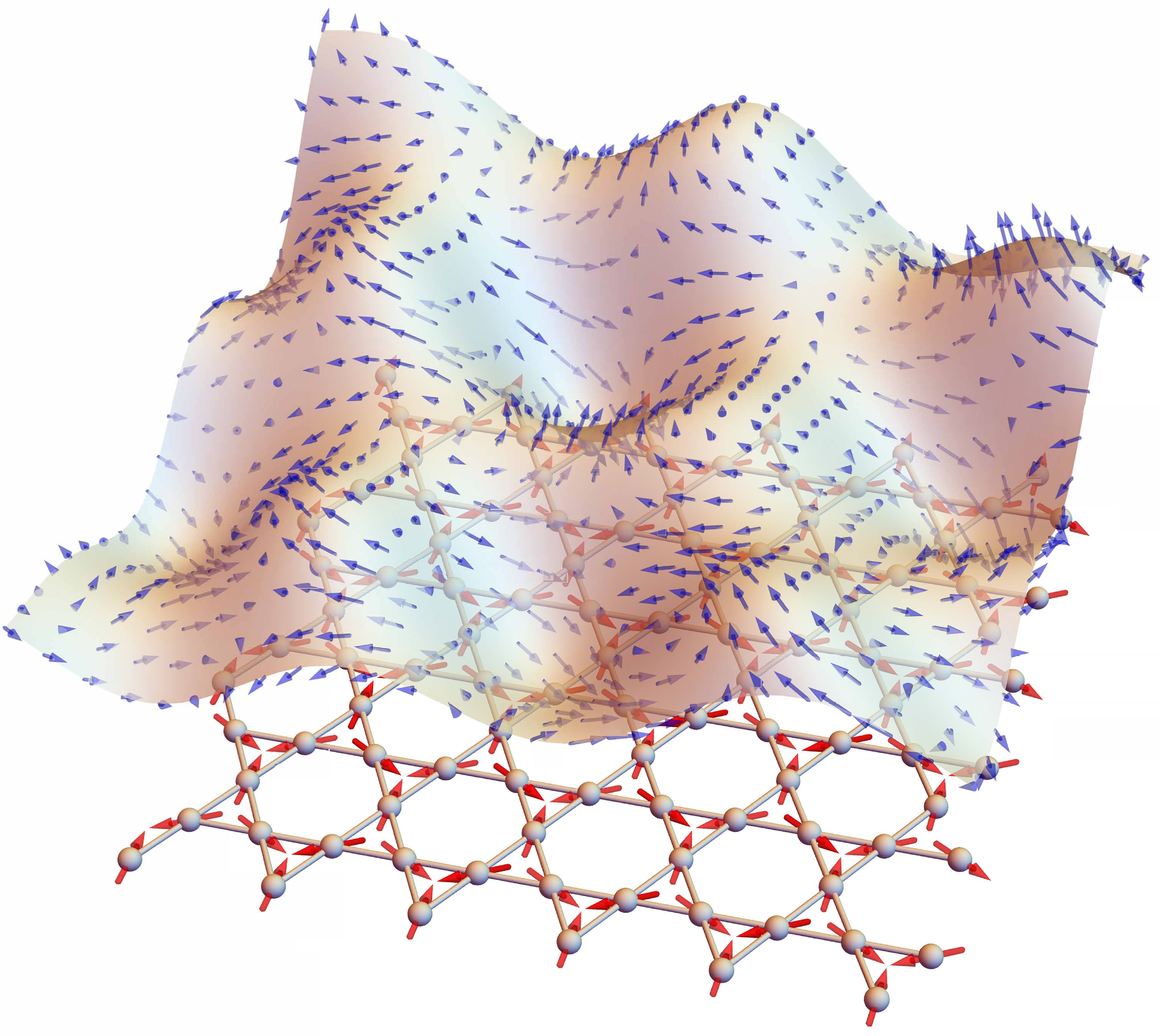}   
    \caption{Schematic illustration of spin densities (blue arrows) created by a spatially nonuniform magnetic octupole for the toy model in Sec.~\ref{sec:model}. The curved surface corresponds to a random $\psi(\mathbf r)$ in Eq.~\eqref{eq:psirM}. The scale of the kagome lattice is exaggerated for illustration purpose.}
    \label{fig:schematic}
\end{figure}

In this work, we resolve the several long-standing issues mentioned above by developing a first-principles framework for spin magnetic multipole moments (SM$^3$) in periodic solids based on a nonlocal spin density entering the macroscopic Maxwell equations. This formulation recasts multipole moments as the long-wavelength expansion coefficients of a well-defined response function, thereby promoting them to intrinsic bulk quantities free of origin ambiguity and directly connected to measurable electromagnetic responses once translational symmetry is weakly broken. It provides a unified definition of arbitrary-order multipoles applicable to both insulating and metallic systems, establishes a transparent link between bulk multipoles and real-space observables such as boundary or texture-induced spin densities (Fig.~\ref{fig:schematic}), and, crucially, introduces a practical computational scheme in which SM$^3$ are obtained as symmetry-constrained coefficients in a small-$q$ expansion of the nonlocal spin susceptibility $\boldsymbol \chi(\mathbf q)$, enabling systematic extraction from first-principles calculations. Applying this approach to representative antiferromagnets, including $\alpha$-$\rm Fe_2O_3$, $\rm Mn_3Sn$, and $\rm Mn_3NiN$, we uncover strong material and symmetry dependence of multipole moments and clarify the role of spin-orbit coupling in shaping their magnitude and observability. More broadly, our work elevates magnetic multipoles from a qualitative symmetry descriptor to a quantitative, predictive framework for complex magnetic materials.

\section{SM$^3$ through nonlocal spin density}\label{sec:formalism}
In this section we introduce SM$^3$ through a physically motivated perspective relevant to extended condensed matter systems. The extended nature of such systems makes it inappropriate to use the usual far-field expansion of electromagnetic potentials to define multipole moments, which unavoidably introduce the unbounded position operators in the resulting expressions. Instead, locally defined multipole moments that enter the macroscopic Maxwell equations in matter are more desirable. Historically, different coarse-graining procedures have been attempted to define the local multipole moments on firm grounds. In Sec.~\ref{ssec:multipole-history} we give a brief review of these approaches, which, despite their limitations, motivate our definition detailed in Sec.~\ref{ssec:SM3-chi}. In Sec.~\ref{ssec:SM3-localspin} we discuss the relation between SM$^3$ and local observables such as local spin densities. Finally, in Sec.~\ref{sec:SM3SOC} we detail the role of SOC in the behavior of SM$^3$.

\subsection{Review of coarse-graining definition of multipole moments}\label{ssec:multipole-history}

Different ways of defining multipole moments by coarse-graining can be found in the historical literature \cite{Jackson:490457, Russakoff1970, Robinson, Robinson1971}. The commonality among these approaches is to introduce a physically motivated scheme for performing average in a confined region about a given position. Taking charge multipoles for example, the starting point is the microscopic Maxwell equations applied to a medium consisting of discrete charge
    \begin{gather}\label{eq:MEmicro}
    \nabla_{\mathbf R}\cdot \mathbf b(\mathbf R) = 0\\\nonumber
 \nabla_{\mathbf R}\times \mathbf e(\mathbf R) = -\partial_t \mathbf b(\mathbf R)\\\nonumber
 \nabla_{\mathbf R}\cdot \mathbf e(\mathbf R) = \frac{1}{4\pi \epsilon_0}\sum_n q_n \delta(\mathbf r_n - \mathbf R) \\\nonumber
 \nabla_{\mathbf R}\times \mathbf b(\mathbf R) = \mu_0 \sum_{n }q_n \dot{\mathbf r}_n \delta(\mathbf r_n - \mathbf R) + \mu_0\epsilon_0\partial_t \mathbf e(\mathbf R)
\end{gather}
where $\mathbf e,\mathbf b$ stands for the microscopic electric and magnetic fields depending on coordinates $\mathbf R$, and $q_n$ is the $n$-th point charge located at position $\mathbf r_n$. 

To arrive at coarse-grained behavior of the microscopic quantities, one introduces a weight function $f(\mathbf s)$ satisfying $\int d^3\mathbf s f(\mathbf s) = 1$ and its Fourier transform denoted by $F(\mathbf k)$. $f(\mathbf s)$ is ideally localized near $\mathbf s=0$ in real space while $F(\mathbf k)$ is localized in momentum space near $\mathbf k=0$. The usage of $f(\mathbf s)$ to control coarse-graining is therefore analogous to the idea of wave packets in formulating semiclassical dynamics of Bloch electrons \cite{Culcer2004}. The coarse-grained version of a microscopic variable $\alpha(\mathbf R)$ is defined as
\begin{eqnarray}\label{eq:fstruncation}
    \bar{\alpha }(\mathbf R) = \int \alpha(\mathbf R - \mathbf s) f(\mathbf s) d^3\mathbf s
\end{eqnarray}
The above equation essentially means that the value of coarse-grained $\alpha$ at position $\mathbf R$ is a weighted average of $\alpha$ in the neighborhood of $\mathbf R$ with a typical spatial range controlled by $f$. Fourier transforming Eq.~\eqref{eq:fstruncation} gives
\begin{eqnarray}
   \bar{\alpha }(\mathbf k) = \alpha(\mathbf k) F(\mathbf k)
\end{eqnarray}
A localized $F(\mathbf k)$, satisfying $F(\mathbf k)\rightarrow 0$ when $k > k_0$, therefore eliminates high-frequency details of $\alpha(\mathbf R)$ that are ignored in a coarse-grained theory. It is pointed out in \cite{Robinson} that $f(\mathbf s)$ does not need to, and often cannot be made positive definite in order to satisfy the above requirement on $F(\mathbf k)$. Therefore it should not be viewed as a probability distribution. 

Applying Eq.~\eqref{eq:fstruncation} to the source terms leads to
\begin{gather}\label{eq:rhoJcg}
    \overline{\sum_n q_n \delta(\mathbf r_n - \mathbf R)} = \sum_n q_n f(\mathbf R - \mathbf r_n) \equiv \bar{\rho}(\mathbf R) \\\nonumber
    \overline{\sum_{n }q_n \dot{\mathbf r}_n \delta(\mathbf r_n - \mathbf R)} = \sum_n q_n \dot{\mathbf r}_n f(\mathbf R - \mathbf r_n) \equiv \bar{\mathbf J}(\mathbf R)
\end{gather}

Due to the static nature of $f(\mathbf s)$ the truncation operation commutes with spatial and temporal derivatives. Applying Eq.~\eqref{eq:fstruncation} on both sides of the microscopic Maxwell equations Eq.~\eqref{eq:MEmicro} then leads to
    \begin{gather}\label{eq:MEmacro}
    \nabla\cdot \mathbf B = 0\\\nonumber
 \nabla\times \mathbf E = -\partial_t \mathbf B\\\nonumber
 \nabla\cdot \mathbf E = \frac{1}{4\pi \epsilon_0}\bar{\rho} \\\nonumber
 \nabla \times \mathbf B = \mu_0 \bar{\mathbf J} + \mu_0\epsilon_0\partial_t \mathbf E
\end{gather}
where $\mathbf B = \bar{\mathbf b}$, $\mathbf E = \bar{\mathbf e}$ and we have omitted the explicit $\mathbf R$ dependence of the various quantities.

The coarse-grained $\rho$ and $\mathbf J$ are now ready to be recast into multipole moments. The key idea is to write the charge density $\rho(\mathbf R)$ as a superposition of ``atomic" contributions:
\begin{eqnarray}\label{eq:rhoRrhou}
    \rho(\mathbf R ) = \sum_n \rho_u(\mathbf R - \mathbf R_n) 
\end{eqnarray}
where $\rho_u$ is the charge density due to each periodic constituent of the crystal. In this way, $\rho(\mathbf R)$ always has the same translation symmetry as the crystal. Apparently, there is arbitrariness in choosing $\rho_u$. The truncation equation for $\rho(\mathbf R)$ then becomes
\begin{eqnarray}\label{eq:rhoutrunc}
    \bar\rho(\mathbf R) &=& \int d^3\mathbf s \sum_n \rho_u(\mathbf R - \mathbf R_n - \mathbf s) f(\mathbf s) \\\nonumber
    &=& \int d^3\mathbf s \sum_n \rho_u( \mathbf s) f(\mathbf R - \mathbf R_n - \mathbf s)
\end{eqnarray}
Due to the smooth dependence of $f$ on its argument, one can expand $f(\mathbf R-\mathbf R_n - \mathbf s)$ about $\mathbf R-\mathbf R_n$, which, upon substitution into Eq.~\eqref{eq:rhoutrunc}, gives
\begin{flalign}\label{eq:rhomultipole}
&\bar\rho(\mathbf R)= \sum_{m=0}^\infty \frac{ (-\nabla_{\mathbf R})^m }{m!}\sum_n\int d^3\mathbf s\rho_u(\mathbf s) \mathbf s^m f(\mathbf R- \mathbf R_n) & \\\nonumber
&\equiv \bar{\rho}_f (\mathbf R) - \nabla_{\mathbf R} \cdot \mathbf P(\mathbf R) + \frac{1}{2}\partial_i\partial_j \mathcal{Q}_{ij}(\mathbf R)  + \dots&
\end{flalign}
The charge multipole moment of order $n$ thus defined is 
\begin{flalign}
    &\mathcal{C}_{j_1j_2\dots j_n} (\mathbf R) = \sum_i \int d^3\mathbf s \rho_u(\mathbf s) s_{j_1} s_{j_2}\dots s_{j_n} f(\mathbf R - \mathbf R_i)& \\\nonumber
    &\equiv \mathcal{C}^u_{j_1j_2\dots j_n} \sum_i  f(\mathbf R - \mathbf R_i) &
\end{flalign}
Namely, a superposition of atomic multipole moments. 

The multipole moments defined in \cite{Robinson,Robinson1971} therefore have two sources of ambiguity. The first is the decomposition $\rho_u$ which determines the atomic multipole moments $\mathcal{C}^u$. The second is the truncation function $f$ which determines the spatial dependence of $\mathcal{C}$. Choosing a specific $\rho_u$ will fix atomic multipole moments $\mathcal{C}^u$, but having a very smooth $f$ will allow the multipole expansion of $\bar{\rho}(\mathbf R)$ to be truncated at low orders. In particular, choosing a constant $f$ makes $\bar{\rho} \propto \int d^3\mathbf r \rho(\mathbf r)$ and become a constant. 

In spite of these limitations, the above coarse-grain procedure is meaningful in the following general sense: The macroscopic electromagnetic fields depend on matter fields non-locally, and multipole moments serve as a device to describe this nonlocality. In the next subsection we follow this spirit to introduce the SM$^3$.

\subsection{Nonlocal spin density and SM$^3$}\label{ssec:SM3-chi}
We start from the Lagrangian of an electronic system interacting with electromagnetic fields:
\begin{eqnarray}
    \mathcal{L} = \mathcal{L}_{\rm EM} + \mathcal{L}_c + \mathcal{L}_{\rm el}
\end{eqnarray}
where $\rm el$ stands for electrons and $c$ stands for the coupling between fields and electrons. In this work we consider only the Zeeman coupling in $\mathcal{L}_c$ relevant to spin multipole moments. Integrating out electrons \cite{Altland2023}, ignoring any frequency dependence since we are concerned with static multipole moments only in this work, and formally expanding the resulting effective Lagrangian to the lowest order in magnetic fields $\mathbf B$ leads to
\begin{eqnarray}
    \mathcal{L}_{\rm eff}(\mathbf r, t) = \mathcal{L}_{\rm EM}(\mathbf r, t) - \int d^3\mathbf r' \chi_j(\mathbf r,\mathbf r') B_j(\mathbf r').
\end{eqnarray}

Since $\boldsymbol \chi$ relates the Lagrangian density at $\mathbf r$ to the Zeeman field at $\mathbf r'$, it can be recognized as a nonlocal spin (magnetization) density. The effective EM Lagrangian therefore becomes nonlocal. One can then perform a Taylor expansion of $\mathbf B$ about $\mathbf r$:
\begin{widetext}
\begin{eqnarray}
    - \int d^3\mathbf r' \chi_j(\mathbf r, \mathbf r') B_j(\mathbf r') &=&-\int d^3\mathbf r' \chi_j(\mathbf r, \mathbf r') \sum_{n=0}^\infty \frac{1}{n!}[(\mathbf r'-\mathbf r)\cdot\nabla_{\mathbf r}]^n B_j(\mathbf r) \\\nonumber
    &=& - \sum_{n=0}^\infty\frac{1}{n!} \left[\partial_{r_{j_1}}\partial_{r_{j_2}}\dots \partial_{r_{j_n}} B^i(\mathbf r)\right] \int d^3\mathbf r' \chi_j(\mathbf r, \mathbf r - \mathbf r')(-\mathbf r')^n \\\nonumber 
    &\equiv& \sum_{n=0}^\infty \frac{1}{n!} \left[\partial_{r_{j_1}}\partial_{r_{j_2}}\dots \partial_{r_{j_n}} B^i(\mathbf r)\right]\mathcal{M}^i_{j_1j_2\dots j_n}(\mathbf r)
\end{eqnarray}
To see that $\mathcal{M}$ in the above equation is indeed the multipole moments, we use the Euler-Lagrange equation for a nonlocal Lagrangian which depends on higher-order gradients:
\begin{eqnarray}\label{eq:ELhigherdev}
    \frac{\partial \mathcal{L}}{\partial \phi^i} + \sum_{l=1}^n \sum_{j_1\leq  \dots \leq j_l} (-1)^l \frac{\partial^l}{\partial_{x_{j_1}}\dots \partial_{x_{j_l}}}\left[\frac{\partial \mathcal{L}}{\partial(\partial_{x_{j_1}}\dots \partial_{x_{j_l}}\phi^i)}\right] = 0.
\end{eqnarray}
The Euler-Lagrange equation leads to the following Ampère-Maxwell law:
\begin{eqnarray}\label{eq:Mampere}
   (\nabla \times \mathbf B - \mu_0\epsilon_0 \partial_t \mathbf E)_i& = & \mu_0\sum_{n=0}^\infty \frac{(-1)^n}{n!}\epsilon_{ijk}\partial_j\partial_{j_1}\partial_{j_2}\dots \partial_{j_n} \mathcal{M}^k_{j_1,j_2,\dots,j_n} \\\nonumber
    &=& \epsilon_{ijk}\mu_0\partial_j \left( M_{\rm loc}^k - \partial_l\mathcal{Q}^k_l + \frac{1}{2} \partial_{l}\partial_m \mathcal{O}^k_{lm}  + \dots \right)\\\nonumber
    &\equiv&\mu_0\left(\nabla \times \mathbf M \right)_i
\end{eqnarray}
\end{widetext}
where the terms $M_{\rm loc}^k - \partial_l\mathcal{Q}^k_l + \frac{1}{2} \partial_{l}\partial_m \mathcal{O}^k_{lm}  + \dots$ stand for multipole contributions to a spatially varying magnetization density $\mathbf M$, in the same way as Eq.~\eqref{eq:rhomultipole}. $\mathbf M_{\rm loc}$, $\mathcal{Q}$, $\mathcal{O}$ respectively stand for the (local) magnetic dipole, quadrupole, and octupole moments. 

To get explicit expressions of $\mathcal{M}$, we consider the case of translation symmetry, when $\boldsymbol \chi (\mathbf r, \mathbf r') = \boldsymbol \chi (\mathbf r-\mathbf r')$. Then for a multipole moment of spatial order $n$ ($n$ spatial indices, which are omitted below for brevity),
\begin{eqnarray}\label{eq:Mfromchi}
 \mathcal{M}^i&=&  -\int d^3\mathbf r' \chi^i(\mathbf r')(-\mathbf r')^n \\\nonumber
        &=&  -\int d^3\mathbf r' \int \frac{d^3\mathbf q}{(2\pi)^3}  \left[(-i\nabla_{\mathbf q})^n \chi^i(\mathbf q) \right] e^{i\mathbf q \cdot \mathbf r'} \\\nonumber 
        &=&  -\lim_{\mathbf q\rightarrow \mathbf 0}\left[ (-i\nabla_{\mathbf q})^n\chi^i(\mathbf q)\right] 
\end{eqnarray}
Namely, it is the $n$-th order derivative of the Fourier transform of the nonlocal spin density evaluated at $\mathbf q=0$. Conversely, $\mathcal{M}$ up to order $k$ serves as a Taylor-series approximant of $\boldsymbol \chi (\mathbf q)$ near $\mathbf q=0$:
\begin{eqnarray}\label{eq:chiTaylor}
    \chi^i(\mathbf q)\approx -\sum_{n=0}^k \frac{i^n}{n!} \mathcal{M}^i_{j_1\dots j_n} q_{j_1}\dots q_{j_n}
\end{eqnarray}
Eq.~\eqref{eq:chiTaylor} is the foundation for obtaining $\mathcal{M}$ from DFT calculations later in this paper.

Explicit formula of $\boldsymbol \chi (\mathbf q)$ can be obtained from the functional derivative of free energy density with respect to Zeeman field, which for non-interacting (mean-field) Hamiltonians become (Appendix~\ref{sec:chiqderivation})
\begin{widetext}
\begin{eqnarray}\label{eq:chiqKubo}
	\chi^i(\mathbf q) = \frac{g\mu_B}{2\hbar }\sum_{mn} \int \frac{d^3\mathbf k}{(2\pi)^3}\frac{(f_{n\mathbf k + \mathbf q} - f_{m \mathbf k})(\epsilon_{m\mathbf k} + \epsilon_{n \mathbf k + \mathbf q} -2\mu)}{\epsilon_{n \mathbf k + \mathbf q} - \epsilon_{m \mathbf k} -  i0^+}  \langle u_{m\mathbf k} | u_{n\mathbf k + \mathbf q}\rangle \langle u_{n \mathbf k + \mathbf q} | s^i | u_{m \mathbf k}\rangle
\end{eqnarray}
\end{widetext}
where $g$ is the $g$-factor, $f_{m\mathbf k} = f(\epsilon_{m\mathbf k})$ is the Fermi-Dirac distribution function evaluated at energy $\epsilon_{m\mathbf k}$, eigenenergy for the $m$-th band at crystal momentum $\mathbf k$; $\mu$ is the chemical potential; $|u_{m\mathbf k}\rangle$ is the periodic part of the Bloch eigenstate $|n\mathbf k\rangle$; $\mathbf s$ is the spin operator. 

Eq.~\eqref{eq:chiqKubo} originates from the perturbed density matrix by the Zeeman field. It applies to any $\mathbf q\neq \mathbf 0$ and vanishes at $\mathbf q=0$. An additional contribution to the uniform part, $\boldsymbol \chi(\mathbf q=\mathbf 0)$, comes from the perturbed free-energy operator and is simply the uniform spin (magnetization) density up to a minus sign: 
\begin{eqnarray}\label{eq:chiqdipole}
    \boldsymbol \chi^i(\mathbf 0) = -\mathbf M_{\rm loc} = \frac{g\mu_B}{\hbar }\sum_{n} \int \frac{d^3\mathbf k}{(2\pi)^3}f_{n\mathbf k} \langle u_{n \mathbf k} | s^i | u_{n \mathbf k}\rangle
\end{eqnarray}

Clearly, for translation-invariant systems $\mathcal{M}$ becomes position independent and drops out of the Maxwell equations Eq.~\eqref{eq:Mampere}. In the next subsection we discuss how they can still be relevant to local experimental observables, particularly the spin density.

\subsection{Relation between SM$^3$ and local spin densities}\label{ssec:SM3-localspin}

Since $\mathcal{M}$ only manifests in Maxwell equations when the system is not translationally invariant, measuring $\mathcal{M}$ requires breaking the translation symmetry in certain way. An obvious choice is the boundary. Assuming $\mathcal{M}$ to be constant inside the system and vanishing outside, characteristic distributions of spin density at the boundary can be used to infer values of $\mathcal{M}$ according to Eq.~\eqref{eq:Mampere}. For example, $\mathcal{Q}$ leads to surface spin densities, while $\mathcal{O}_{lm}$ with $l\neq m$ corresponds to edge spin densities. However, the closer to the boundary, the more different $\boldsymbol \chi(\mathbf r, \mathbf r')$ becomes from the bulk $\boldsymbol \chi(\mathbf r-\mathbf r')$, meaning the local $ \mathcal{M}(\mathbf r)$ can be very different from that in the bulk. Such a subtlety on the relation between multipole moments and surface spins has been appreciated in the literature \cite{Andreev1996, Belashchenko2010,chen2019spin,Spaldin2021a}, though not in the language of $\boldsymbol \chi$. Alternatively, one may treat the boundary as a perturbation that modifies $\boldsymbol \chi$ as a response. Except for ideal cases such as smooth confinement, real boundaries are rarely equivalent to a small change in the bulk Hamiltonian, which also explains the less ideal correspondence between bulk $\mathcal{M}$ and boundary spin densities.

However, the above discussion also suggests that, when the degree of translation symmetry breaking is weak, there can be a direct connection between bulk $\mathcal{M}$ and local spin densities. Below we discuss such a connection.

The nonlocal spin density $\boldsymbol \chi$ is a static susceptibility between a generalized force conjugate to the free-energy density $\hat{F}(\mathbf r)$ and the Zeeman field \cite{Qin2011,daido_2020}. As first proposed by Luttinger \cite{Luttinger1964} to study thermal transport driven by a temperature gradient, which is not a mechanical force, in the Kubo linear response framework, one can represent the effect of temperature by a dimensionless scalar field $\psi(\mathbf r)$ that can be loosely understood as a gravitational field. Although the idea is motivated by the Tolman-Ehrenfest relation \cite{Tolman1930,Tolman1930a,Klein1949} demonstrated for classical gasses and fluids, Luttinger made the equivalence more rigorous in electronic systems by showing that, if the Hamiltonian is perturbed by
\begin{eqnarray}
    \delta \hat{H} = \int d^3\mathbf r\left[ \hat{N}(\mathbf r) \phi(\mathbf r) + \hat{H}(\mathbf r) \psi(\mathbf r)\right]
\end{eqnarray}
$\beta \equiv 1/(k_B T)$ and $\beta\mu$ will have to be adjusted as
\begin{eqnarray}\label{eq:phipsiLuttinger}
\phi + \frac{1}{\beta}\delta (\beta \mu) = 0,\quad \psi - \frac{1}{\beta}\delta \beta = 0
\end{eqnarray}
in order for the system to stay in equilibrium. In other words, the current driven by $-\beta^{-1}\delta (\beta \mu)$ is compensated by that driven by $\phi$ through the same response function, and that driven by $-\beta^{-1}\delta\beta  = -\delta \ln \beta$ is compensated by that driven by $\psi$ through the same response function. Thus the response to $\delta \ln \beta$ can be obtained by that to $\psi$, and the response to $\beta^{-1}\delta(\beta \mu)$ can be obtained by that to $\phi$. Note that the second equation in Eq.~\eqref{eq:phipsiLuttinger} is essentially the Tolman-Ehrenfest relation $\psi = \delta \ln \beta = - \delta \ln T$. The first one is
\begin{eqnarray}
    \phi + \mu \delta \ln \beta + \delta \mu = \phi +\mu\psi +\delta \mu = 0
\end{eqnarray}
In the canonical ensemble considered in \cite{Luttinger1964}, current driven by $\mu \delta \ln \beta + \delta \mu$ is only compensated by that by $\phi$. In other words, had $\phi = 0$, $\psi$ alone would not create another current in the ``number channel" due to $\mu\delta \ln \beta$. This means that any current driven by $\mu\psi$ in the number channel through the same response function as that to $\phi$ is compensated by the current driven by $\delta \mu$. In a grand canonical ensemble, $\mu$ is fixed. Therefore the current created by $\mu\psi$ in the number channel cannot be compensated without $\phi$. Therefore $\phi^\psi = -\mu\psi$ must be present to cancel such a current. In other words, to ensure that the equivalence Eq.~\eqref{eq:phipsiLuttinger} still holds in the grand canonical ensemble, $\psi$ must be introduced into $\delta\hat{H}$ as
\begin{eqnarray}
    \delta \hat{H} &=& \int d^3\mathbf r\left[ \hat{N}\phi^\psi(\mathbf r) + \hat{H}(\mathbf r)\psi(\mathbf r) \right] \\\nonumber
    &=& \int d^3\mathbf r\left[ \hat{H}(\mathbf r)-\mu \hat{N}(\mathbf r)\right ]\psi(\mathbf r) = \int d^3\mathbf r \hat{F}(\mathbf r)\psi(\mathbf r)
\end{eqnarray}

The free energy operator perturbed by both $\psi$ and a Zeeman field is
\begin{eqnarray}
    \hat{F}\left[\psi, \mathbf B\right] = \int d^3\mathbf r [1+\psi(\mathbf r)]\left[\hat{F}_0(\mathbf r) + \frac{g\mu_B}{\hbar}\hat{\mathbf s}(\mathbf r) \cdot \mathbf B(\mathbf r) \right]
\end{eqnarray}
Following through the same procedure that leads to Eq.~\eqref{eq:Mampere}, it is straightforward to see that the $\mathbf r$-dependent $\mathcal{M}$ is 
\begin{eqnarray}\label{eq:psirM}
    \mathcal{M}(\mathbf r) = \psi(\mathbf r) \mathcal{M}
\end{eqnarray}
where the $\mathbf r$-independent $\mathcal{M}$ on the right hand side is the bulk value defined in Eq.~\eqref{eq:Mfromchi}.

The above result also makes it clear why the multipole contributions to the local spin magnetization density $\mathbf M$ generated by $\psi$ must be subtracted in order to get the true thermal-magnetization response (magnetization induced by a temperature gradient) \cite{Shitade2019}. Let the (equilibrium) magnetization density in Eq.~\eqref{eq:Mampere} $\mathbf M_{\rm eq} = \mathbf M_{\rm loc} + \widetilde{\mathbf M} =  (1+\psi)\mathbf M_0 + \widetilde{\mathbf M}$, where $\mathbf M_0$ is the magnetization density in the absence of $\psi$, and $\widetilde{\mathbf M}$ contains all contributions from higher-order multipoles due to the nontrivial gradients of $\psi$. For a translationally invariant system subject to a time-dependent $\psi$, one has
\begin{eqnarray}
    \mathbf M (\mathbf r, \mathbf t) &=& {\rm Tr}[(\rho_0 + \rho')(1+\psi)\hat{\mathbf M}]\\\nonumber
    &=&(1+\psi)\mathbf M_0  + \delta \mathbf M(\mathbf r, \mathbf t) \\\nonumber
    &=& \mathbf M_{\rm eq} +\left[\delta \mathbf M(\mathbf r, \mathbf t) - \widetilde{\mathbf M}\right]
\end{eqnarray}
where $\rho_0$ is the density matrix without $\psi$, $\rho'$ is the perturbation to $\rho$ due to $\psi$, $\mathbf M_0$ is the ground state magnetization in the absence of $\psi$, $\delta \mathbf M(\mathbf r, t)$ is the usual Kubo linear response to $\psi(\mathbf r, t)$. The nonequilibrium part is therefore the terms in the last brackets. For a linear response to $\nabla \psi \rightarrow \nabla \ln T$, one therefore needs to subtract the quadrupole contribution to $\widetilde{\mathbf M}$. 

Based on the above discussion, the gravitational field $\psi(\mathbf r)$ serves as a physical realization of the weight function $f(\mathbf r)$ in Sec.~\ref{ssec:multipole-history}. We next discuss in practice how different sources of inhomogeneity can be related to $\psi(\mathbf r)$. Consider an arbitrary source of inhomogeneity $A_i(\mathbf r)$ that may have multiple components labeled by $i$. One generally expects there to be a crossed term in the perturbed free energy
\begin{eqnarray}
    \delta F = \int d^3\mathbf r\int d^3\mathbf r' A_i(\mathbf r) \chi^A_{ij}(\mathbf r-\mathbf r') B_j(\mathbf r')
\end{eqnarray}
One can therefore define multipole moments specific to this source of inhomogeneity through the nonlocal spin density $ A_i(\mathbf r) \chi^A_{ij}(\mathbf r-\mathbf r')$. One can formally introduce a $\psi(\mathbf r)$ so that $A_i(\mathbf r) \chi^A_{ij}(\mathbf r-\mathbf r') = \psi(\mathbf r) \chi_j(\mathbf r-\mathbf r')$, but the relation between $\psi(\mathbf r)$ and $\mathbf A(\mathbf r)$ is in general undetermined. However, if the effect of $\mathbf A$ as a Hamiltonian density term $\delta \hat{H}_{A} (\mathbf r)$ can be approximated by a rescaling of the free-energy density:
\begin{eqnarray}
    \delta \hat{H}_{A}(\mathbf r) \approx \psi_{A} (\mathbf r) \hat{F}(\mathbf r)
\end{eqnarray}
one can write $A_i(\mathbf r) \chi^A_{ij}(\mathbf r-\mathbf r') \approx \psi_{A}(\mathbf r) \chi_j(\mathbf r-\mathbf r')$. Such an approximation may be applicable to strain fields. For example, a positive strain $\epsilon_{xx} = \Delta L_x/L_x$ corresponds to a volume expansion and therefore a decrease of all densities by a similar fraction. One can therefore approximate
\begin{eqnarray}
    \psi_\epsilon (\mathbf r) \approx - \bar{\epsilon} (\mathbf r)
\end{eqnarray}
where $\bar{\epsilon}$ is the isotropic part of the strain. The resulting SM$^3$ is
\begin{eqnarray}\label{eq:MapproxA}
    \mathcal{M}^\epsilon(\mathbf r)\approx - \bar{\epsilon} (\mathbf r) \mathcal{M}
\end{eqnarray}

Although Eq.~\eqref{eq:MapproxA} is a very crude approximation, it can help understand the typical size of inhomogeneity-induced spin densities in AFM. To give an example, considering a strain field of a magnitude $1\%$ and correlation length $10^2 ~\rm \AA$, we have a gradient of order $10^{-4} ~\rm \AA^{-1}$. For a material with a quadrupole moment density $\sim 10^{-2} ~\mu_B\rm \AA^{-2}$ the induced magnetization density is $\sim 10^{-6} ~ \mu_B\rm \AA^{-3}$, which is comparable to the net magnetization in canted AFM.

\subsection{SM$^3$ and spin-orbit coupling}\label{sec:SM3SOC}
As can be seen from Eq.~\eqref{eq:chiqKubo}, the existence of SM$^3$ does not require spin-orbit coupling (SOC). Indeed,  dipole moments trivially exist in ferromagnets without the need of SOC. However, in the absence of SOC, which couples point group operations in real (orbital) space and spin space, certain components of SM$^3$ may be forbidden due to the higher symmetry of the magnetic order. Consequently, such components are expected to be small when SOC is weak. The net magnetic dipole moments of canted weak ferromagnets \cite{DZYALOSHINSKII1992579, Moriya1960} is a good example. Symmetry constraints on physical properties of magnetic materials for which SOC can be treated as a perturbation have received surging interests recently in the context of altermagnets and spin space groups. In this subsection we discuss a few characteristic behavior of SM$^3$ in such a regime.

We start from the no-SOC case. Considering a mean-field Hamiltonian with the spin (magnetization) density field entering the Hamiltonian through a scalar product: $J\mathbf M(\mathbf r)\cdot \boldsymbol \sigma$, a global rotation of $\mathbf M\rightarrow \mathbf M' = R \mathbf M$, where $R$ is a rotation matrix, is equivalent to a unitary transformation: $JR\mathbf M(\mathbf r)\cdot \boldsymbol \sigma = J\mathbf M(\mathbf r)\cdot U_R \boldsymbol \sigma U_R^\dag$. In the absence of SOC, if $U_R$ commutes with other terms of the Hamiltonian, the density matrix satisfies $\rho[R\mathbf M] = U_R \rho[\mathbf M] U_R^\dag$. Thus for any physical observable $\hat{O}$ that commutes with $U_R$, we have $O[R\mathbf M] = {\rm Tr}(\rho[R\mathbf M] \hat{O}) = O[\mathbf M]$. The free-energy density functional $F_{\mathbf r}[\mathbf M]$ (we use $\mathbf r$ in subscripts to avoid confusing it with arguments of a normal function) depending on the spin magnetization field $\mathbf M(\mathbf r)$ is such a quantity. Namely, $F_{\mathbf r}[\mathbf M] = F_{\mathbf r}[R\mathbf M]$. In the presence of a Zeeman field, the relation becomes $F_{\mathbf r}[\mathbf M, \mathbf B] = F_{\mathbf r}[R\mathbf M, R \mathbf B]$ by using a similar argument as above. As a result, the nonlocal spin density satisfies
\begin{eqnarray}\label{eq:chiRtran}
    \boldsymbol \chi[\mathbf M](\mathbf r, \mathbf r') =\frac{\delta F_{\mathbf r}}{\delta \mathbf B(\mathbf r')}\bigg |_{\mathbf B = 0} =  R^{-1}\boldsymbol \chi[R\mathbf M](\mathbf r, \mathbf r')
\end{eqnarray}
Namely, rigid rotation of all ordered spins in such systems amounts to performing the same rotation on the spin index of $\boldsymbol \chi$. We thus have, for arbitrary multipole order:
\begin{eqnarray}
    \mathcal{M}^i[R\mathbf M] = R_{ij}\mathcal{M}^j[\mathbf M]
\end{eqnarray}
where $i,j$ are spin indices.

It is worth considering the case that the translation symmetry is broken by a position-dependent spin rotation $R(\mathbf r)$ on scales much larger than the magnetic unit cell, relevant to domain walls or other smooth magnetic textures at temperatures much lower than the transition temperature for the uniform order.
If 
\begin{eqnarray}\label{eq:localRapprox}
    F_{\mathbf r}[R\mathbf M_0]\approx F_{\mathbf r}[R_{\mathbf r}\mathbf M_0]
\end{eqnarray}
[the subscript $\mathbf r$ here is understood as a fixed parameter, i.e. $R_{\mathbf r_0}(\mathbf r) \equiv R(\mathbf r_0)$ is a constant matrix], $\mathbf M_0$ being the unrotated magnetization field, then 
\begin{eqnarray}
    \boldsymbol \chi(\mathbf r, \mathbf r') = \frac{\delta F_{\mathbf r}[R_{\mathbf r} \mathbf M]}{\delta \mathbf B(\mathbf r')} = R(\mathbf r)\boldsymbol \chi(\mathbf r - \mathbf r').
\end{eqnarray}
Namely, $R(\mathbf r)$ now plays the role of $\psi(\mathbf r)$ in Eq.~\eqref{eq:psirM} and connects the bulk SM$^3$ for translationally invariant systems to the spatially varying values. The local spin density at the texture is
\begin{eqnarray}\label{eq:localMR}
M^{i} = R_{ij}M^j_{\rm loc} - \mathcal{M}_l^j\partial_l R_{ij} + \frac{1}{2} \mathcal{M}_{lm}^j \partial_l\partial_m R_{ij} + \dots
\end{eqnarray}
Eq.~\eqref{eq:localMR} makes it possible to use local magnetic probes near magnetic textures to infer the values of bulk multipole moments.

The error in the approximation Eq.~\eqref{eq:localRapprox} can be estimated from the functional Taylor expansion up to the lowest order in spatial gradients of $R$:
\begin{widetext}
\begin{eqnarray}\label{eq:FrRcorrection}
    F_{\mathbf r}[R\mathbf M_0]-F_{\mathbf r}[R_{\mathbf r}\mathbf M_0] &\approx &\int d^3\mathbf r'\frac{\delta F_{\mathbf r}[R\mathbf M_0]}{\delta R_{ij}(\mathbf r')}\bigg |_{\mathbf R = \mathbf R_{\mathbf r}} [R(\mathbf r') - R(\mathbf r)]_{ij} + O(\Delta R^2) \\\nonumber
    &\approx & \int d^3\mathbf r'\frac{\delta F_{\mathbf r}[R\mathbf M_0]}{\delta R_{ij}(\mathbf r')}\bigg |_{\mathbf R = \mathbf R_{\mathbf r}} (\mathbf r'- \mathbf r)\cdot \nabla R_{ij}(\mathbf r)\\\nonumber
    &\equiv & \int d^3\mathbf r' \boldsymbol{\mathcal{J}}_{ij}(\mathbf r'-\mathbf r)\cdot \nabla R_{ij}(\mathbf r) 
\end{eqnarray}
\end{widetext}
where $\boldsymbol{\mathcal{J}}_{ij}$ is the conjugate variable of $\nabla R_{ij}(\mathbf r)$. Since the latter can be understood as an SO(3) gauge field, $\mathcal{J}$ has the physical meaning of nonlocal spin currents. (Note that $\boldsymbol{\mathcal{J}}$ has time-reversal symmetry as expected for spin currents). Since equilibrium spin currents typically arise from SOC or long-wavelength spin order, and $\boldsymbol{\mathcal{J}}(\mathbf r'-\mathbf r)$ is evaluated in the translationally invariant state, we expect Eqs.~\eqref{eq:localRapprox} and \eqref{eq:localMR} to generally hold when SOC is weak. 

We next discuss constraints on SM$^3$ when SOC is negligible. For collinear AFM such as $\alpha$-$\rm Fe_2O_3$ without SOC, rotating all spins about an axis normal to the collinear ordering direction (denoted by $\hat{n}_{\rm cl}$) by $\pi$ followed by time reversal is always a symmetry, as well as rotating all spins about $\hat{n}_{\rm cl}$ \cite{Liu2022}. As a result, the direction of $\boldsymbol \chi$, according to Eq.~\eqref{eq:chiRtran}, can only be along or opposite to $\hat{n}_{\rm cl}$. Similarly, $\mathcal{M}^i$ ($i$ being the spin index) of any order is nonzero only if $\hat{n}_{\rm cl}$ has a nonzero projection on the $i$-axis, and the spin density from a magnetic texture Eq.~\eqref{eq:localMR} only has components perpendicular to $\hat{n}_{\rm cl}$.

Another nontrivial consequence of zero SOC for collinear AFM is that, according to Eq.~\eqref{eq:chiqKubo}, $\boldsymbol \chi(\mathbf q) = ( \chi^\uparrow -  \chi^\downarrow)\hat{n}_{\rm cl}$, where $\chi^\uparrow$ and $\chi^\downarrow$ are the contributions from the spin-up and spin-down bands, respectively. Up to a constant prefactor, $\chi^{\uparrow,\downarrow}$ can be understood as the spin-resolved nonlocal charge density. Since equilibrium charge multipoles can be defined in a similar manner as SM$^3$, one can formally write for collinear AFM without SOC
\begin{eqnarray}
    \mathcal{M} = (\mathcal{C}^\uparrow - \mathcal{C}^\downarrow)\hat{n}_{\rm cl}
\end{eqnarray}
where $\mathcal{C}$ stands for the charge multipole (up to a dimensional prefactor) of spin-up or spin-down electrons with the same spatial order as $\mathcal{M}$. It also follows that the spin-up and spin-down bands cannot be completely degenerate in order for $\mathcal{M}$ to be nonzero. Namely, among collinear AFM, only altermagnets can have finite SM$^3$ in the absence of SOC.

It is known that \cite{daido_2020}, due to the periodic nature of the integrand in Eq.~\eqref{eq:chiqKubo}, $\mathcal{C}$ of odd spatial order identically vanish. Consequently, for collinear AFM without SOC, $\mathcal{M}$ of odd spatial order always vanish. This applies to, for example, the quadrupole moment of $\rm Cr_2 O_3$ in the absence of SOC. However, we note that $\mathcal{M}$ calculated from Eq.~\eqref{eq:chiqKubo} for translationally invariant systems is only an intermediary for the actual $\mathcal{M}(\mathbf r)$ entering the macroscopic Maxwell equations Eq.~\eqref{eq:Mampere}. The odd-spatial-order $\mathcal{M}(\mathbf r)$ in such systems must be evaluated in the presence of the inhomogeneity. Effectively, doing so amounts to considering nonlinear responses to the inhomogeneity, with the resulting translationally invariant coefficient corresponding to the response of nonlocal spin or charge densities to gradients, similar to Eq.~\eqref{eq:FrRcorrection}. We will discuss this point in more detail in Sec.~\ref{sec:discussion}.

For coplanar but noncollinear AFM structures such as $\rm Mn_3Sn$, rotating all spins about the normal direction of the ordering plane (denoted by $\hat{n}_{\rm cp}$) by $\pi$ followed by time reversal is a symmetry. As a result, $\boldsymbol \chi$ must be perpendicular to $\hat{n}_{\rm cp}$, so is the spin component of $\mathcal{M}$ at all orders. Consequently, magnetic textures formed by rotations within the ordering plane do not produce any local spins along $\hat{n}_{\rm cp}$. 

A finite SOC changes the above conclusions in the following ways: (1) more independent components appear in SM$^3$, due to the typically lower symmetry of magnetic point group than that of spin point group of a given system; (2) Eq.~\eqref{eq:chiRtran} requires corrections in powers of the SOC strength, making $\boldsymbol \chi$ and the spin components of $\mathcal{M}$ not simply follow the global rotations on $\mathbf M(\mathbf r)$; (3) For spatially varying rotations, additional corrections due to Eq.~\eqref{eq:FrRcorrection} and higher-order terms become significant, making the local spin densities not trivially related to the bulk SM$^3$ through Eq.~\eqref{eq:localMR}. 

If considering translationally invariant systems only, points (1) and (2) above can be examined in more detail by a formal expansion of $\boldsymbol \chi$ or $\mathcal{M}$ in powers of SOC \cite{Liu2025, Liu2026}. More specifically, the expansion parameter is the linear coupling matrix between spin and orbital angular momentum $C_{Ij}$, where $I$ is the index for orbital angular momentum (time-reversal odd, inversion even), and $j$ is the spin index. Taking $\mathcal{M}$ of arbitrary order for example, leaving its spatial indices implicit, we have
\begin{eqnarray}\label{eq:MSOCexp}
    \mathcal{M}_i = \mathcal{M}^{(0)}_i + \mathcal{M}^{(1)}_{iIj} C_{Ij} + \mathcal{M}^{(2)}_{iIjKl} C_{Ij} C_{Kl} + \dots
\end{eqnarray}
The coefficients $\mathcal{M}^{(1)}_{iIj}$ etc. are then constrained by the spin point group of the given system. New components of $\mathcal{M}$ that become finite only due to SOC can be found from the above $\mathcal{M}^{(n>0)}$ with a trivial $C_{Ij} = \delta_{Ij}$. Such an analysis can also indicate the power-law dependence of the given component on the SOC strength. More discussions on general symmetry constraints for $\boldsymbol \chi$ and SM$^3$ will be given in Sec.~\ref{sec:symmetry}.

\section{Method for extracting SM$^3$ from $\chi (\mathbf q)$}\label{sec:implementation}

Although explicit formulas for general-order SM$^3$ can in principle be derived from Eq.~\eqref{eq:Mfromchi}, they become excessively complicated with increasing order (Appendix~\ref{sec:SM3explicit}) and do not offer much insight. In contrast, the discussion in the previous section has made it clear that the major role of SM$^3$ is to collectively determine the behavior of nonlocal spin density $\boldsymbol \chi$. Therefore, SM$^3$ can be obtained as coefficients of a truncated-Taylor-series approximant of $\boldsymbol \chi (\mathbf q)$ as in Eq.~\eqref{eq:chiTaylor}, by the criteria that (1) the approximant best fits the $\boldsymbol \chi (\mathbf q)$ at small $\mathbf q$ and (2) the coefficients are consistent with the symmetry of the system under study. In this section we present details on extracting SM$^3$ from $\boldsymbol \chi (\mathbf q)$ in the above scheme which can be applied to first-principles calculations.

\subsection{Symmetry constraints}\label{sec:symmetry}
We start by discussing the symmetry of $\boldsymbol \chi$ for translationally invariant systems. Consider a magnetic point group operation $O$ that transforms general tensor fields as
\begin{eqnarray}\label{eq:Ttensortran}
    O[T_{ijk\dots}(\mathbf r)] = a[O,T] e[O,T] R^O_{ia}R^O_{jb} R^O_{kc}\dots  T_{abc\dots} ((R^O)^{-1}\mathbf r)
\end{eqnarray}
where $R^O$ is a $3\times 3$ rotation matrix standing for the (proper or improper) rotation part of $O$, $a$ is the time-reversal constant and $e$ is the spatial inversion constant. The rules of determining $a[O,T]$ and $e[O,T]$ are as follows
\begin{widetext}
    \begin{eqnarray}
&&a[O,T]:~\begin{cases}
    T\ \text{is time-reversal-even} & a[O,T] = +1\\
    T\ \text{is time-reversal-odd} & \begin{cases}
        O\ \text{contains time reversal} & a[O,T] = -1\\
        O\ \text{does not contain time reversal} & a[O,T] = +1
    \end{cases}
\end{cases}\\\nonumber
&&e[O,T]:~\begin{cases}
    T\ \text{is a pseudotensor} & e[O,T] = \det(R^O)\\
    T\ \text{is a normal tensor} & e[O,T] = +1\\
\end{cases}
\end{eqnarray}

Since $\boldsymbol \chi$ is a time-reversal-odd pseudotensor, $a[O,\boldsymbol \chi] = a[O]$, $e[O,\boldsymbol \chi] = e[O] = \det(R^O)$. $\chi_j$ therefore transforms as
\begin{eqnarray}
    \chi_j(\mathbf r) \rightarrow \chi'_j(\mathbf r) = a[O]\det(R^O)(R^O\chi)_j((R^O)^{-1}\mathbf r)
\end{eqnarray}
which leads to, when $O$ is a symmetry,
\begin{eqnarray}
    \chi_j(\mathbf r) = a[O]\det(R^O)(R^O\chi)_j((R^O)^{-1}\mathbf r)
\end{eqnarray}
Fourier transforming both sides, we can get
\begin{eqnarray}\label{eq:chiqsymm}
    \chi_j(\mathbf q)  = a[O]\det(R^O)(R^O\chi)_j((R^O)^{-1}\mathbf q)
\end{eqnarray}
Namely, $\chi(\mathbf q)$ transforms in the same way as $\chi(\mathbf r)$. Additionally, since $\boldsymbol \chi(\mathbf r)$ is real, we also have
\begin{eqnarray}\label{eq:chiqcc}
\boldsymbol \chi^*(\mathbf q) = \boldsymbol \chi(-\mathbf q)
\end{eqnarray}
Taylor-expanding both sides of Eq.~\eqref{eq:chiqsymm} leads to
\begin{eqnarray}\label{eq:chiqTaylorsymm}
    \left(\chi^{(n)}_{\mathbf q=0}\right)^i_{j_1,j_2,\dots,j_n} 
    = a[O]\det(R^O) R^O_{i\alpha} R^O_{j_1\beta_1} R^O_{j_2\beta_2}\dots R^O_{j_n\beta_n}  \left(\chi^{(n)}_{\mathbf q=0}\right )^\alpha_{\beta_1,\beta_2,\dots,\beta_n}
\end{eqnarray}
\end{widetext}
where $\chi^{(n)}_{\mathbf q=0}$ is the $n$-th order $q$-derivative of $\boldsymbol \chi(\mathbf q)$ at $\mathbf q=0$. So $\chi^{(n)}_{\mathbf q=0}$ transforms as a time-reversal-odd rank-$(n+1)$ pseudotensor. Eq.~\eqref{eq:chiqcc} additionally requires
\begin{eqnarray}\label{eq:chiqTaylorcc}
     \chi^{(n)*}_{\mathbf q=0} = (-1)^{n} \chi^{(n)}_{\mathbf q=0}
\end{eqnarray}
Namely, odd orders of $\chi^{(n)}_{\mathbf q=0}$ are purely imaginary while even orders are purely real. Finally, the nature of partial derivatives leads to
\begin{eqnarray}\label{eq:chiqTaylorperm}
     \left(\chi^{(n)}_{\mathbf q=0}\right)^i_{j_1,j_2,\dots,j_n} = \left(\chi^{(n)}_{\mathbf q=0}\right)^i_{P\{j_1,j_2,\dots,j_n\}}
\end{eqnarray}
where $P\{j_1,j_2,\dots,j_n\}$ means permutation of all the spatial indices. 

Eqs.~\eqref{eq:chiqTaylorsymm}, \eqref{eq:chiqTaylorcc}, and \eqref{eq:chiqTaylorperm} are the symmetry constraints on $\chi^{(n)}_{\mathbf q=0}$, i.e., $(n+1)$-th order SM$^3$ according to Eq.~\eqref{eq:chiTaylor}. 

In the absence of SOC, the spin point group operation $O$ generally contains three parts: spatial rotation $R^O_R\in \rm O(3)$, spin rotation $R^O_S \in \rm SO(3)$, and time reversal. (To avoid confusion, here we do not absorb time reversal into $R^O_S$, which makes $R^O_S\in \rm O(3)$). Then Eq.~\eqref{eq:Ttensortran} stills holds, provided that one regards $R^{O}$ as either real- or spin-space rotations included in the given $O$, depending on the nature of the indices that they are acting on. Moreover, the $e$ factor only accounts for the spatial indices of the tensor and becomes trivial for $\boldsymbol \chi$ and SM$^3$. More specifically, if $O$ is a symmetry,
\begin{eqnarray}
    \chi_j(\mathbf r) = a[O](R_S^O\chi)_j((R_R^O)^{-1}\mathbf r)
\end{eqnarray}
and
\begin{eqnarray}\label{eq:MnoSOCTaylorsymm}
    \mathcal{M}^i_{j_1,j_2,\dots,j_n} 
    = aR^S_{i\alpha} R^R_{j_1\beta_1} R^R_{j_2\beta_2}\dots R^R_{j_n\beta_n}  \mathcal{M}^\alpha_{\beta_1,\beta_2,\dots,\beta_n}
\end{eqnarray}
where we have skipped the coefficients' dependence on $O$. Finally, the coefficients for SOC expansion Eq.~\eqref{eq:MSOCexp} satisfy (taking 1st order in the expansion for example, recovering all spatial indices)
\begin{widetext}
\begin{eqnarray}
    \left(\mathcal{M}^{(1)}_{iIk}\right)_{j_1,\dots, j_n} =  {\rm det}(R^R) a R^S_{i\alpha} R^R_{IJ} R^S_{k\gamma} R^R_{j_1\beta_1}\dots R^R_{j_n\beta_n} \left(\mathcal{M}^{(1)}_{\alpha J\gamma}\right)_{\beta_1,\dots, \beta_n}
\end{eqnarray}    
\end{widetext}

\subsection{Symmetry-constrained fitting of $ \chi (\mathbf q)$}
We next discuss how to extract SM$^3$ based on Eq.~\eqref{eq:Mfromchi}, using $\boldsymbol \chi (\mathbf q)$ calculated on a momentum space grid.

\subsubsection{Calculate $\chi(\mathbf q)$ on a grid}
Consider a general Bloch Hamiltonian $H_{\mathbf k}$, with $\mathbf k\in Z$, $Z$ being a Monkhorst-Pack mesh:
\begin{eqnarray}
    Z = \{ \mathbf k| \mathbf k = \sum_{l=1}^d \frac{n_l}{N_l} \mathbf b_l,~0\leq n_l < N_l\}.
\end{eqnarray} 
We first diagonalize all $H_{\mathbf k}$ to get eigenvalues $\epsilon_{n\mathbf k}$ and eigenvectors $|u_{n\mathbf k}\rangle$. 
\begin{eqnarray}
    H_{\mathbf k} |u_{n\mathbf k}\rangle  = \epsilon_{n\mathbf k} |u_{n\mathbf k}\rangle
\end{eqnarray}
We then create a mapping $\mathbf k\rightarrow \mathbf k' = \mathbf k + \mathbf q$, $\mathbf k'\in Z$ modulo reciprocal lattice vectors, while $\mathbf q$ only includes a subset of $Z$. For example, 
\begin{eqnarray}\label{eq:WSqmesh}
    \mathbf q \in Z_{\mathbf q} \equiv Z\cap \{\mathbf q | |\mathbf q|\leq q_{\rm cut}\}
\end{eqnarray}
where $q_{\rm cut}$ is a spherical cutoff reciprocal lattice vector. One can further reduce the number of $\mathbf q$ points using the symmetry of the system under study. Suppose $O$ is a symmetry operation of the system's magnetic point group $G$, we have
\begin{eqnarray}
    \mathbf q' = R^O\mathbf q \in Z_{\mathbf q}
\end{eqnarray}
which splits $Z_{\mathbf q}$ into distinct orbits defined by 
\begin{eqnarray}
    G\cdot \mathbf q = \{ R^O\mathbf q, O\in G\}
\end{eqnarray}
where for each $\mathbf q'\in G\cdot \mathbf q$ we have
\begin{eqnarray}\label{eq:chiqporbit}
    \chi^i(\mathbf q') = a[O]\det(R^O)[R^O\chi^i(\mathbf q)]
\end{eqnarray}
Therefore, only one representative $\mathbf q$ needs to be picked for each $G\cdot \mathbf q$. The number of $\mathbf q$ points needed is then equal to the number of orbits in $Z_{\mathbf q}$. We denote the subset of $Z_{\mathbf q}$ formed by picking one $\mathbf q$ in each orbit by $Z'_{\mathbf q}$.

For each $\mathbf q \in Z'_{\mathbf q}$, we find $\mathbf k+\mathbf q$ modulo reciprocal lattice vectors through the map defined above, and calculate the following elements of the multi-dimensional arrays $M_{mn\mathbf k}^{\mathbf q}, \mathbf S_{mn\mathbf k}^{\mathbf q}$
\begin{eqnarray}\label{eq:MSmats}
   M_{mn\mathbf k}^{\mathbf q} = \langle u_{m\mathbf k}| u_{n\mathbf k+\mathbf q}\rangle\\\nonumber
   \mathbf S_{mn\mathbf k}^{\mathbf q} = \langle u_{m\mathbf k}| \mathbf s|u_{n\mathbf k+\mathbf q}\rangle
\end{eqnarray}
Then we can calculate $\boldsymbol \chi (\mathbf q)$ according to Eq.~\eqref{eq:chiqKubo}, which can be stored in a $3\times N_{\mathbf q}$ array.

Note that Eq.~\eqref{eq:chiqKubo} does not contain the dipole contribution Eq.~\eqref{eq:chiqdipole} and vanishes exactly at $\mathbf q = 0$. The dipole order contribution $\mathbf M_{\rm loc}$ is always determined by Eq.~\eqref{eq:chiqdipole}.

\subsubsection{Fit $\chi(\mathbf q)$ to Taylor-series approximant}

The next step is to fit $\boldsymbol \chi_{\mathbf q}$ with the following polynomial
\begin{eqnarray}\label{eq:chiqfit}
    \chi_{\mathbf q}^i \approx -\sum_{n=0}^{n_{\rm max}} \frac{i^n}{n!} \mathcal{M}^i_{j_1, j_2, \dots, j_n} q_{j_1}q_{j_2}\dots q_{j_n}
\end{eqnarray}
where $n_{max}+1$ is the highest multipole order for the fitting. The number of unknowns before applying symmetry constraints is $N_{\mathcal{M}} = \sum_{n=0}^{n_{\rm max}}3^{n_{\rm max}+1}$. Using Eq.~\eqref{eq:chiqTaylorperm} reduces it to
\begin{eqnarray}\label{eq:NMbeforesym}
    N_{\mathcal{M}} &=& 3\sum_{n=0}^{n^{\rm max}} \sum_{a_n=0}^{n}\sum_{b_n = 0}^{n-a_n}\sum_{c_n = 0}^{n-a_n-b_n} 1\\\nonumber
    &=& 3 (1 + 3 +  6 + 10 + \dots)
\end{eqnarray}
Namely, there are at most 3 dipole, 9 quadrupole, 18 octupole, and 30 hexadecapole components. Moreover, the dipole moment is uniquely fixed by $\boldsymbol \chi_{\mathbf q=0}$ as mentioned above. To further reduce the number of unknowns we need to use Eq.~\eqref{eq:chiqTaylorsymm} for magnetic point group or Eq.~\eqref{eq:MnoSOCTaylorsymm} for spin point group. In particular, when there is inversion symmetry, the quadrupole and hexadecapole vanish identically, we only need to get the up to 18 octupole moment components from the fitting. This applies to, e.g., $\rm Mn_3X$ and $\alpha$-$\rm Fe_2O_3$. In contrast, for $\rm Cr_2O_3$ the quadrupole moment is the lowest order contribution.  

Here we follow \cite{Zhu2025} to get the symmetry-reduced $\mathcal{M}$. From Eq.~\eqref{eq:chiqTaylorsymm} we have
\begin{eqnarray}
     \mathcal{M}^i_{j_1,\dots,j_n} 
    = a[O]\det(R^O)R^O_{i\alpha} R^O_{j_1\beta_1} \dots R^O_{j_n\beta_n}  \mathcal{M}^\alpha_{\beta_1,\dots,\beta_n}
\end{eqnarray}
which can be viewed as an eigenequation
\begin{eqnarray}
     L_O \mathcal{M}_{n+1} =\mathcal{M}_{n+1}
\end{eqnarray}
where 
\begin{eqnarray}
    L_O \equiv a[O]\det(R^O) \left(\otimes_{l=1}^{n+1} R^O\right)
\end{eqnarray}
in the vector space of dimension $3^{n+1}$. In addition to $L_O\in G$, Eq.~\eqref{eq:chiqTaylorperm} also dictates
\begin{eqnarray}
    L_P \mathcal{M}_{n+1} =\mathcal{M}_{n+1}
\end{eqnarray}
where $L_P$ is the matrix form of permutation operations on the spatial indices of $\mathcal{M}$, which can be obtained as kronecker products of elementary representation of permutation operations on $n$ quantities and the 3D identity matrix. We then successively go through $O\in G$ and $P$, and solve for each operation the kernel or null space
\begin{eqnarray}
    \ker(L-\mathbb{I}) = \{ \mathcal{M}_{n+1}| (L- \mathbb{I}) \mathcal{M}_{n+1} = 0 \}
\end{eqnarray}
The resulting $\ker(L-\mathbb{I})$ is similar to eigenvectors and has the shape of $3^{n+1}\times N_{\ker(L-\mathbb{I})}$, where $N_{\ker(L-\mathbb{I})}$ is the number of elements in the kernel and also that of free nonzero parameters resulting from the symmetry constraint corresponding to $L$. Apparently only the generators of $G$ and $P$ need to be considered in this procedure. To see how to iterate over the set of generators, suppose after a step $i$ we have a $3^{n+1}\times N_i$ ($0<N_i\leq 3^{n+1}$) rectangular matrix $V_{i}$ defining the common null space of operations considered in prior steps. We need to find vectors $v$ that satisfies
\begin{eqnarray}
  v\in V_i\cap \ker(L_{O_{i+1}} - \mathbb{I})
\end{eqnarray}
This can be done by noticing that $v$ is generally written as 
\begin{eqnarray}
    v = V_i x 
\end{eqnarray}
where $x$ is a $N_i\times 1$ column vector. We then need to find
\begin{eqnarray}
    L_{O_{i+1}} v = v
\end{eqnarray}
or consequently
\begin{eqnarray}
    V_i^T L_{O_{i+1}} V_i x = x 
\end{eqnarray}
This is equivalent to finding 
\begin{eqnarray}
    x\in \ker(V_i^T L_{O_{i+1}}V_i -\mathbb{I}_i)
\end{eqnarray}
where $\mathbb{I}_i$ is an $N_i\times N_i$ identity matrix. $\ker(V_i^T L_{O_{i+1}}V_i -\mathbb{I}_i)$ is a $N_i\times N_{i+1}$ matrix where $N_{i+1}\leq N_i$ is the dimension of the kernel. We then have
\begin{eqnarray}
    V_{i+1} = V_{i} \ker(V_i^T L_{O_{i+1}}V_i -\mathbb{I}_i)
\end{eqnarray}
At the end of the procedure, we have the symmetry-constrained multipole
\begin{eqnarray}\label{eq:Msymmreduced}
    \mathcal{M}_{n+1} = V_{N_{\rm gen}}\cdot (c_1,c_2,\dots, c_{N_{N_{\rm gen}}})^T
\end{eqnarray}
where $N_{\rm gen}$ is the number of generators in $G$ and $P$ combined, $N_{N_{\rm gen}}$ is the dimension of the kernel after the last step of the procedure, $c_1, c_2,\dots$ are the independent parameters that fully characterize $ \mathcal{M}_{n+1}$ which will be obtained from fitting in the next step. The number of fitting parameters is therefore reduces from Eq.~\eqref{eq:NMbeforesym} to 
\begin{eqnarray}
    N_{\mathcal{M}}^{\rm sym} = N_{N_{\rm gen}}^1 + N_{N_{\rm gen}}^2 + N_{N_{\rm gen}}^3 + \dots
\end{eqnarray}
where $N_{N_{\rm gen}}^l$ means the dimension of the common kernel of all generators of $G$ and $P$ for the $l$-th order multipole. 

As the last step, we substitute Eq.~\eqref{eq:Msymmreduced} into Eq.~\eqref{eq:chiqfit}, so that the right hand side is a polynomial of $q_{x,y,z}$ up to $n^{\rm max}$ order with unknown coefficients $c_i$ as defined in Eq.~\eqref{eq:Msymmreduced}. This can be recast into a linear algebra problem
\begin{eqnarray}
    \chi = Q C + \epsilon
\end{eqnarray}
where $\chi$ is a $3\times N_{\mathbf q}$ column vector with $N_{\mathbf q}$ the dimension of $Z'_{\mathbf q}$; $C$ is a $N_{\mathcal{M}}^{\rm sym}\times 1$ column vector containing all fitting parameters; $Q$ is a $3N_{\mathbf q}\times N_{\mathcal{M}}^{\rm sym}$ matrix containing all the coefficients on the right hand side of Eq.~\eqref{eq:chiqfit} evaluated at each $\mathbf q$; $\epsilon$ is a $3N_{\mathbf q}\times 1$ column vector standing for errors. The best fitting is achieved by
\begin{eqnarray}
    C = (Q^TQ)^{-1}Q^T\chi
\end{eqnarray}
From which we get all multipole moments up to order $n+1$
\begin{eqnarray}
    \mathcal{M} = V C
\end{eqnarray}
In practice, since $\chi_{\mathbf q=0}$ is fixed explicitly by the total spin, one only needs to fit $\chi'(\mathbf q)\equiv \chi (\mathbf q) - \chi_{\mathbf q=0}$ to get $\mathcal{M}_{n+1}$ for $n>0$. Also since $\chi (\mathbf q)$ is complex, to ensure $C$ to be real one can define $\chi$ as a $3\times (N_{\mathbf q}-1) \times 2$ column vector that stores the real and imaginary parts of $\chi (\mathbf q)$ separately, so that $Q$ is purely real.

Before ending this subsection, we note that the error of the above linear regression method has two sources. The first is that in the numerical calculation of $\boldsymbol \chi(\mathbf q)$ which is approximately uncorrelated, so that the fitting error can be straightforwardly propagated to individual fitted multipole components. The second is caused by the truncation of the Taylor series and is expected to have certain systematic correlation, making the error propagation more complex. In general, we expect a relatively large fitting error in spite of converged values of $\boldsymbol \chi(\mathbf q)$ and sufficient number of $\mathbf q$ points to indicate the need to consider higher-order multipoles beyond the truncated order.

\subsection{Model example}\label{sec:model}

\begin{figure}[ht]
	\centering
	\subfloat[]{\includegraphics[width=0.38\linewidth]{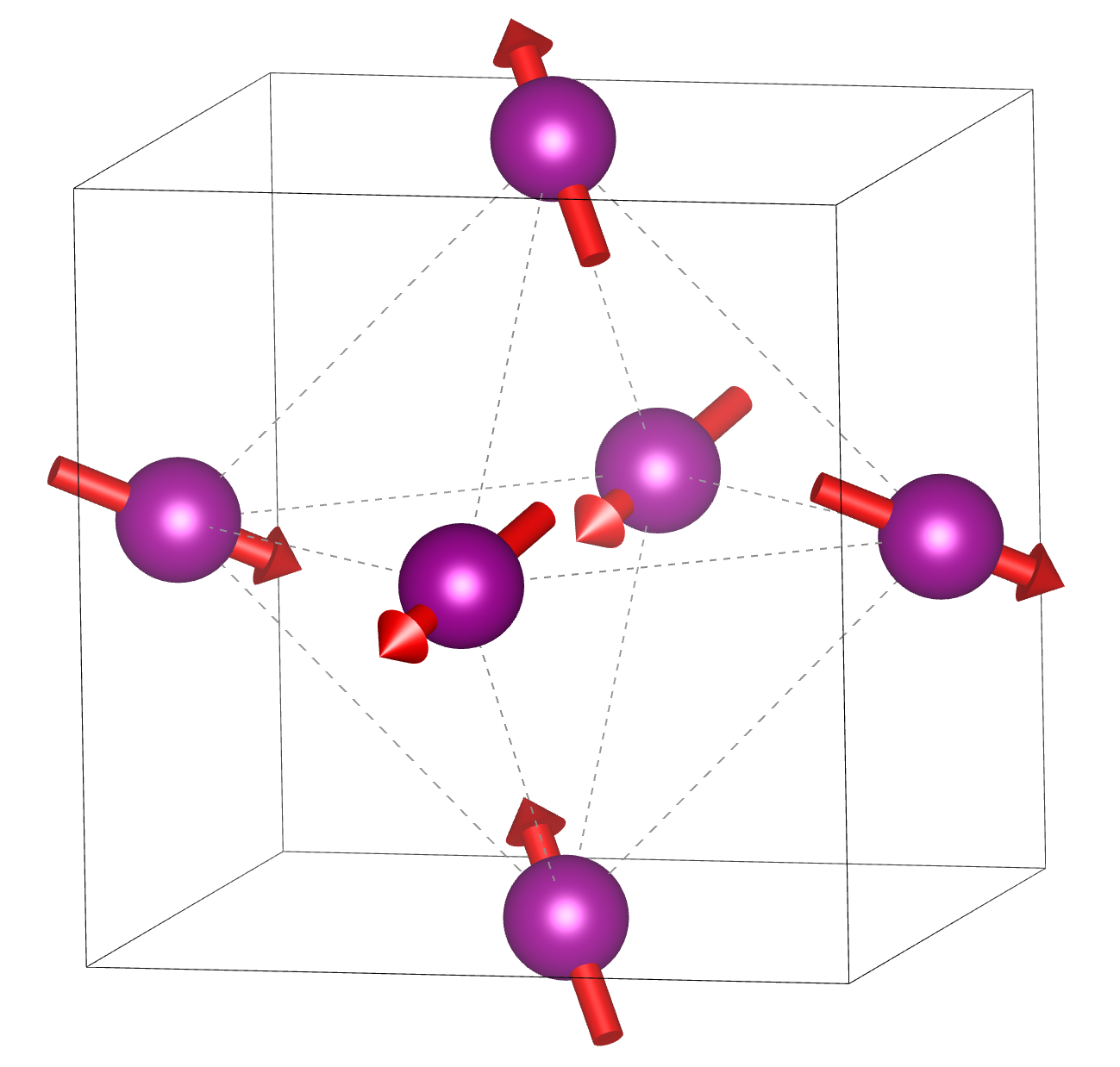}} 
        \subfloat[]{\includegraphics[width=0.62\linewidth]{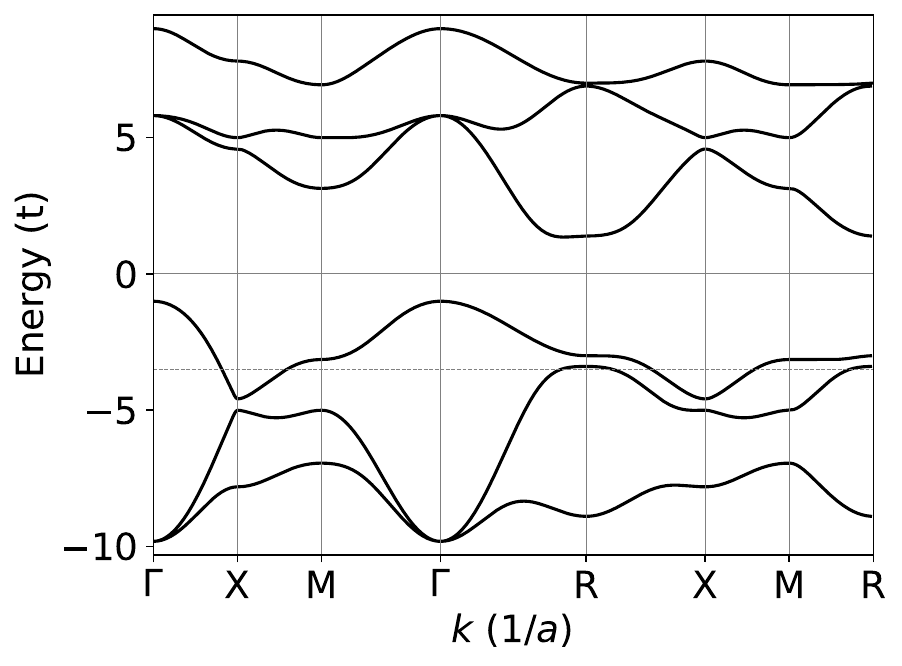}}
	\caption{(a) Crystal structure and magnetic order of the toy model. (b) Band structure of the model. $t = 1, t_{\rm SO} = 0.5, J = -5$. The horizontal solid and dashed lines represent $\mu = 0$ (insulator) and $\mu = -3.5$ (metal), respectively.}
	\label{fig:model}
\end{figure}

In this subsection we illustrate the use of the above scheme by using a toy model motivated by the structure of cubic Mn$_3X$, with a single $s$-orbital on each face-center site on the fcc lattice, as illustrated in Fig.~\ref{fig:model} (a). The model Hamiltonian is \cite{chen_2022}:
\begin{flalign}\label{eq:TBmodel}
	&H =  -t\sum_{\langle i j\rangle \alpha} c_{i\alpha}^\dag c_{j\alpha} + \imath t_{\rm so} \sum_{\langle ij\rangle \alpha\beta} (\hat{r}_{ij}\times \hat{e}_{ij})\cdot \boldsymbol{\sigma}_{\alpha\beta} c_{i\alpha}^\dag c_{j\beta} &\\\nonumber
	&- J\sum_{i\alpha\beta} \hat{n}_i \cdot \boldsymbol{\sigma}_{\alpha\beta} c_{i\alpha}^\dag c_{i\beta}&
\end{flalign}
where $i,j$ label lattice sites, $\langle \rangle $ means nearest neighbor, $\alpha,\beta$ label spin, $t >0$ is the spin-independent hopping amplitude and is chosen as the energy unit, $t_{\rm so}$ is the spin-orbit coupling strength, $\hat{r}_{ij}$ is a unit vector along the position vector $\mathbf r_j - \mathbf r_i$, $\hat{e}_{ij}$ is a unit vector standing for the electric field or electric dipole moment direction at the center of the nearest-neighbor $ij$ bond, $J$ is the strength of a local exchange field along $\hat{n}_i$, denoting the vector directions in Fig.~\ref{fig:model} (a).

The magnetic space group of the model is R-$\rm 3m'$. The corresponding magnetic point group is -$\rm 3m'$ whose generators can be chosen as a 3-fold improper rotation about the [111] direction normal to all the local spins in Fig.~\ref{fig:model} (a), and a mirror plane defined by any pair of parallel spins in the figure followed by a time reversal. This symmetry forbids odd-spatial-order magnetic multipole moments, such as quadrupole and hexadecapole, but allows even-spatial order moments. Since the dipole moment is not relevant to the fitting procedure, we focus on the octupole moment $\mathcal{M}^i_{jk}$, where $i$ is the spin index, and $j,k$ are the spatial indices. The 27 components of the octupole moment only depend on four free parameters due to the constraints by permutation and the -$\rm 3m'$ group, which we chose to be $\mathcal{M}^x_{xx}, \mathcal{M}^x_{xy}, \mathcal{M}^x_{yy}, \mathcal{M}^x_{yz}$. All other components can be obtained from them by permuting the indices due to the $C_3$ symmetry about $[111]$.

The band structure of the model is illustrated in Fig.~\ref{fig:model} (b). We consider two cases: (i) insulator, by setting $\mu = 0$, and (ii) metal, with $\mu = -3.5$.

\begin{figure}[ht]
	\centering
	\subfloat[]{\includegraphics[width=0.6\linewidth]{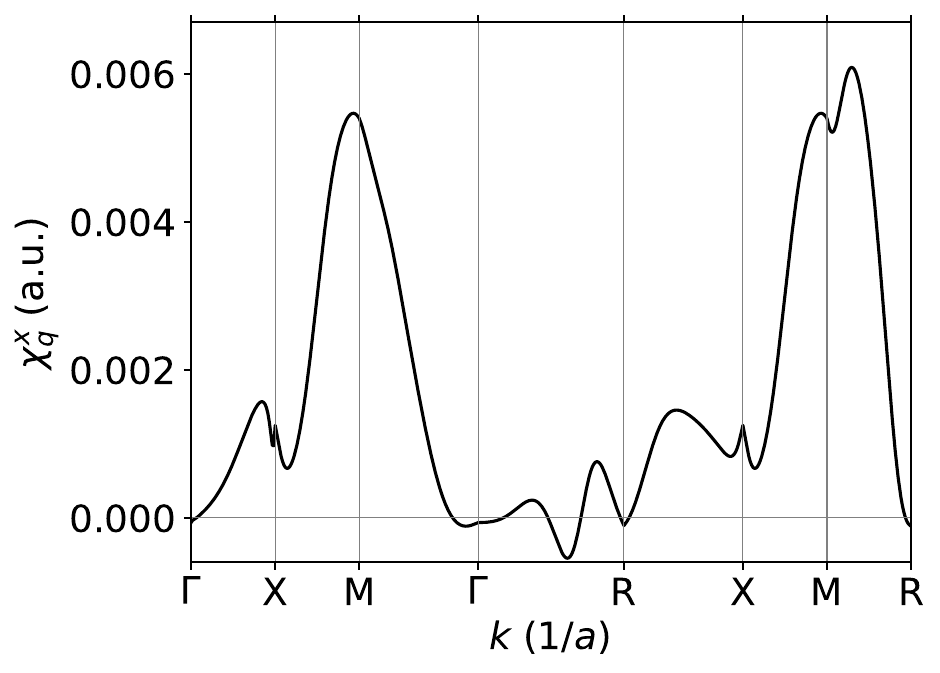}} \\
	\subfloat[]{\includegraphics[width=0.6\linewidth]{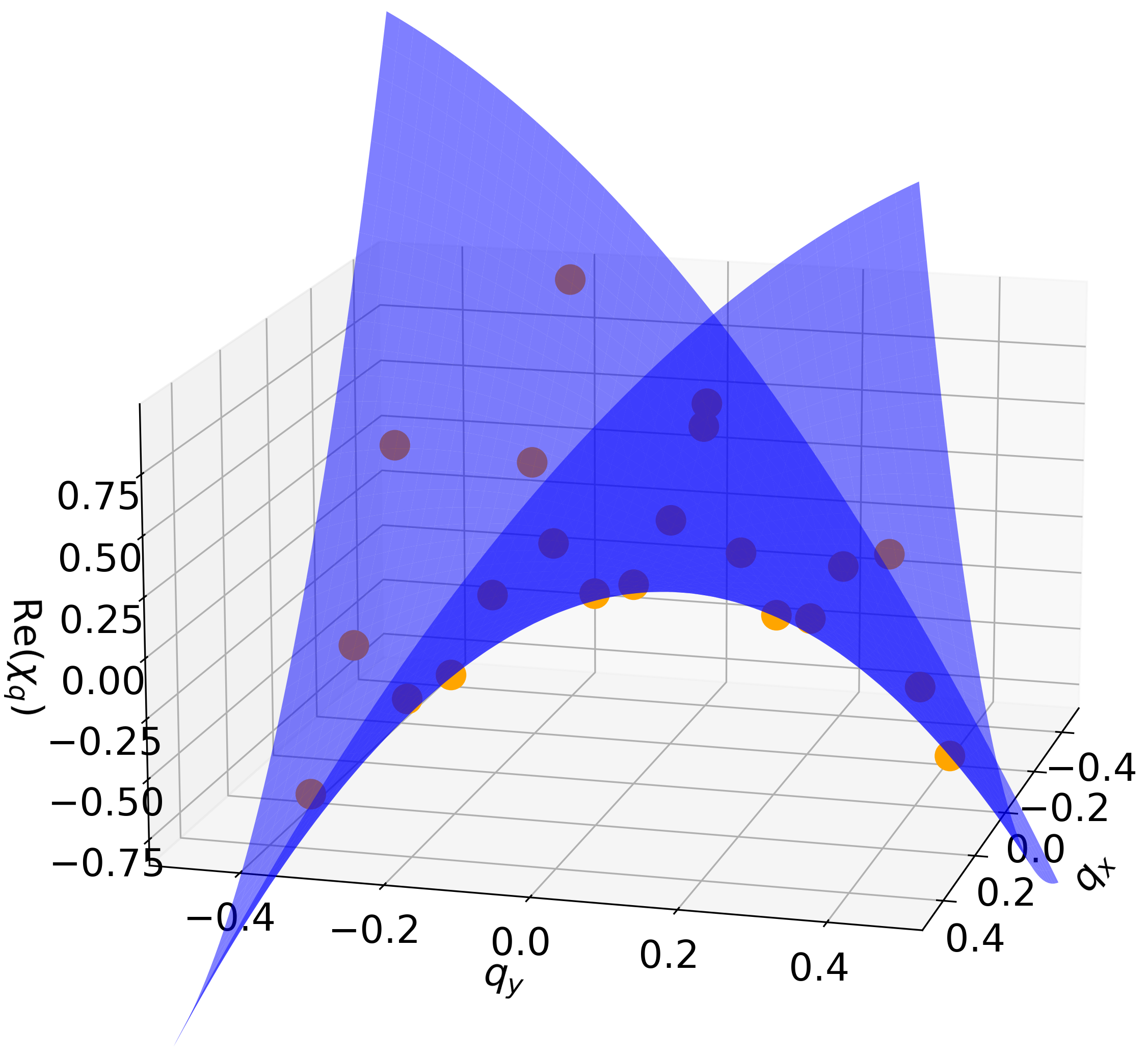}}
	\caption{(a) Integrand of $\chi^{x}(\mathbf q)$ plotted along the high-symmetry path as in Fig.~\ref{fig:model} (b) for the insulating case and an arbitrarily chosen $\mathbf q = (0.02,0.08,-0.03)$ (in units of $1/a$). (b) $\chi^x (\mathbf q)$ calculated on a grid near the Brillouin zone center (orange dots) and its approximant (blue surface) using the fitted octupole moments plotted on the $k_y = 0$ plane.}
	\label{Fig:modelinsulator}
\end{figure}

In the insulating case, the integrand of $\boldsymbol \chi (\mathbf q)$ is a smooth function over the whole Brillouin zone, since only cross-gap terms contribute to it. An example of $\boldsymbol \chi (\mathbf q)$ integrand for an arbitrarily chosen $\mathbf q$ point close to the zone center plotted along a high-symmetry line is shown in Fig.~\ref{Fig:modelinsulator} (a). As a result, $\boldsymbol \chi(\mathbf q)$ converges quickly with increasing $k$-mesh density. Table~\ref{tab:modelinsulator} lists the independent components of the octupole moment calculated using a $31\times 31\times 31$ mesh and a spherical cutoff of $q_{\rm cut} = 0.5 a^{-1}$. The results agree very well with that calculated using the explicit formula of octupole moments Eq.~\eqref{eq:octupoleallT0}. 

\begin{table}[ht]
    \centering
    \renewcommand{\arraystretch}{1.5} 
     \setlength{\tabcolsep}{6pt}
    \begin{tabular}{c | c | c | c | c}
    \hline\hline
       $\mathcal{M}^i_{jk}$  & $\mathcal{M}^x_{xx}$ & $\mathcal{M}^x_{xy}$ & $\mathcal{M}^x_{yy}$ & $\mathcal{M}^x_{yz}$ \\\hline
       Fitting   &  -4.205  & 7.251 & 5.899  & 2.853\\
       Error ($\times 10^{-3}$) & 1.67 & -0.02 & -0.41 & 0.31 \\\hline
       Explicit  &  -4.247  & 7.302 & 5.929  & 2.866 \\ \hline\hline
    \end{tabular}
    \caption{Independent components of the octupole moment for the model Eq.~\eqref{eq:TBmodel} using parameters of Fig.~\ref{fig:model} and $\mu = 0$ (insulator), in units of $10^{-3} \mu_B a^{-1}$. The first row is the result from fitting $\boldsymbol \chi (\mathbf q)$ using $q_{\rm cut} = 0.5 a^{-1}$ and a $31\times 31\times 31$ $k$-mesh. The second row is the fitting error. The third row is the result obtained by using the explicit formula Eq.~\eqref{eq:octupoleallT0} and a $k$-mesh of $21\times 21\times 21$.}
    \label{tab:modelinsulator}
\end{table}

In the metallic case, the denominator in the integrand of $\boldsymbol \chi (\mathbf q)$ can vanish near the Fermi surface, especially near $\mathbf q = 0$, which in principle makes $\boldsymbol \chi (\mathbf q)$ more difficult to converge with $k$-mesh density. However, it should be noted that $\boldsymbol \chi (\mathbf q)$ itself is a smooth function of $\mathbf q$ in the first Brillouin zone, especially in the range of $\sim 0.1 a^{-1}$, e.g., at wavelengths an order of magnitude larger than the lattice constant. (We do not consider exotic situations such as nested Fermi surfaces in this work, which deserves future investigation.) Therefore, a good fit of $\boldsymbol \chi(\mathbf q)$ near the zone center can be achieved with $\mathbf q$ points not too close to $\Gamma$, for which the integrand is reasonably smooth. 

\begin{figure}[ht]
	\centering
	\subfloat[]{\includegraphics[width=0.5\linewidth]{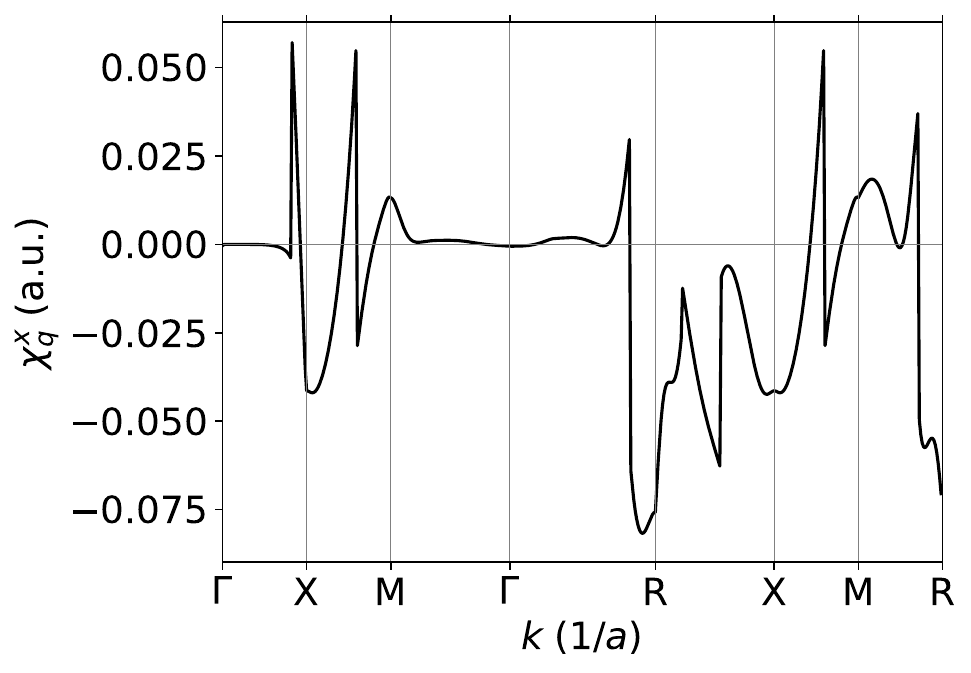}} 
    \subfloat[]{\includegraphics[width=0.5\linewidth]{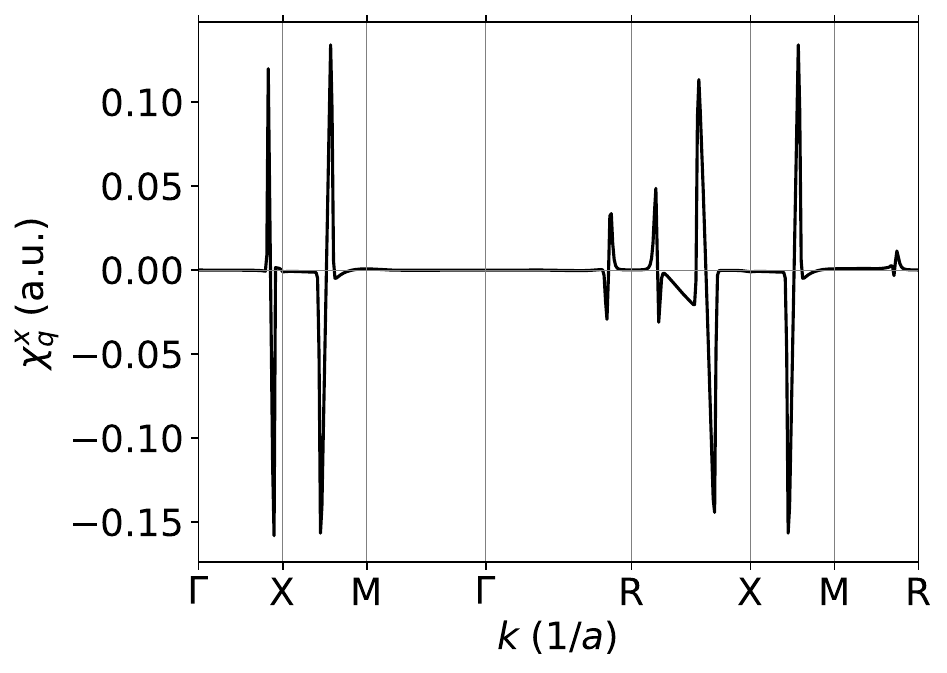}} \\
	\subfloat[]{\includegraphics[width=0.6\linewidth]{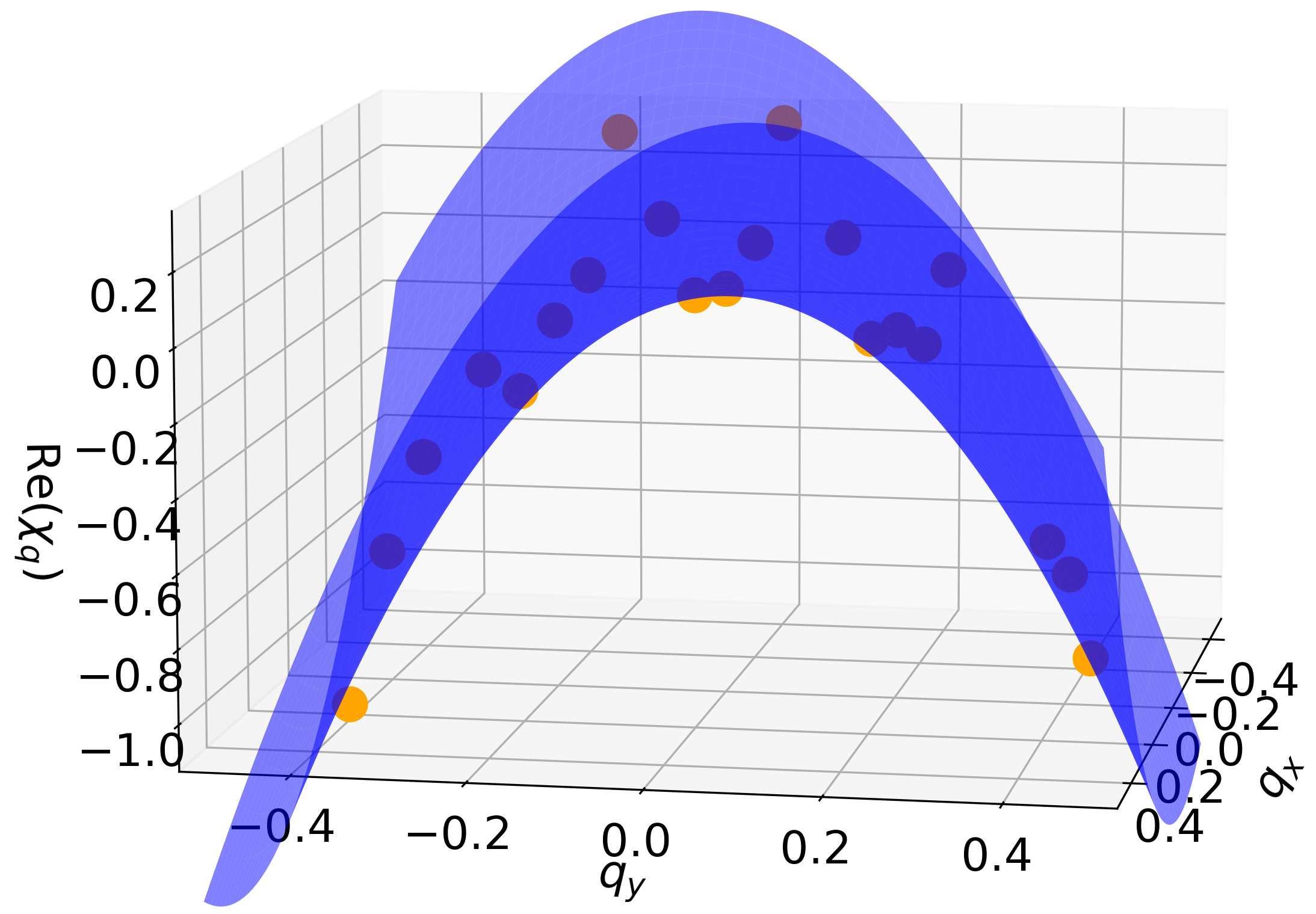}}
	\caption{(a) Integrand of $\chi^{x}(\mathbf q)$ plotted along the high-symmetry path as in Fig.~\ref{fig:model} (b) for the metallic case and a smallest nonzero $\mathbf q$ on a $31\times 31\times 31$ mesh. (b) Same as (a) but for a smallest nonzero $\mathbf q$ on a $150\times 150\times 150$ mesh and $k_B T = 0.01$. (c) $\chi^x (\mathbf q)$ calculated on the $31\times 31\times 31$ mesh near the Brillouin zone center (orange dots) and its approximant (blue surface) using the fitted octupole moments plotted on the $k_y = 0$ plane.}
	\label{Fig:modelmetal}
\end{figure}

As an explicit example, in Fig.~\ref{Fig:modelmetal} (a) we plot the integrand of $\chi^x(\mathbf q)$ for a $\mathbf q$ that is closest to $\Gamma$ on a $31\times 31\times 31$ mesh, i.e. $q\sim 0.2 a^{-1}$. The plot only shows some mild peaks along the high symmetry path. Using such a mesh and $q_{\rm cut} = 0.5 a^{-1}$, i.e., same as the insulating case above, we got the octupole moments listed in the first two rows of Table~\ref{tab:modelmetal}. The calculated vs. fitted $\boldsymbol \chi(\mathbf q)$ are shown in Fig.~\ref{Fig:modelmetal} (c). 

In comparison, Fig.~\ref{Fig:modelmetal} (b) plots the $\chi^x(\mathbf q)$ integrand for the smallest nonzero $\mathbf q$ on a $150\times 150\times 150$ mesh with a $k_B T = 0.01$ smearing. Even with the smearing, the integrand has much sharper peaks than that in Fig.~\ref{Fig:modelmetal} (a). Nonetheless, the denser $k$-mesh allows us to characterize $\boldsymbol \chi (\mathbf q)$ within a smaller $q_{\rm cut} = 0.1$. What is intriguing is that the obtained octupole moments (last two rows in Table~\ref{tab:modelmetal}) do not have much difference from that calculated using the coarse mesh. Such a behavior makes it possible to use a relatively coarse mesh to calculate SM$^3$ in real materials in Sec.~\ref{sec:results}.

\begin{table}[ht]
    \centering
    \renewcommand{\arraystretch}{1.5} 
     \setlength{\tabcolsep}{6pt}
    \begin{tabular}{c | c | c | c | c}
    \hline\hline
       $\mathcal{M}^i_{jk}$  & $\mathcal{M}^x_{xx}$ & $\mathcal{M}^x_{xy}$ & $\mathcal{M}^x_{yy}$ & $\mathcal{M}^x_{yz}$ \\\hline
       $31\times 31\times 31$   &  -10.845  & 2.526 & 4.316  & 1.280\\
       Error & 0.760 & -0.008 & -0.072 & 0.054 \\\hline
       $150\times 150\times 150$   &  -10.846  & 2.605 & 4.838  & 1.874\\
       Error & 0.554 & -0.008 & -0.042 & 0.062 \\ \hline\hline
    \end{tabular}
    \caption{Independent components of the octupole moment for the model Eq.~\eqref{eq:TBmodel} using parameters of Fig.~\ref{fig:model} and $\mu = -3.5$ (metal), in units of $10^{-3} \mu_B a^{-1}$. The first two rows are results using a $31\times 31\times 31$ mesh. The last two rows are that using a $150\times 150\times 150$ and $k_B T = 0.01$.}
    \label{tab:modelmetal}
\end{table}

\section{DFT implementation}
In this section we discuss technical details of implementing the above scheme in DFT codes. The key steps are: (1) Calculating the overlap matrices $M_{mn\mathbf k}^{\mathbf q}$ and $\mathbf S_{mn\mathbf k}^{\mathbf q}$ in Eq.~\eqref{eq:MSmats} using Kohn-Sham eigenfunctions, (2) calculating $\boldsymbol \chi(\mathbf q)$ on a set of $\mathbf q$ points close to $\Gamma$, and (3) extracting $\mathcal{M}$ by fitting. The last two steps are not unique to DFT and are already detailed in Sec.~\ref{sec:implementation}. Here we only focus on step (1) and note that the procedure is very similar to that in \texttt{pw2wannier90} included in Quantum ESPRESSO \cite{Giannozzi2009, Giannozzi2017}. 

We consider plane-wave basis and norm-conserving pseudopotentials for simplicity. The plane-wave convention is
\begin{eqnarray}\label{eq:psiQE}
\psi_{n\mathbf{k}}(\mathbf{r}) =
\frac{1}{\sqrt{\Omega}}\sum_{G,p}
C^{(n,\mathbf{k})}_{G,p}
\, e^{i(\mathbf{k}+\mathbf{G})\cdot\mathbf{r}} \chi_p,
\end{eqnarray}
where $\Omega$ is the unit cell volume, and $p$ is the spin index. It is normalized such that
\begin{eqnarray}
    \int_{\rm uc} d^3\mathbf r \psi^\dag_{m\mathbf k}(\mathbf r)\psi_{n\mathbf k'}(\mathbf r) = \delta_{mn}\delta_{\mathbf k\mathbf k'}
\end{eqnarray}
The periodic part is then
\begin{eqnarray}
    u_{n\mathbf k}(\mathbf r) &=& \frac{1}{\sqrt{\Omega}} \sum_{\mathbf G, p}C^{(n,\mathbf k)}_{\mathbf G,p} e^{i\mathbf G \cdot \mathbf r} \chi_p
\end{eqnarray}
The coefficients $C^{(n,\mathbf k)}_{\mathbf G,p}$ are therefore the $|u_{n\mathbf k}\rangle$ in the plane wave basis $|\mathbf G, p\rangle$. 
\begin{eqnarray}
    \langle \mathbf G, p| u_{n\mathbf k}\rangle = C^{(n,\mathbf k)}_{\mathbf G,p}
\end{eqnarray}

Since the $G$-list is local to each $\mathbf k$ and varies from $k$ point to $k$ point (due to the plane-wave cutoff scheme), one needs to find the intersection of $\mathcal{G}_{\mathbf k}\equiv \{\mathbf G_{\mathbf k}\}$ and $\mathcal{G}_{\mathbf k+\mathbf q}$ when calculating $M_{mn\mathbf k}^{\mathbf q}$ and $\mathbf S^{\mathbf q}_{mn\mathbf k}$. More explicitly, if $\mathbf k + \mathbf q$ is outside the $\mathbf k$ mesh so that $\mathbf k+\mathbf q - \mathbf G_{\mathbf k,\mathbf q} = \mathbf k_{\mathbf q}$, we need $|u_{n\mathbf k+\mathbf q}\rangle = |u_{n\mathbf k_{\mathbf q} + \mathbf G_{\mathbf k,\mathbf b}}\rangle$. Since 
\begin{eqnarray}\label{eq:ukGtran}
    |u_{n\mathbf k+\mathbf G}\rangle = e^{-i\mathbf G\cdot \mathbf r} |u_{n\mathbf k}\rangle
\end{eqnarray}
one can get
\begin{eqnarray}\label{eq:CkqGtran}
 C^{(n,\mathbf k+\mathbf q)}_{\mathbf G,p}  = C^{(n,\mathbf k_{\mathbf q})}_{\mathbf G + \mathbf G_{\mathbf k,\mathbf q},p}
\end{eqnarray}
Note that Eq.~\eqref{eq:ukGtran} also needs to be considered in tight-binding Hamiltonians.

To deal with the problem that $\mathcal{G}_{\mathbf k}\neq\mathcal{G}_{\mathbf k_{\mathbf q}}$ in general, we first find
\begin{eqnarray}
    \mathcal{G}_{\cap} \equiv \mathcal{G}_{\mathbf k}\cap \mathcal{G}_{\mathbf k_{\mathbf q}}
\end{eqnarray}
by looping through $\mathcal{G}_{\mathbf k_{\mathbf q}}$. Then we extract the wavefunction coefficients $C_{\mathbf k}$ and $C_{\mathbf k_{\mathbf q}}$ in $\mathcal{G}_{\cap}$. Finally,
\begin{eqnarray}
    M_{\mathbf k}^{\mathbf q} &=& C_{\mathbf k}^\dag C_{\mathbf k_{\mathbf q}}\\\nonumber
    \mathbf S_{\mathbf k}^{\mathbf q} &=& C_{\mathbf k}^\dag (\mathbb{I}_{\mathcal{G}_{\cap}} \otimes \mathbf \sigma) C_{\mathbf k_{\mathbf q}}
\end{eqnarray}
When $\mathbf G_{\mathbf k_{\mathbf q}} \neq 0$, we just need to replace
\begin{eqnarray}
    C_{\mathbf G,p}^{\mathbf k+\mathbf q} = C_{\mathbf G + \mathbf G_{\mathbf k_{\mathbf q}},p}^{\mathbf k_{\mathbf q}}
\end{eqnarray}

In reality not all eigenfunctions within the plane-wave cutoff are computed in DFT, but only those from the lowest eigenenergy up to a few bands above the Fermi energy are retained. Such a truncation leads to errors in applying Eq.~\eqref{eq:chiqKubo}, so convergence versus number of bands must be checked. In the calculations detailed in Sec.~\ref{sec:results}, we test the convergence of $\boldsymbol \chi(\mathbf q)$ versus band truncation on a coarse $k$-mesh, such as $8\times 8\times 8$. Convergence is considered reached when $|\boldsymbol \chi(\mathbf q)|$ changes less than $10^{-7}$ in atomic units ($\mu_B/a_0^3$). For metals, we found $\boldsymbol \chi(\mathbf q)$ to be more sensitive to temperature than to the imaginary broadening in the denominator based on the minimal model in Sec.~\ref{sec:model}. However, due to the behavior mentioned near the end of Sec.~\ref{sec:model}, we always keep zero temperature and use a relatively small broadening $\eta = 10^{-8}$ Hartree in our calculations. 

In addition to the SM$^3$ specific points above, our DFT calculations in Sec.~\ref{sec:results} are performed using Quantum ESPRESSO \cite{Giannozzi2009, Giannozzi2017} and fully-relativistic optimized norm-conserving Vanderbilt pseudopotentials ONCVPSP \cite{Hamann2013}. Unless noted otherwise, the self-consistent calculations use a wavefunction cutoff of $100~\rm Ry$, $10\times 10\times 10$ $k$-mesh, convergence threshold of $10^{-8}$ Ry, and with the magnetic symmetry of the calculated material enforced so that the resulting potential has the correct symmetry.

\section{SM$^3$ in some representative AFM}\label{sec:results}

\subsection{$\alpha$-Fe$_2$O$_3$}

\begin{figure}[ht]
	\centering
	\subfloat[]{\includegraphics[width=0.25\linewidth]{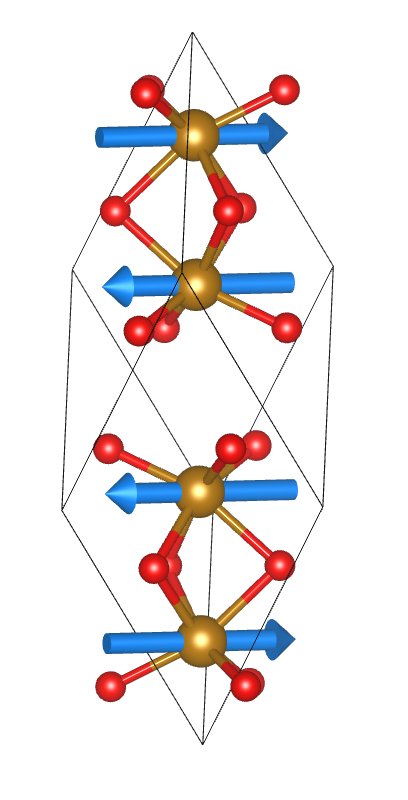}}
    \subfloat[]{\includegraphics[width=0.6\linewidth]{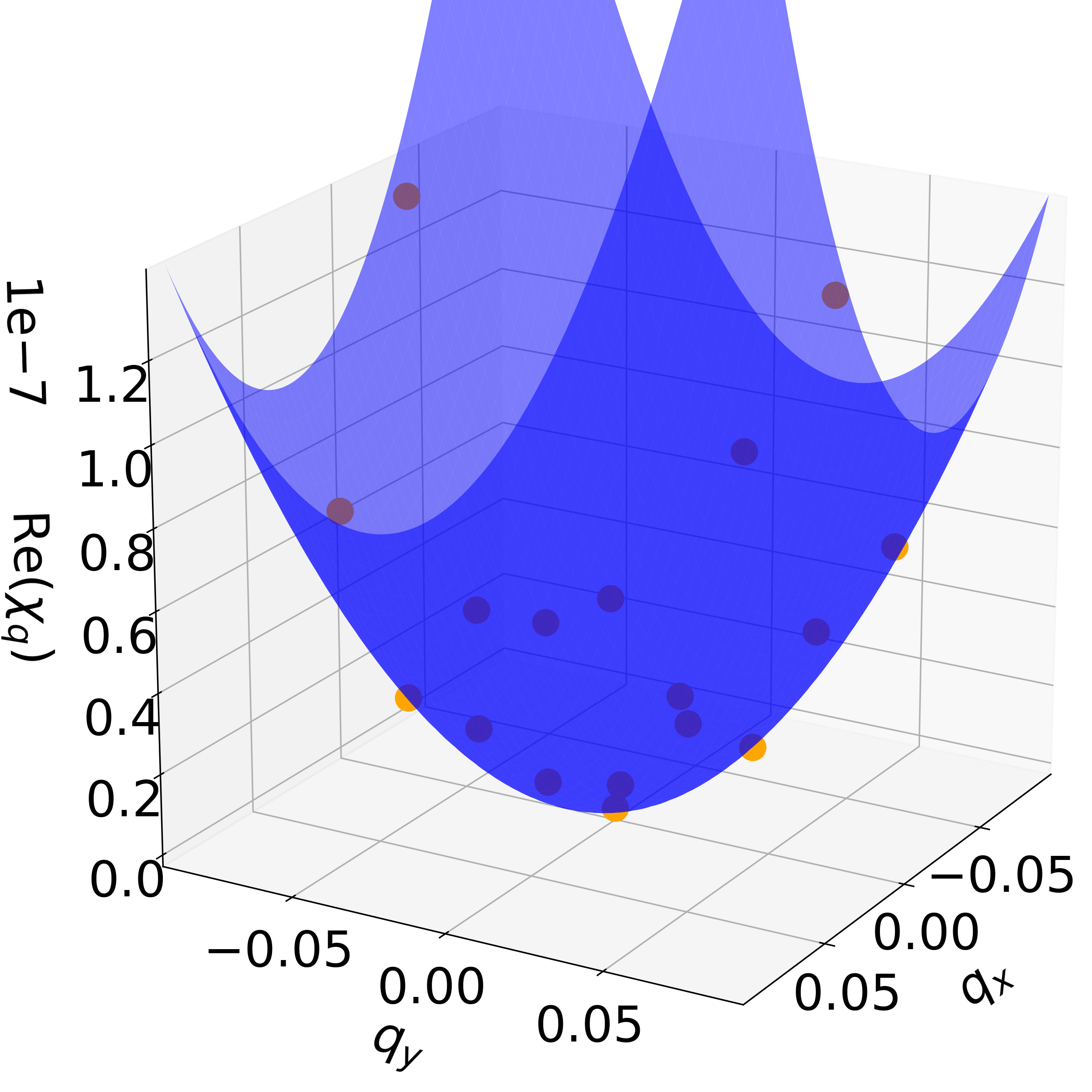}}
	\caption{(a) Structure and magnetic order of $\alpha$-$\rm Fe_2O_3$ in the canted antiferromagnet state. (b) $\chi^x (\mathbf q)$ calculated on a $20\times 20\times 20$ mesh near the Brillouin zone center (orange dots) and its approximant (blue surface) using the fitted octupole moments plotted in the $(010)$ plane. } 
	\label{fig:fe2o3}
\end{figure}

As one of the original weak ferromagnets, $\alpha$-Fe$_2$O$_3$ (hematite) directly inspired the discovery of Dzyaloshinskii-Moriya (DM) interaction \cite{DZYALOSHINSKII1992579, Moriya1960}. More recently, new developments in AFM spintronics have revived interests in hematite due to its several benefits combining AFM, nonzero net magnetization, and easy-plane anisotropy. The structure and magnetic order of hematite in its weak ferromagnetic phase is shown in Fig.~\ref{fig:fe2o3} (a). In the paramagnetic state, hematite has the space group symmetry R$\bar{3}$c (No. 167). The four Fe residing on the three-fold axis (chosen as $\hat{z}$) have their local magnetic moments arranged in a $(+--+)$ manner, where the sign is relative to the $\hat{y}$ axis that the Fe moments are nearly aligned with. Due to the DM vector pointing along $\hat{z}$ with opposite signs for the bonds between Fe$_{1,2}$ and Fe$_{3,4}$, the DM interaction-induced canting adds up to a nonzero net magnetization along $\hat{x}$.

From the symmetry perspective, the collinear antiferromagnetic order reduces the space group symmetry to C2/c.1 (No. 15.85). Its corresponding magnetic point group is 2/m.1 (No. 5.1.12), which has inversion, a 2-fold axis along $\hat{x}$, and a mirror plane perpendicular to $\hat{x}$. The inversion symmetry forbids odd-spatial-order SM$^3$ but allows even-spatial-order ones such as dipole and octupole. The only symmetry-allowed dipole component is along $\hat{x}$, while for octupole there are 8 independent components.

\begin{table}[ht]
    \centering
    \renewcommand{\arraystretch}{1.5} 
     \setlength{\tabcolsep}{4pt}
    \begin{tabular}{c | c | c | c | c | c | c | c | c }
    \hline\hline
       $\mathcal{M}^i_{jk}$  & $\mathcal{M}^y_{xz}$ & $\mathcal{M}^y_{xy}$ & $\mathcal{M}^x_{yy}$ & $\mathcal{M}^x_{xx}$ & $\mathcal{M}^x_{zz}$ & $\mathcal{M}^x_{yz}$ & $\mathcal{M}^z_{xz}$ & $\mathcal{M}^z_{xy}$ \\\hline
       Value   & 0.95  & -0.56 &  4.11 & 4.01 & 6.32  & 0.02 & 2.83 & -2.85\\
       Error  & -0.35  & -0.02 & 0.29 & -0.72 & -0.55 & 0.19 & -0.04 & -0.60 \\ \hline\hline
    \end{tabular}
    \caption{Independent components of the octupole moment for $\alpha$-$\rm Fe_2O_3$, in units of $10^{-5}~\mu_{\rm B} a_0^{-1}$ calculated using a $20\times 20\times 20$ mesh and 500 bands. Without SOC, only $\mathcal{M}^y_{xz}$ and $\mathcal{M}^y_{xy}$ are nonzero.} 
    \label{tab:Fe2O3}
\end{table}

The symmetry-allowed octupole moments calculated using a $20\times 20\times 20$ $k$-mesh, 500 bands, and $q_{\rm cut} = 0.09~a_0^{-1}$ (20 irreducible $q$ points within the cutoff) are listed in Table~\ref{tab:Fe2O3}. It is interesting to see that the largest components ($\sim 10^{-4}~\mu_B\rm \AA^{-1}$) are those due to SOC (spin index different from $y$ where the local moments are along). Since the unit cell volume of $\alpha$-$\rm Fe_2O_3$ is about $100~\rm \AA^3$, the octupole moment per unit cell is on the order of $10^{-2} \mu_B\rm \AA^2$. This is 2-3 orders of magnitude smaller than that naively estimated using the geometric arrangement of point-like Fe dipole moments within the unit cell, which is not surprising due to the SOC origin of such components. 

One thing worth noticing is that, although the arrangement of the four Fe moments illustrated in Fig.~\ref{fig:fe2o3} (a) appears to have an octupole component of $\mathcal{M}^y_{zz}$, the symmetry of the periodic structure forbids such a component. This is one example that one should not naively perform a classical multipole expansion using atomic dipoles within a unit cell of a periodic crystal. On the other hand, had $\mathcal{M}^y_{zz}$ not be forbidden by symmetry, it would likely to be a dominant component. $\rm Cr_2 O_3$, for example, has a magnetic structure very similar to hematite, but with the four Co moments in the unit cell arranged as $(+-+-)$. The quadrupole moment $\mathcal{M}^y_z$ is not forbidden by symmetry. The calculated value using the naive unit-cell approach by Dzyaloshinskii \cite{DZYALOSHINSKII1992579} turns out to be of the same order of magnitude as that inferred from experimental measurements of the far field \cite{Astrov1996}, which can be viewed as evidence supporting the above hypothesis. We also comment based on the discussion in Sec.~\ref{sec:SM3SOC} that any octupole-induced local spin magnetization at magnetic domain walls of $\alpha$-Fe$_2$O$_3$ is likely to be orders of magnitude smaller than the canting-induced net dipole moment.

\subsection{Mn$_3$Sn}

Mn$_3$Sn is one of the first Mn$_3X$ anomalous Hall AFM confirmed experimentally \cite{Kubler_2014,Nakatsuji_2015}. Many of its unconventional properties have been discussed in terms of cluster octupole moments, defined through a symmetry-adapted classical multipole expansion of the spin densities within the unit cell \cite{Suzuki2017}. In this subsection we calculate its SM$^3$ from DFT.

\begin{figure}[ht]
	\centering
	\subfloat[]{\includegraphics[width=0.25\linewidth]{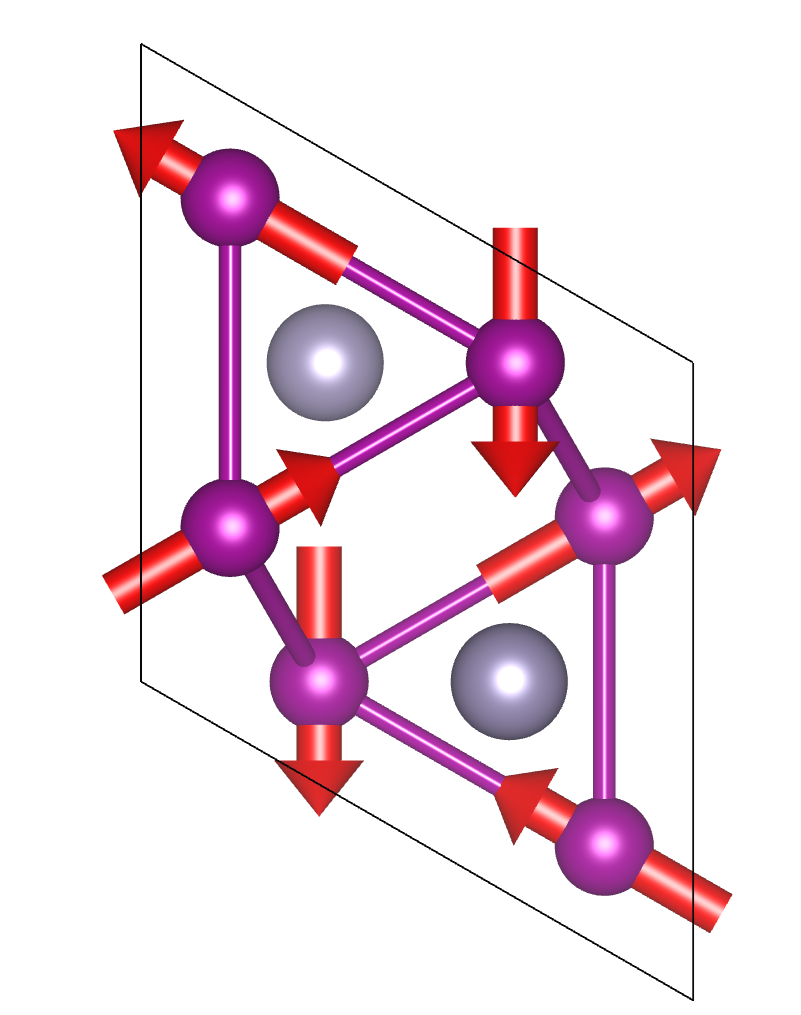}}\quad
    \subfloat[]{\includegraphics[width=0.6\linewidth]{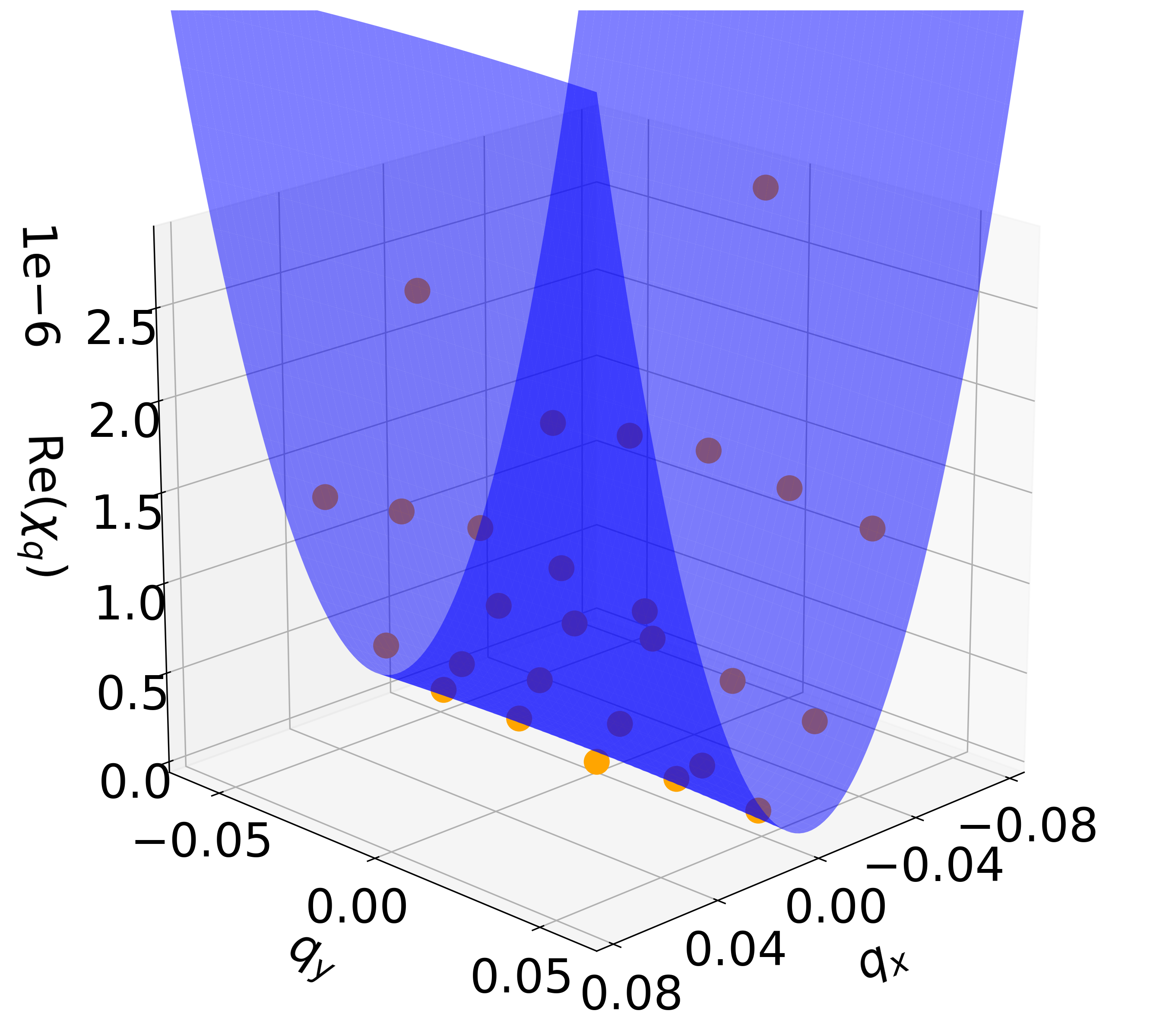}}
	\caption{(a) Structure and magnetic order of $\rm Mn_3Sn$. (b) $\chi^x (\mathbf q)$ calculated on the $30\times 30\times 30$ mesh near the Brillouin zone center (orange dots) and its approximant (blue surface) using the fitted octupole moments plotted on the $(100)$ plane. } 
	\label{fig:mn3sn}
\end{figure}

The magnetic unit cell of $\rm Mn_3Sn$ is illustrated in Fig.~\ref{fig:mn3sn} (a). It has six Mn atoms sitting in two AB-stacked (0001) kagome layers, forming a hexagonal lattice. The Mn moments in each kagome layer form an inverse triangular structure with zero net moment if not considering canting, which breaks any rotation symmetry about the [0001] direction (taken as $\hat{z}$) and allows the AHE vector as well as the weak net magnetization to have a nonzero in-plane component. More specifically, the magnetic order in Fig.~\ref{fig:mn3sn} (a) reduces the symmetry of the nonmagnetic structure from $\rm P 6_3/m m c$ (No. 194) to $\rm Cmc'm'$ (No. 63.463). The corresponding magnetic point group is $\rm m'm'm$ (No. 8.4.27), which can be generated by the following operations: (1) reflection by a mirror plane perpendicular to $\hat{y}$ [vertical direction in Fig.~\ref{fig:mn3sn} (a)]; (2) reflection by a mirror plane perpendicular to $\hat{x}$, followed by time reversal; (3) reflection by the $xy$ plane followed by time reversal. It is the last symmetry that forbids any $z$ components of the AHE vector or net magnetization. (1) and (2) allow a net magnetization along $\hat{y}$. Due to the inversion symmetry of this structure, odd-spatial-order multipoles are forbidden. We therefore focus on the octupole moment as the lowest-order nontrivial SM$^3$. 

\begin{table}[ht]
    \centering
    \renewcommand{\arraystretch}{1.5} 
     \setlength{\tabcolsep}{6pt}
    \begin{tabular}{c | c | c | c | c | c  }
    \hline\hline
       $\mathcal{M}^i_{jk}$  & $\mathcal{M}^y_{yy}$ & $\mathcal{M}^y_{xx}$ & $\mathcal{M}^y_{zz}$ & $\mathcal{M}^x_{xy}$ & $\mathcal{M}^z_{yz}$ \\\hline
       Value   & 2.484  & -2.463 & 0.032 & 2.456 &  -0.166\\
       Error  & -0.013  & 0.004 & 0.051 & -0.055 & -0.003 \\ \hline\hline
    \end{tabular}
    \caption{Independent components of the octupole moment for $\rm Mn_3Sn$, in units of $10^{-3} \mu_B a_0^{-1}$, calculated using a $30\times 30\times 30$ mesh. Without SOC, $\mathcal{M}^z_{yz}$ in the table is forbidden.} 
    \label{tab:Mn3Sn}
\end{table}

Table~\ref{tab:Mn3Sn} lists values of the symmetry-allowed octupole components as well as the fitting errors for $\rm Mn_3Sn$, calculated using a $30\times 30
\times 30$ mesh, 500 bands, and $q_{\rm cut} = 0.07~a_0^{-1}$ (29 irreducible $q$ points within the cutoff). Remarkably, the dominant octupole components are nearly two orders of magnitude larger than that of hematite. These components all have their spin indices in the $xy$ plane, and are therefore allowed even without SOC, which only forbids $\mathcal{M}^z_{yz}$ in the table. Since the unit cell volume of $\rm Mn_3 Sn$ is about $850~ a_0^3$, the dominant components are $\sim 0.6~\mu_B\rm \AA^2$ per unit cell, which is more comparable to that naively estimated from the classical expansion using atomic dipoles in a unit cell. 

Since the dominant octupole components are not due to SOC, it is meaningful to discuss the local spin densities induced by them near magnetic domain walls. Consider a planar N\'{e}el wall defined by a $y$-dependent rotation matrix about $\hat{z}$ acting on spins $R_z(\theta)$, with $\theta = [1+\tanh(y/\lambda)]\pi/2$, where $\lambda$ is the domain wall width. The maximum value of the octupole-induced local spin density is on the order of $\mathcal{M}^{y}_{yy}/\lambda^2\sim 10^{-8}~\mu_B/\rm \AA^3$ using $\lambda\sim 10^3~\rm \AA$ \cite{Tsukamoto2025}. Such a spin density is about $10^{-4}$ times the weak magnetization of Mn$_3$Sn. Since the latter produces $\sim 10^{-3}$ T stray field detected by NV magnetometers \cite{Li2023,Tsukamoto2025}, we expect a sensitivity of $10^{-7}$ T is needed to resolve the octupole-induced spin density, which is within the limit of modern single-spin NV magnetometers \cite{Huxter2022}. It is also worth mentioning that the SOC-free $\mathcal{M}$ components rotate in the same direction as the underlying spin structure, in contrast to the opposite rotation of the canting-induced weak magnetization. Moreover, the opposite signs of $\mathcal{M}^{y}_{yy}$ and $\mathcal{M}^{y}_{xx}$ suggest that the octupole-induced local spin densities have opposite signs at $xz$- and $yz$-plane domain walls. Separately, the $\mathcal{M}^x_{xy}$ component corresponds to staggered corner spins pointing along $\pm \hat{x}$ for an (0001) film of rectangular shape \cite{Tahir2023}, with a typical size of $\sim 1~\mu_B$ for a 0.1 $\mu$m thick film. 

\subsection{Mn$_3$NiN}

\begin{figure}[ht]
	\centering
	\subfloat[]{\includegraphics[width=0.3\linewidth]{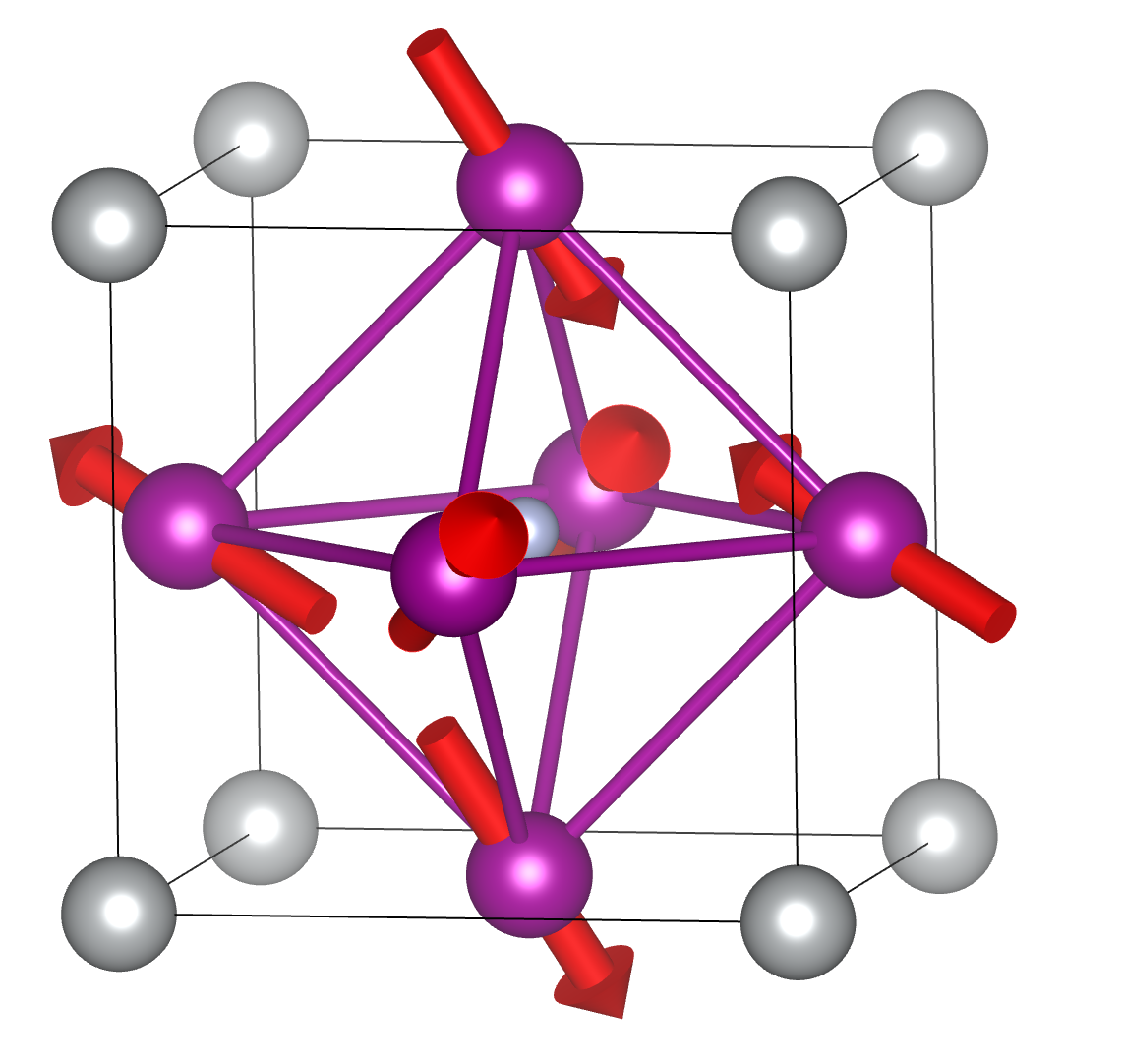}}\quad
    \subfloat[]{\includegraphics[width=0.6\linewidth]{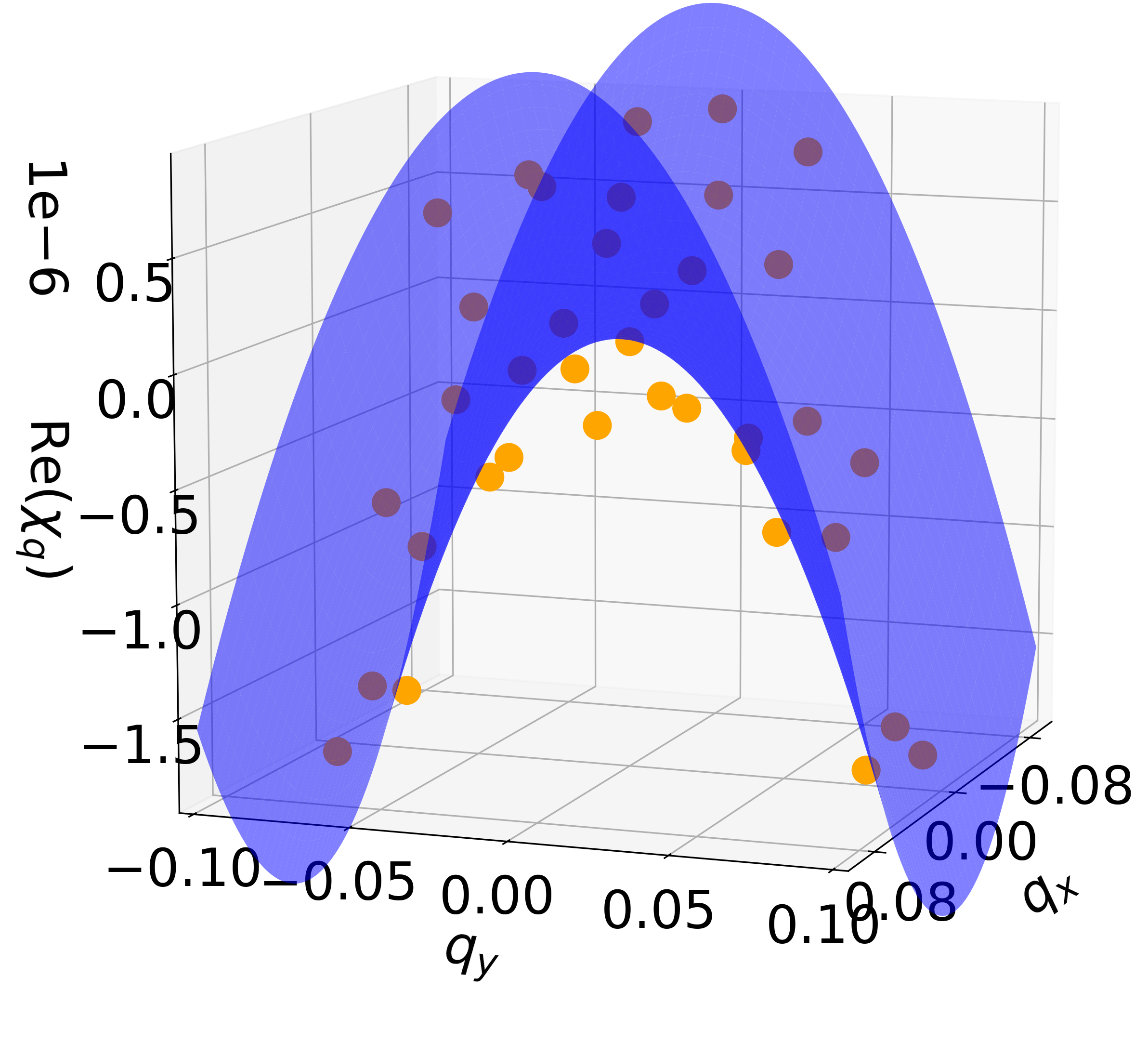}}
	\caption{(a) Structure and magnetic order of $\rm Mn_3NiN$ in the $\Gamma_{4g}$ phase. (b) $\chi^x (\mathbf q)$ calculated on the $30\times 30\times 30$ mesh near the Brillouin zone center (orange dots) and its approximant (blue surface) using the fitted octupole moments plotted on the $(100)$ plane. } 
	\label{fig:mn3nin}
\end{figure}

$\rm Mn_3NiN$ as an AFM antiperovskite nitride has received a lot of attention recently due to its many unusual properties \cite{Takenaka2005, Takenaka2014, Ding2011, Matsunami2014, Boldrin2018}, in particular the AHE \cite{Zhou2019, Gurung2019, Boldrin2019, Zhao2019}. Its magnetic structure at higher temperatures, known as the $\Gamma_{4g}$ phase, is essentially identical to that of cubic Mn$_3X$ noncollinear AFM as illustrated in Fig.~\ref{fig:mn3nin} (a). However, as temperature decreases the local Mn spins gradually rotate together about the [111] direction towards the $\Gamma_{5g}$ structure, in which each spin is parallel to the cubic cell face that the corresponding Mn is in. The $\Gamma_{5g}$ structure forbids the AHE, and it was found that Cu doping is essential to stabilize $\Gamma_{4g}$ at lower temperatures \cite{Zhao2019}. In this subsection we calculate the spin magnetic octupole of $\Gamma_{4g}$ $\rm Mn_3NiN$, since the quadrupole is also forbidden due to its inversion symmetry.

\begin{table}[ht]
    \centering
    \renewcommand{\arraystretch}{1.5} 
     \setlength{\tabcolsep}{6pt}
    \begin{tabular}{c | c | c | c | c   }
    \hline\hline
       $\mathcal{M}^i_{jk}$  & $\mathcal{M}^x_{xy}$ & $\mathcal{M}^x_{xz}$ & $\mathcal{M}^z_{xx}$ & $\mathcal{M}^z_{zz}$  \\\hline
       Value   & -3.92  & -3.76 & 0.07 & 0.33 \\
       Error  & 0.28  & 0.30 & 0.24 & 0.26 \\ \hline\hline
    \end{tabular}
    \caption{Independent octupole components for the $\Gamma_{4g}$ phase of $\rm Mn_3NiN$, in units of $10^{-4} \mu_B a_0^{-1}$, calculated using a $30\times 30\times 30$ mesh. The other nonzero components are $\mathcal{M}^y_{xx} = - \mathcal{M}^y_{yy} = \mathcal{M}^x_{xy}$, $\mathcal{M}^y_{yz} = \mathcal{M}^x_{xz}$, and $ \mathcal{M}^z_{yy} =  \mathcal{M}^z_{xx}$. Without SOC, $\mathcal{M}^z$ components are forbidden.} 
    \label{tab:Mn3NiN}
\end{table}

We consider a cubic parent structure, which makes $\rm Mn_3NiN$ have the same symmetry as the model in Sec.~\ref{sec:model}. For convenience, we choose $\hat{z}$ to be normal to the kagome planes that the Mn spins are parallel to, and $\hat{x}$ to be along a nearest Mn-Mn bond within a kagome plane. The independent octupole components calculated using $30\times 30\times 30$ $k$-mesh, 600 bands, and $q_{\rm cut} = 0.1~a_0^{-1}$ (25 irreducible $q$ points) are listed in Table~\ref{tab:Mn3NiN}. Since the magnetic structure is coplanar, the $\mathcal{M}^z$ components are nonzero only because of SOC. Indeed, such components are at least one order of magnitude smaller than the others. The size of the SOC-free components is in between that of $\alpha$-$\rm Fe_2O_3$ and $\rm Mn_3Sn$. Together, the orders of magnitude differences in the above three materials suggest that the variation of SM$^3$ across AFM materials can be significant. 

Since the dominant octupole components are not due to SOC, it is instructive to consider the local spin densities created by them due to spatial textures of coherent rotations of the Mn spins. This is particularly relevant to Mn$_3$NiN, since at low temperatures $\lesssim 100~\rm K$ the $\Gamma_{4g}$ and $\Gamma_{5g}$ states are nearly degenerate, as well as a continuum of states between them. We therefore consider a periodic rotation $R_{[111]}(\theta)$ with $\theta$ linearly changing with position. Since Mn$_3$NiN is typically grown with $[001]$ orientation, we temporarily switch to the cubic axes coordinates and use $\theta = 2\pi\frac{x}{\lambda}$. The octupole-induced spin density is found to have a constant magnitude $\sim 2\times 10^{-8}~\mu_B/\rm \AA^3$ for $\lambda = 10^3~\rm \AA$, lie in the (111) plane, and rotate together with the local Mn spins. Such a signature makes the $\Gamma_{5g}$ phase, which has strictly zero net magnetization, visible under a magnetometer. Also note that for films, the above contribution to the area spin density scales linearly with film thickness, in contrast to that due to uncompensated surface spins which is thickness independent. 

\section{Discussion and Conclusion}\label{sec:discussion}

Although we only consider spin magnetism in this work, the formalism can be straightforwardly generalized to orbital magnetism. In ordinary magnetic materials, where exchange interaction is the driving force behind spin ordering, orbital magnetism is usually perceived as a weak effect. However, we note that this does not have to be the case for multipole moments defined through the nonlocal densities. At the dipole order, the orbital magnetization in the modern theory has been shown to be comparable to the net spin magnetization in Mn$_3X$ \cite{chen_2020}, since both can be viewed as responses to the generally weak spin-orbit coupling. However, there are counterexamples that orbital magnetization can exist in the absence of spin-orbit coupling \cite{Shindou2001,Martin2008,Zhou2016} and can therefore dominate over net spin magnetization. It is therefore possible that the orbital magnetic multipole moments (OM$^3$) can be comparable to SM$^3$ if the former does not rely on spin-orbit coupling. 

Our formalism also directly applies to charge multipoles. The resulting charge multipoles have a similar meaning as SM$^3$, i.e., their spatial variation gives rise to local charge densities in the macroscopic Maxwell equations. To address the apparent issue that odd-order multipoles defined in this way identically vanish, we follow our discussion in Sec.~\ref{sec:SM3SOC} and point out that one needs to consider at least a nonlinear response for such orders. Namely,
\begin{eqnarray}
&&-\int d^3\mathbf r' \rho(\mathbf r, \mathbf r')\phi(\mathbf r')\\\nonumber
&\approx& -\int d^3\mathbf r' \psi(\mathbf r) \rho^{(1)}(\mathbf r-\mathbf r')\phi(\mathbf r') \\\nonumber
&-&\int d^3\mathbf r' \psi(\mathbf r)\int d^3\mathbf r'' \psi(\mathbf r'') \rho^{(2)}(\mathbf r-\mathbf r', \mathbf r''-\mathbf r')\phi(\mathbf r')
\end{eqnarray}
where $\rho(\mathbf r,\mathbf r')$ is the nonlocal charge density, $\phi(\mathbf r)$ is the electrostatic potential, $\rho^{(1)}(\mathbf r-\mathbf r')$ is similar to $\boldsymbol \chi(\mathbf r - \mathbf r')$ for SM$^3$, while $\rho^{(2)}(\mathbf r-\mathbf r', \mathbf r''-\mathbf r')$ is the next order in the functional derivative of the free energy with respect to $\psi$. After some algebra, one can find the polarization
\begin{flalign}
 & \mathbf P(\mathbf r) \approx \mathbf P^{(1)}(\mathbf r) + \mathbf P^{(2)}(\mathbf r)&\\\nonumber
 & \mathbf P^{(1)}(\mathbf r) =- \psi(\mathbf r) \int d^3\mathbf r' \rho^{(1)} (\mathbf r') \mathbf r' = 0&\\\nonumber
  & \mathbf P^{(2)}(\mathbf r) = -\psi(\mathbf r) \int d^3\mathbf r''\psi(\mathbf r'')\int d^3\mathbf r' \rho^{(2)}(\mathbf r' , \mathbf r''-\mathbf r + \mathbf r')  \mathbf r'    &
\end{flalign}
The nonzero contribution, $\mathbf P^{(2)}$, involves an additional integral over $\psi$. This expression has the same spirit as the definition of polarization as an integral of the adiabatic current, in that the polarization is introduced through its well-behaved derivative with respect to an arbitrary parameter. Also note that $\psi(\mathbf r'')$ can be replaced by any other perturbation, with $\rho^{(2)}$ replaced by the corresponding nonlinear susceptibility. The same treatment can be generalized to any odd-order charge multipoles, or odd-spatial-order SM$^3$ in the absence of SOC. 

Within spin density functional theory (SDFT), the SM$^3$ in this work should also receive self-consistent-field corrections \cite{chen_2020}. Namely, besides the bare Zeeman field $\mathbf B(\mathbf r)$, the mean-field Hamiltonian is also perturbed by an exchange field due to the perturbed spin density responding to the Zeeman field. 
\begin{eqnarray}\label{eq:dFSDFT}
    \delta F(\mathbf q) &=& \boldsymbol \chi(\mathbf q) \cdot [\mathbf B(\mathbf q) + \mathbf B_{\rm ind}(\mathbf q)] \\\nonumber
    &\approx& \boldsymbol \chi(\mathbf q) \cdot [\mathbf B(\mathbf q) + J\mathbf S_{\rm ind}(\mathbf q)]
\end{eqnarray}
where the induced Zeeman field is assumed to be proportional to the perturbed spin density $\mathbf S_{\rm ind}(\mathbf q)$ throught a constant factor $J$ for simplicity. In general the two are related through a $\mathbf q$-dependent kernel. When self-consistency is reached, we expect
\begin{eqnarray}
    \mathbf S_{\rm ind}(\mathbf q) = \overleftrightarrow{\chi}_s(\mathbf q) \cdot [\mathbf B(\mathbf q) + \mathbf B_{\rm ind}(\mathbf q)]
\end{eqnarray}
where $\overleftrightarrow{\chi}_s(\mathbf q)$ is the spin magnetic susceptibility calculated using the unperturbed Kohn-Sham Hamiltonian. We then have
\begin{eqnarray}
  \mathbf B_{\rm ind}(\mathbf q) = J\left[1- J\overleftrightarrow{\chi}_s(\mathbf q)\right]^{-1}\cdot \overleftrightarrow{\chi}_s(\mathbf q) \cdot \mathbf B(\mathbf q)
\end{eqnarray}
Substituting this into Eq.~\eqref{eq:dFSDFT}, we get the complete nonlocal spin density within SDFT:
\begin{eqnarray}
    \boldsymbol \chi^{\rm SDFT}(\mathbf q) = \boldsymbol \chi(\mathbf q) \cdot \left[1 + J\left[1- J\overleftrightarrow{\chi}_s(\mathbf q)\right]^{-1}\cdot \overleftrightarrow{\chi}_s(\mathbf q)\right]
\end{eqnarray}

Multipole expansion generally requires localized objects. In periodic crystals the normal charge and spin densities are unbounded in space, inherently plaguing any multipole expansions directly based on them. In contrast, susceptibilities or response functions always have a sense of locality or ``nearsightedness" that makes them friendly objects for multipole expansion. The nonlocal spin density focused on in this work is one such example. In this context, experimental measurements of SM$^3$ can be broadly regarded as probing the $\mathbf q$ dependence of crossed susceptibilities between the Zeeman field and other perturbing fields $A$, as discussed at the end of \ref{ssec:SM3-localspin}. The manifestation of the multipole moments will be local spin magnetization induced by a nonuniform $A$. We expect our work to inspire systematic experimental and computational characterization, comparison, and verification of magnetic multipole moments as an intrinsic quantity across antiferromagnetic materials, and to foster new technologically relevant advances based on this concept.

\begin{acknowledgements} 

HC thanks Qian Niu and Chunhui Du for helpful discussion. HC acknowledges support by NSF grant DMR-2531960. DX is supported by DOE Award No.~DE-SC0012509. This work utilized the Alpine high performance computing resource at the University of Colorado Boulder. Alpine is jointly funded by the University of Colorado Boulder, the University of Colorado Anschutz, Colorado State University, and the National Science Foundation (award 2201538). This work also used Bridges-2 HPC at Pittsburgh Supercomputing Center through allocation PHY260075 from the Advanced Cyberinfrastructure Coordination Ecosystem: Services \& Support (ACCESS) program \cite{Boerner2023}, which is supported by U.S. National Science Foundation grants \#2138259, \#2138286, \#2138307, \#2137603, and \#2138296.

\end{acknowledgements} 
\appendix
\section{Derivation of $\chi(\mathbf q)$ expression}\label{sec:chiqderivation}
To calculate $\boldsymbol \chi (\mathbf q)$, we first define the free-energy density (in this work we consider zero temperature only, at which the grand potential is ${\rm Tr}[\rho \hat{F}]$, where $\hat{F} = \hat{H} - \mu \hat{N}$)
\begin{eqnarray}
	\hat{F}(\mathbf r) \equiv \frac{1}{2}\left\{\hat{F}, P_{\mathbf r} \right\} 
\end{eqnarray}
 where $P_{\mathbf r} = |\mathbf r \rangle \langle \mathbf r | $. The density matrix perturbed by a Hamiltonian term $\hat{H}_E$ up to its first order is
\begin{widetext}
\begin{eqnarray}
	\delta \rho  = \frac{1}{N^2}\sum_{m\mathbf k\neq  n\mathbf k'} \frac{f_{m\mathbf k} - f_{n \mathbf k'}}{\epsilon_{m \mathbf k} - \epsilon_{n \mathbf k'} - i0^+ } \langle m\mathbf k| H_E | n \mathbf k'\rangle |m\mathbf k\rangle \langle n \mathbf k' |
\end{eqnarray}
where the $1/N^2$ factor is because of the normalization condition:
\begin{eqnarray}\label{eq:normalization}
	\langle n \mathbf k | m \mathbf k'\rangle &=& \sum_{\mathbf R} e^{i (\mathbf k' - \mathbf k)\cdot \mathbf R} \langle u_{n \mathbf k} | e^{i (\mathbf k' - \mathbf k)\cdot \mathbf R}  | u_{m \mathbf k'} \rangle = N \delta_{\mathbf k,\mathbf k'} \langle u_{n \mathbf k} | u_{m \mathbf k} \rangle   = N \delta_{\mathbf k,\mathbf k'} \delta_{nm} \\\nonumber
	&\rightarrow&  \frac{(2\pi)^3}{V_{\rm uc}} \delta_{nm} \delta(\mathbf k -\mathbf k').
\end{eqnarray}
In our case the perturbation $H_E$ is the Zeeman energy ($g>0$):
\begin{eqnarray}
	H_E = \frac{g\mu_B}{\hbar}\int d^3\mathbf r' \mathbf s\cdot \mathbf B(\mathbf r') P_{\mathbf r'}
\end{eqnarray}
We therefore get
\begin{eqnarray}\label{eq:dFpertb}
 \delta \langle F(\mathbf r )\rangle &=& {\rm Tr}[\delta \rho \hat{F}(\mathbf r)] \\\nonumber
 &=& \frac{g\mu_B}{2\hbar N^2} \int d^3\mathbf r' B_i(\mathbf r') \sum_{m\mathbf k\neq  n\mathbf k'} \frac{(f_{m\mathbf k} - f_{n \mathbf k'})(\epsilon_{n\mathbf k'} + \epsilon_{m \mathbf k} -2\mu)}{\epsilon_{m \mathbf k} - \epsilon_{n \mathbf k'} - i0^+ }\langle m\mathbf k|  s_i P_{\mathbf r'} | n \mathbf k'\rangle  \langle n \mathbf k' |P_{\mathbf r}|m\mathbf k\rangle 
\end{eqnarray}
The matrix elements of $P_{\mathbf r}$ is
\begin{eqnarray}
	 \langle n \mathbf k' |P_{\mathbf r}|m\mathbf k\rangle =  e^{i(\mathbf k - \mathbf k')\cdot \mathbf r} u^\dag_{n \mathbf k'} (\mathbf r) u_{m \mathbf k}(\mathbf r)
\end{eqnarray}
Similarly
\begin{eqnarray}
	\langle m\mathbf k|  s_i P_{\mathbf r'} | n \mathbf k'\rangle = u^\dag_{m \mathbf k} (\mathbf r')s_i u_{n \mathbf k'}(\mathbf r')  e^{i(\mathbf k' - \mathbf k)\cdot \mathbf r'}
\end{eqnarray}
We can now carry out the Fourier transform:
\begin{eqnarray}
	\delta F (\mathbf q) \equiv \int d^3\mathbf r e^{-i\mathbf q \cdot \mathbf r }\delta \langle F(\mathbf r )\rangle
\end{eqnarray}
Since $\delta \langle F(\mathbf r )\rangle$ depends on $\mathbf r$ only through $ \langle n \mathbf k' |P_{\mathbf r}|m\mathbf k\rangle$, it is sufficient to consider
\begin{eqnarray}\label{eq:Prmat}
	\int d^3\mathbf r e^{-i\mathbf q \cdot \mathbf r }\langle n \mathbf k' |P_{\mathbf r}|m\mathbf k\rangle = N \delta_{\mathbf k - \mathbf q, \mathbf k'} \langle u_{n\mathbf k-\mathbf q} | u_{m\mathbf k}\rangle
\end{eqnarray}
On the other hand, the integration over $\mathbf r'$ can also be performed
\begin{eqnarray}
\int d^3\mathbf r' B_i(\mathbf r') \langle m\mathbf k|  s_i P_{\mathbf r'} | n \mathbf k'\rangle = \frac{1}{V_{\rm uc}} B^i_{\mathbf k - \mathbf k'} \langle u_{m \mathbf k} | s_i | u_{n \mathbf k'}\rangle \\\nonumber
\end{eqnarray}
Taken together
\begin{eqnarray}
\delta F ({\mathbf q}) = \frac{g\mu_B}{2\hbar V} \sum_{mn\mathbf k} \frac{(f_{m\mathbf k} - f_{n \mathbf k-\mathbf q})(\epsilon_{n\mathbf k-\mathbf q} + \epsilon_{m \mathbf k} -2\mu)}{\epsilon_{m \mathbf k} - \epsilon_{n \mathbf k-\mathbf q} -  i0^+}  \langle u_{n\mathbf k-\mathbf q} | u_{m\mathbf k}\rangle \langle u_{m \mathbf k} | s_i | u_{n \mathbf k-\mathbf q}\rangle B^i_{\mathbf q}
\end{eqnarray}
Therefore
\begin{eqnarray}\label{eq:chiqKuboapx}
	\chi^i(\mathbf q) = \frac{\delta F (\mathbf q)}{B_{\mathbf q}^i} = \frac{g\mu_B}{2\hbar }\sum_{mn} \int \frac{d^3\mathbf k}{(2\pi)^3}\frac{(f_{m\mathbf k} - f_{n \mathbf k-\mathbf q})(\epsilon_{n\mathbf k-\mathbf q} + \epsilon_{m \mathbf k} -2\mu)}{\epsilon_{m \mathbf k} - \epsilon_{n \mathbf k-\mathbf q} -  i0^+}  \langle u_{n\mathbf k-\mathbf q} | u_{m\mathbf k}\rangle \langle u_{m \mathbf k} | s_i | u_{n \mathbf k-\mathbf q}\rangle
\end{eqnarray}

Note that the above $\chi^i(\mathbf q)$ exactly vanishes at $\mathbf q=0$, since $\hat{F} (\mathbf q=0)$ is $\hat{F}$, but $\delta\rho$ is off-diagonal in the eigenstate basis of $\hat{F}$. To get the $\mathbf q=0$ contribution, we note that $\hat{F}$ in the presence of the Zeeman field should be modified. Namely,
\begin{eqnarray}
     \delta\langle F(\mathbf r)\rangle = {\rm Tr}[(\rho+\delta \rho) (\hat{F}(\mathbf r)+H_{\rm Zeeman}(\mathbf r))] - {\rm Tr}[\rho \hat{F}(\mathbf r)]
\end{eqnarray}
where $H_{\rm Zeeman}(\mathbf r)$ is the Zeeman energy density
\begin{eqnarray}
    H_{\rm Zeeman}(\mathbf r) = \frac{g\mu_B}{\hbar}\mathbf s\cdot \mathbf B(\mathbf r) P_{\mathbf r}
\end{eqnarray}
At linear order, the total $\delta\langle F(\mathbf r)\rangle$ is
\begin{eqnarray}\label{eq:deltaFcomplete}
    \delta\langle F(\mathbf r)\rangle = {\rm Tr}[\delta \rho \hat{F}(\mathbf r)] +  {\rm Tr}[\rho H_{\rm Zeeman}(\mathbf r)] 
\end{eqnarray}
After Fourier tranform, the second term in Eq.~\eqref{eq:deltaFcomplete} vanishes for any $\mathbf q\neq 0$ modulo reciprocal lattice vectors because of translation symmetry, but contributes to the $\mathbf q=0$ term
\begin{eqnarray}\label{eq:Mdipole}
    \chi^i({\mathbf q=0}) = \frac{\partial {\rm Tr}[\rho H_{\rm Zeeman}] }{\partial B_i} = \frac{g\mu_B}{\hbar} \sum_{n}\int[d\mathbf k] f_{n\mathbf k} \langle u_{n\mathbf k} | s_i | u_{n\mathbf k} \rangle
\end{eqnarray}

\section{Explicit formulas of SM$^3$}\label{sec:SM3explicit}
\subsection{General multipole order}
We separate the summation in the expression of $\mathcal{M}$ into $m=n$ (intra-band) and $m\neq n$ (inter-band) terms. The inter-band term is simply
\begin{eqnarray}
	&&({\rm interband}) =\\\nonumber
	 &&-\frac{g\mu_B}{2\hbar } \lim_{\mathbf q \rightarrow 0} {\rm Re}\left[i^{l-1} \partial_{\mathbf q}^{l-1} \sum_{m\neq n} \int [d\mathbf k]\frac{(f_{m\mathbf k+\mathbf q} - f_{n \mathbf k})(\epsilon_{n\mathbf k} + \epsilon_{m \mathbf k+\mathbf q} -2\mu)}{\epsilon_{m \mathbf k + \mathbf q} - \epsilon_{n \mathbf k} }  \langle u_{m\mathbf k + \mathbf q}| u_{n\mathbf k} \rangle \langle u_{n \mathbf k} | s_i |   u_{m \mathbf k + \mathbf q}\rangle \right]
\end{eqnarray}
where we have dropped the $i0^+$ from the denominator since $\epsilon_{m \mathbf k} \neq \epsilon_{n \mathbf k+\mathbf q}$ when $\mathbf q\rightarrow 0$ and $m\neq n$. The $\mathbf q$-derivative can then be distributed into the various $\mathbf q$-dependent quantities in the integrand based on the generalized Leibniz rule
\begin{eqnarray}
	(fg)^{(n)} = \sum_{k=0}^n \frac{n!}{k!(n-k)!} f^{(n-k)}g^{(k)},
\end{eqnarray}
which for three differentiation variables becomes
\begin{eqnarray}\label{eq:DxyzAB}
	(\partial^{n_x}_{x} \partial^{n_y}_{y} \partial^{n_z}_{z}) (AB) =\sum_{i = 0}^{n_x} \sum_{j=0}^{n_y} \sum_{k=0}^{n_z} \frac{n_x!}{i!(n_x-i)!}  \frac{n_y!}{j!(n_y-j)!} \frac{n_z!}{k!(n_z-k)!}  A^{(n_x-i,n_y-j,n_z-k)}B^{(i,j,k)} 
\end{eqnarray}
Here we would like to further distinguish two types of terms in the interband contribution. Note that whenever a $\partial_{\mathbf q}$ acts on $f_{m\mathbf k+\mathbf q}$, the term is nonzero only on the Fermi surface. We therefore have the Fermi-sea-only interband contribution:
\begin{eqnarray}
&&\begin{pmatrix}
	{\rm interband} \\
	{\rm Fermi~sea}
\end{pmatrix} =\\\nonumber
&&-\frac{g\mu_B}{2\hbar } \lim_{\mathbf q \rightarrow 0} {\rm Re}\left[i^{l-1} \sum_{m \neq n} \int [d\mathbf k](f_{m\mathbf k} - f_{n \mathbf k}) \partial_{\mathbf q}^{l-1} \left(\frac{\epsilon_{n\mathbf k} + \epsilon_{m \mathbf k+\mathbf q} -2\mu}{\epsilon_{m \mathbf k + \mathbf q} - \epsilon_{n \mathbf k} }  \langle u_{m\mathbf k + \mathbf q}| u_{n\mathbf k} \rangle \langle u_{n \mathbf k} | s_i |   u_{m \mathbf k + \mathbf q}\rangle \right)\right]
\end{eqnarray}
To simplify the result we define
\begin{eqnarray}
	P_{m\mathbf k} \equiv |u_{m\mathbf k}\rangle \langle u_{m\mathbf k}|
\end{eqnarray}
Therefore, using Eq.~\eqref{eq:DxyzAB} and denoting the combinatorial coefficients by $C_{j_{x,y,z}}$, we have
\begin{eqnarray}
 (\partial^{n_x}_{x} \partial^{n_y}_{y} \partial^{n_z}_{z}) \frac{P_{m \mathbf k+\mathbf q} (\epsilon_{m\mathbf k+\mathbf q} + \epsilon_{n\mathbf k} - 2\mu) }{\epsilon_{m \mathbf k+\mathbf q} - \epsilon_{n \mathbf k}} 
	= \sum_{j_{x,y,z}} C_{j_{x,y,z}}  P_{m \mathbf k}^{(n_x-j_x,n_y-j_y,n_z-j_z)}\left[\left( 1 + 2\frac{\epsilon_{n\mathbf k} - \mu}{\epsilon_{m \mathbf k'} - \epsilon_{n \mathbf k}} \right)^{(j_x,j_y,j_z)'}\Big|_{\mathbf k' = \mathbf k}\right]
\end{eqnarray}
The constant term in the parentheses only contributes to the term with $j_x = j_y = j_z = 0$. We therefore rewrite the above term into
\begin{eqnarray}
&&(\partial^{n_x}_{x} \partial^{n_y}_{y} \partial^{n_z}_{z}) \frac{P_{m \mathbf k+\mathbf q} (\epsilon_{m\mathbf k+\mathbf q} + \epsilon_{n\mathbf k} - 2\mu) }{\epsilon_{m \mathbf k+\mathbf q} - \epsilon_{n \mathbf k}} \\\nonumber
&=& 2(\epsilon_{n\mathbf k} - \mu) \sideset{}{'}\sum_{j_{x,y,z}} C_{j_{x,y,z}}  P_{m \mathbf k}^{(n_x-j_x,n_y-j_y,n_z-j_z)}\left[\left( \frac{1}{\epsilon_{m \mathbf k'} - \epsilon_{n \mathbf k}} \right)^{(j_x,j_y,j_z)'}\Big|_{\mathbf k' = \mathbf k}\right] \\\nonumber
&+&  \frac{\epsilon_{n\mathbf k} + \epsilon_{m \mathbf k} -2\mu}{\epsilon_{m \mathbf k} - \epsilon_{n \mathbf k} } P_{m \mathbf k}^{(n_x,n_y,n_z)}
\end{eqnarray}
where the primed sum means the term with $j_x = j_y = j_z = 0$ is excluded. 

We therefore have
\begin{eqnarray}\label{eq:interbandsea}
\begin{pmatrix}
	{\rm interband} \\
	{\rm Fermi~sea}
\end{pmatrix} &=&-\frac{g\mu_B}{2\hbar }\times{\rm Re}\Bigg \{ i^{l-1} \sum_{m\neq n} \int [d\mathbf k](f_{m\mathbf k} - f_{n \mathbf k}) \times \\\nonumber
&& \Bigg [\sideset{}{'}\sum_{j_{x,y,z}} C_{j_{x,y,z}}  \langle u_{n \mathbf k} | s_i P_{m \mathbf k}^{(n_x-j_x,n_y-j_y,n_z-j_z)} | u_{n\mathbf k} \rangle \left(2\frac{\epsilon_{n\mathbf k} - \mu}{\epsilon_{m \mathbf k'} - \epsilon_{n \mathbf k}} \right)^{(j_x,j_y,j_z)'}\Big|_{\mathbf k' = \mathbf k} \\\nonumber
&& +   \frac{\epsilon_{n\mathbf k} + \epsilon_{m \mathbf k} -2\mu}{\epsilon_{m \mathbf k} - \epsilon_{n \mathbf k} } \langle u_{n\mathbf k} | s_i P_{m \mathbf k}^{(n_x,n_y,n_z)} | u_{n\mathbf k}\rangle \Bigg]\Bigg \}
\end{eqnarray}
Eq.~\eqref{eq:interbandsea} therefore only has cross-gap matrix elements and is free from issues due to degeneracies in the occupied or unoccupied states.

The Fermi surface contribution to the interband term is 
\begin{eqnarray}\label{eq:interbandsurf}
&&\begin{pmatrix}
	{\rm interband} \\
	{\rm Fermi~surface}
\end{pmatrix} =-\frac{g\mu_B}{2\hbar }\times \\\nonumber
&&{\rm Re} \sum_{m\neq n} \int [d\mathbf k]\sideset{}{'}\sum_{j_{x,y,z}} C_{j_{x,y,z}}  f_{m\mathbf k}^{(j_x,j_y,j_z)} \left[ i^{l-1} \frac{\langle u_{n\mathbf k}| s_i P_{m \mathbf k'} | u_{n \mathbf  k}\rangle (\epsilon_{m\mathbf k'} + \epsilon_{n\mathbf k} - 2\mu) }{\epsilon_{m \mathbf k'} - \epsilon_{n \mathbf k}}  \right]^{(n_x-j_x,n_y-j_y,n_z-j_z)'}\Bigg|_{\mathbf k' = \mathbf k}
\end{eqnarray}

We next turn to the intraband term:
\begin{eqnarray}
	&&({\rm intraband})\\\nonumber
&=& -\frac{g\mu_B}{2\hbar } \lim_{\mathbf q \rightarrow 0} {\rm Re}\left[i^{l-1} \partial_{\mathbf q}^{l-1} \sum_{m} \int [d\mathbf k]\frac{(f_{m\mathbf k+\mathbf q} - f_{m \mathbf k})(\epsilon_{m\mathbf k} + \epsilon_{m \mathbf k+\mathbf q} -2\mu)}{\epsilon_{m \mathbf k + \mathbf q} - \epsilon_{m \mathbf k}  +  i0^+ }  \langle u_{m\mathbf k + \mathbf q}| u_{m\mathbf k} \rangle \langle u_{m \mathbf k} | s_i |   u_{m \mathbf k + \mathbf q}\rangle \right]
\end{eqnarray}
If it were not for the $q-$derivative one could simply use the rule of limits and rewrite the ratio in the above equation as
\begin{eqnarray}
	\lim_{\mathbf q\rightarrow 0}\frac{(f_{m\mathbf k+\mathbf q} - f_{m \mathbf k})(\epsilon_{m\mathbf k} + \epsilon_{m \mathbf k+\mathbf q} -2\mu)}{\epsilon_{m \mathbf k + \mathbf q} - \epsilon_{m \mathbf k} }  =  2(\epsilon_{m \mathbf k}-\mu)f'_{m\mathbf k}
\end{eqnarray}
However, the $\mathbf q$-derivative precedes the limit and must be done first. We need to replace the ratio by a Taylor expansion up to the order of $q^{l-1}$. For example, up to the 3rd order in $\mathbf q$ we have (omitting the $m\mathbf k$ subscript for simplicity)
\begin{eqnarray}\label{eq:limtaylor3rd}
	&&\frac{(f_{m\mathbf k+\mathbf q} - f_{m \mathbf k})(\epsilon_{m\mathbf k} + \epsilon_{m \mathbf k+\mathbf q} -2\mu)}{\epsilon_{m \mathbf k + \mathbf q} - \epsilon_{m \mathbf k} } \\\nonumber
	&=&2(\epsilon - \mu) f^{(1)}\\\nonumber
	&+&\partial_i \epsilon  \left( f^{(1)}+ (\epsilon - \mu) f^{(2)} \right ) q_i \\\nonumber
	&+& \left[\frac{1}{2} \partial_{ij} \epsilon \left( f^{(1)}  + (\epsilon - \mu) f^{(2)}\right ) + \frac{1}{6}\partial_i \epsilon \partial_j \epsilon \left( 3f^{(2)} +2(\epsilon-\mu)f^{(3)} \right)  \right]q_i q_j \\\nonumber
	&+& \left[\frac{1}{6} \partial_{ijk}\epsilon \left( f^{(1)} + (\epsilon - \mu) f^{(2)} \right) + \frac{1}{6} \partial_{i}\epsilon \partial_{jk} \epsilon \left( 3f^{(2)} + 2(\epsilon-\mu) f^{(3)}  \right) + \frac{1}{12}\partial_{i} \epsilon \partial_{j} \epsilon \partial_{k} \epsilon \left( 2 f^{(3)} + (\epsilon-\mu) f^{(4)} \right) \right] q_i q_j q_k \\\nonumber
	&+& O(q^4)
\end{eqnarray} 
Therefore the intra-band term vanishes in insulators. If defining 
\begin{eqnarray}
	G_{\mathbf k'} \equiv G_{\mathbf k', \mathbf k} \equiv \frac{(f_{m\mathbf k'} - f_{m \mathbf k})(\epsilon_{m\mathbf k} + \epsilon_{m \mathbf k'} -2\mu)}{\epsilon_{m \mathbf k'} - \epsilon_{m \mathbf k} }
\end{eqnarray}
the Fermi-surface-only intraband term can be written as
\begin{eqnarray}\label{eq:intraband}
&&({\rm intraband})\\\nonumber
&=& -\frac{g\mu_B}{2\hbar } {\rm Re}\left[i^{l-1} \sum_{m} \int [d\mathbf k]\sideset{}{'}\sum_{j_{x,y,z}} C_{j_{x,y,z}} \langle u_{m \mathbf k} | s_i P_{m\mathbf k}^{(n_x-j_x,n_y-j_y,n_z-j_z)} |   u_{m \mathbf k}\rangle G_{\mathbf k'}^{(j_x,j_y,j_z)'}\Big |_{\mathbf k' = \mathbf k}\right]
\end{eqnarray}
where the $j_x = j_y = j_z = 0$ term is excluded because it vanishes at zero temperature.

\subsection{Spin quadrupole moment}
In this subsection we compare the thermodynamic spin quadrupole moment with that in \cite{gao_2018,shitade_2019}. For definiteness we consider the component $\mathcal{M}_y^x$.

First consider the intraband contribution. Since there is only one derivative, it is either acting on $P_{m\mathbf k}$ or $G_{\mathbf k'}$. Only the former is relevant since $\langle u_{m \mathbf k} | [s^x, P_{m\mathbf k}] | u_{m \mathbf k}\rangle = 0$. Therefore we only need to consider the first line in Eq.~\eqref{eq:limtaylor3rd}. But at zero temperature the second term there vanishes. Therefore (ignoring the factor $-g\mu_B/\hbar$ below)
\begin{eqnarray}
	({\rm intraband}) &=& i\int \frac{d^3\mathbf k}{(2\pi)^3}  \sum_m f_{m\mathbf k} \langle u_{m \mathbf k} | s^x \partial_{y} P_{m \mathbf k}| u_{m \mathbf k}\rangle \\\nonumber
	&=&\int \frac{d^3\mathbf k}{(2\pi)^3}  \sum_m f_{m\mathbf k} \langle u_{m \mathbf k} | s^x (i\partial_y - \mathcal{A}_y)| u_{m \mathbf k}\rangle \\\nonumber
	&=& -{\rm Im}\int \frac{d^3\mathbf k}{(2\pi)^3}  \sum_{n\neq m} f_{m\mathbf k}\frac{s^x_{mn} (\hbar v_y)_{nm}}{\epsilon_{m\mathbf k} - \epsilon_{n \mathbf k}}
\end{eqnarray}

Next consider the interband contribution. Again the derivative can act on either $P_{n\mathbf k}$ or the inverse energy difference. However, since $P_{n\mathbf k} | u_{m \mathbf k}\rangle = 0$ for $n\neq m$, only the former is nonzero. Therefore
\begin{eqnarray}
	({\rm interband}) &=& 2i\int \frac{d^3\mathbf k}{(2\pi)^3}  \sum_{n\neq m} f_{m\mathbf k} (\epsilon_{m\mathbf k} - \mu) \langle u_{m \mathbf k} | s^x (\partial_{y} P_{n \mathbf k})| u_{m \mathbf k}\rangle \frac{1}{\epsilon_{m\mathbf k} - \epsilon_{n\mathbf k}}\\\nonumber
	&=&2{\rm Im}\int \frac{d^3\mathbf k}{(2\pi)^3}  \sum_{n\neq m} f_{m\mathbf k} (\epsilon_{m\mathbf k} - \mu)\frac{s^x_{mn} (\hbar v_y)_{nm}}{(\epsilon_{m\mathbf k} - \epsilon_{n \mathbf k})^2}
\end{eqnarray}

Taken together
\begin{eqnarray}
	\mathcal{M}_y^x = {\rm Im}\int \frac{d^3\mathbf k}{(2\pi)^3}  \sum_{n\neq m} f_{m\mathbf k} (\epsilon_{m\mathbf k} + \epsilon_{n\mathbf k} - 2\mu)\frac{s^x_{mn} (\hbar v_y)_{nm}}{(\epsilon_{m\mathbf k} - \epsilon_{n \mathbf k})^2}
\end{eqnarray}
which is consistent with \cite{gao_2018,shitade_2019}. The above formula can be rewritten in a way that only cross-gap matrix elements contribute
\begin{eqnarray}
	\mathcal{M}_y^x = {\rm Im}\int \frac{d^3\mathbf k}{(2\pi)^3}  \sum_{n\neq m} \frac{f_{m\mathbf k} - f_{n\mathbf k}}{2} (\epsilon_{m\mathbf k} + \epsilon_{n\mathbf k} - 2\mu)\frac{s^x_{mn} (\hbar v_y)_{nm}}{(\epsilon_{m\mathbf k} - \epsilon_{n \mathbf k})^2}
\end{eqnarray}

\subsection{Spin octupole moment}
Without loss of generality we consider the $\mathcal{M}_{xy}^z$ component. For later convenience we introduce the following notation when making use of the Leibniz rule (below $\partial_{x,y,z}$ means $\partial_{k_{x,y,z}}$ unless noted otherwise):
\begin{eqnarray}
	(\partial_x \partial_y)(AB) &=& (\partial_x\partial_y A) B + (\partial_x A) (\partial_y B) +  (\partial_y A) (\partial_x B) + A (\partial_x\partial_y B) \\\nonumber
	&\equiv & (xy,) + (x,y) + (y, x) + (,xy)
\end{eqnarray}

We next calculate the interband and intraband contributions separately. 

\subsubsection{Interband, Fermi sea contribution}
We use the following equivalent form of Eq.~\eqref{eq:interbandsea}:
\begin{eqnarray}
		\begin{pmatrix}
			{\rm interband} \\
			{\rm Fermi~sea}
		\end{pmatrix} &=&-\frac{g\mu_B}{2\hbar }{\rm Re}\Bigg \{ i^{\ell-1} \sum_{m\neq n} \int [d\mathbf k](f_{m\mathbf k} - f_{n \mathbf k}) \times \\\nonumber
		&& \Bigg [2(\epsilon_{m\mathbf k} - \mu)\sideset{}{'}\sum_{j_{x,y,z}} C_{j_{x,y,z}} \sum_l s^i_{ml\mathbf k} \left(P_{n \mathbf k}^{(n_x-j_x,n_y-j_y,n_z-j_z)}\right)_{lm\mathbf k} \left(\frac{1}{\epsilon_{m \mathbf k} - \epsilon_{n \mathbf k'}} \right)^{(j_x,j_y,j_z)'}\Big|_{\mathbf k' = \mathbf k} \\\nonumber
		&& +   \frac{\epsilon_{m\mathbf k} + \epsilon_{n \mathbf k} -2\mu}{\epsilon_{m \mathbf k} - \epsilon_{n \mathbf k} } s^i_{ml\mathbf k} \left(P_{n \mathbf k}^{(n_x,n_y,n_z)} \right)_{lm \mathbf k} \Bigg]\Bigg \}
\end{eqnarray}
where to avoid confusion we have replaced the multipole order $l$ by $\ell$ so that it is different from the newly introduced dummy index. 

For octupole moments we need to calculate the following derivatives explicitly
\begin{eqnarray}
	\partial_{x'} \left(\frac{1}{\epsilon_{m \mathbf k} - \epsilon_{n\mathbf k'}}\right)\Bigg |_{\mathbf k' = \mathbf k} &=& \frac{\partial_x \epsilon_{n\mathbf k}}{(\epsilon_{m\mathbf k} - \epsilon_{n \mathbf k})^2} \\\nonumber
	\partial_{x'} \partial_{y'} \left(\frac{1}{\epsilon_{m \mathbf k} - \epsilon_{n\mathbf k'}}\right)\Bigg |_{\mathbf k' = \mathbf k} &=& \frac{\partial_x \partial_y \epsilon_{n\mathbf k}}{(\epsilon_{m\mathbf k} - \epsilon_{n \mathbf k})^2}  +  \frac{2(\partial_x \epsilon_{n\mathbf k}) (\partial_y \epsilon_{n\mathbf k})}{(\epsilon_{m\mathbf k} - \epsilon_{n \mathbf k})^3}
\end{eqnarray}	
The derivatives of $\epsilon_{n\mathbf k}$ can either be obtained directly from Wannier interpolation, or by first interpolating $\epsilon_{n\mathbf k}$ on the fine $k$-mesh and then using FFT or finite difference. 

On the other hand, we need to calculate $\langle u_{l\mathbf k}| \partial_x \partial_y P_{n\mathbf k} | u_{m\mathbf k}\rangle$, $\langle u_{l\mathbf k}| \partial_x P_{n\mathbf k} | u_{m\mathbf k}\rangle$, and $\langle u_{l\mathbf k}|\partial_y P_{n\mathbf k} | u_{m\mathbf k}\rangle$. Our goal is to transform them into the following quantities implemented in Wannier90 \cite{IbanezAzpiroz2018, Lihm2021, Mostofi2014}:
\begin{eqnarray}\label{eq:rarab}
	r^a_{\mathbf k nm} &\equiv& (1-\delta_{nm})A^a_{\mathbf k nm},\\\nonumber
	r^{a;b}_{\mathbf k nm}  &\equiv&  \partial_b r^a_{\mathbf k nm} - i (A^b_{\mathbf k nn} - A^b_{\mathbf k mm})r^a_{\mathbf k  nm}
\end{eqnarray}
Below we neglect the $\mathbf k$ subscript for brevity. Also repeated indices are not summed over unless noted otherwise. The first derivative can be immediate obtained as
\begin{eqnarray}\label{eq:pxP}
	(\partial_x P_{n})_{lm} & \equiv & \langle u_{l} | \partial_x P_{n} | u_{m}\rangle = -iA_{ln}^x \delta_{nm} +i A_{nm}^x\delta_{ln} \\\nonumber
	&=& -i\delta_{nm} r^x_{ln} + i\delta_{ln} r^x_{ nm} \\\nonumber
	&=& ir^x_{ lm} (\delta_{ln} - \delta_{nm})
\end{eqnarray}
For the second derivative, we have
\begin{eqnarray}
	\langle u_{l} | \partial_x \partial_y P_{n} | u_{m}\rangle = \partial_y (\langle u_{l} | \partial_x P_{n} | u_{m}\rangle) - \langle \partial_y u_{l} | \partial_x P_{n} | u_{m}\rangle - \langle u_{l} | \partial_x P_{n} | \partial_y u_{m}\rangle 
\end{eqnarray}
where the first term is, using Eq.~\eqref{eq:pxP}
\begin{eqnarray}
	\partial_y (\langle u_{l} | \partial_x P_{n} | u_{m}\rangle) = i\partial_y r^x_{lm} (\delta_{ln} - \delta_{nm})
\end{eqnarray}
the second term is
\begin{eqnarray}
	\langle \partial_y u_{l\mathbf k} | \partial_x P_{n\mathbf k} | u_{m\mathbf k}\rangle &=& \sum_{q} \langle \partial_y u_{l\mathbf k}| u_{q\mathbf k} \rangle \langle u_{q\mathbf k}| \partial_x P_{n\mathbf k} | u_{m\mathbf k}\rangle \\\nonumber
	&=&\sum_{q} iA^y_{lq} ir^x_{\mathbf k qm} (\delta_{qn} - \delta_{nm}) \\\nonumber
	&=& -r^y_{ln} r^x_{nm}  + \sum_q r^y_{lq} r^x_{qm}\delta_{nm} - r^x_{lm} (\delta_{ln} - \delta_{nm}) A^y_{ll} 
\end{eqnarray}
and the third term is
\begin{eqnarray}
\langle u_{l} | \partial_x P_{n} | \partial_y u_{m}\rangle &=& \sum_q \langle u_{l} | \partial_x P_{n}| u_{q} \rangle \langle u_{q} | \partial_y u_{m}\rangle \\\nonumber
&=& \sum_{q}  ir^x_{ lq} (\delta_{ln} - \delta_{nq}) (-iA^y_{qm}) \\\nonumber
&=& -r^x_{ln}r^y_{nm} +  \sum_q r^x_{lq} r^y_{qm} \delta_{ln} + r^x_{lm} (\delta_{ln} - \delta_{nm})A^y_{mm}
\end{eqnarray}
Putting them together, we get
\begin{eqnarray}\label{eq:pxpyP}
(\partial_x \partial_y P_{n })_{lm} &\equiv & \langle u_{l } | \partial_x \partial_y P_{n } | u_{m }\rangle \\\nonumber
&=& i r^{x;y}_{lm} (\delta_{ln} - \delta_{nm}) + r^y_{ln} r^x_{nm} + r^x_{ln} r^y_{nm} - (r^y r^x)_{lm}\delta_{nm}- (r^x r^y)_{lm}\delta_{ln}
\end{eqnarray}
Since the result must be symmetric under $x\leftrightarrow y$, we obtain after symmetrization:
\begin{eqnarray}\label{eq:pxpyPsym}
(\partial_x \partial_y P_{n })_{lm}  = \frac{1}{2} \left[i r^{x;y}_{lm} (\delta_{ln} - \delta_{nm}) + 2r^x_{ln} r^y_{nm} - (r^x r^y)_{lm}(\delta_{ln} +\delta_{nm}) +(x\leftrightarrow y) \right]
\end{eqnarray}
Depending on the wannierisation process, Eq.~\eqref{eq:pxpyP} may not be exactly symmetric under $x,y$ permutation. If this is the case Eq.~\eqref{eq:pxpyPsym} needs to be used.

Cleaning things up, we get
\begin{eqnarray}\label{eq:octupoleintersea}
	(xy,) &=& \frac{g\mu_B}{2\hbar}{\rm Re}\int[d\mathbf k] \sum_{m\neq n} \frac{(f_{m}-f_n)(\epsilon_{m} + \epsilon_n - 2\mu) }{\epsilon_{m } -\epsilon_{n }}\sum_{l}s^z_{ml }(\partial_x \partial_y P_{n })_{lm} \\\nonumber
	(x,y) &=& \frac{g\mu_B}{\hbar}{\rm Re}\int[d\mathbf k]\sum_{m\neq n} \frac{(f_{m}-f_n) (\epsilon_m - \mu) \partial_y \epsilon_{n }}{(\epsilon_{m } - \epsilon_{n  })^2} \sum_{l} s^z_{ml } (\partial_x P_{n })_{lm} \\\nonumber
	(y,x) &=& \frac{g\mu_B}{\hbar}{\rm Re}\int[d\mathbf k]\sum_{m\neq n} \frac{(f_{m}-f_n) (\epsilon_m - \mu) \partial_x \epsilon_{n }}{(\epsilon_{m } - \epsilon_{n  })^2} \sum_{l} s^z_{ml } (\partial_y P_{n })_{lm} \\\nonumber
	(,xy) &=& 0
\end{eqnarray}
where the last line is because $P_{n \mathbf k} | u_{m\mathbf k}\rangle = 0$ when $n\neq m$; $(\partial_x \partial_y P_{n })_{lm}$ is given by Eq.~\eqref{eq:pxpyP} or \ref{eq:pxpyPsym}, $(\partial_x P_n)_{lm}$ is given by Eq.~\eqref{eq:pxP}, which also applies to $(\partial_y P_n)_{lm}$ with $x\rightarrow y$. For insulators, Eq.~\eqref{eq:octupoleintersea} is the only contribution.  

\subsubsection{Interband, Fermi surface contribution}
We use the following equivalent form of Eq.~\eqref{eq:interbandsurf}:
\begin{eqnarray}\label{eq:interbandsurf}
	\begin{pmatrix}
		{\rm interband} \\
		{\rm Fermi~surface}
	\end{pmatrix} &=&\frac{g\mu_B}{2\hbar }{\rm Re} \sum_{m\neq n;l} \int [d\mathbf k]\sideset{}{'}\sum_{j_{x,y,z}} C_{j_{x,y,z}}\times  \\\nonumber
	&&  \left[ i^{\ell-1} \frac{ s^i_{ml\mathbf k}\left( P_{n \mathbf k'}\right)_{lm\mathbf k} (\epsilon_{n\mathbf k'} + \epsilon_{m\mathbf k} - 2\mu) }{\epsilon_{m \mathbf k} - \epsilon_{n \mathbf k'} }  \right]^{(n_x-j_x,n_y-j_y,n_z-j_z)'}\Bigg|_{\mathbf k' = \mathbf k}f_{n\mathbf k}^{(j_x,j_y,j_z)} 
\end{eqnarray}
Since the primed sum excludes $(xy,0)$, while $(,xy)=0$ for the same reason as that in Eq.~\eqref{eq:octupoleintersea}, we only need to consider $(x,y)$ and $(y,x)$:
\begin{eqnarray}\label{eq:octupoleintersurf}
(x,y) &=& -\frac{g\mu_B}{2\hbar}{\rm Re}\int[d\mathbf k]\sum_{m\neq n} f^{(1)}_n  \frac{(\epsilon_m + \epsilon_n - 2\mu) \partial_y \epsilon_{n }}{\epsilon_{m } - \epsilon_{n  }} \sum_{l} s^z_{ml } (\partial_x P_{n })_{lm} \\\nonumber
(y,x) &=& -\frac{g\mu_B}{2\hbar}{\rm Re}\int[d\mathbf k]\sum_{m\neq n} f^{(1)}_n  \frac{(\epsilon_m + \epsilon_n - 2\mu) \partial_x \epsilon_{n }}{\epsilon_{m } - \epsilon_{n  }} \sum_{l} s^z_{ml } (\partial_y P_{n })_{lm}
\end{eqnarray}
At zero temperature the above results can be simplified further, since $f_n^{(1)} = -\delta(\epsilon_n - \mu)$, which makes $f_n^{(1)} \frac{\epsilon_m + \epsilon_n - 2\mu}{\epsilon_{m } - \epsilon_{n  }} = f_n^{(1)} $. We therefore get, for $T=0$ K,
\begin{eqnarray}\label{eq:octupoleintersurfT0}
	(x,y) &=& \frac{g\mu_B}{2\hbar}{\rm Re}\int[d\mathbf k]\sum_{m\neq n} \delta(\epsilon_n -\mu)  \partial_y \epsilon_{n } \sum_{l} s^z_{ml } (\partial_x P_{n })_{lm} \\\nonumber
	(y,x) &=& \frac{g\mu_B}{2\hbar}{\rm Re}\int[d\mathbf k]\sum_{m\neq n} \delta(\epsilon_n -\mu)   \partial_x \epsilon_{n } \sum_{l} s^z_{ml } (\partial_y P_{n })_{lm}
\end{eqnarray}

\subsubsection{Intraband contribution}
We use the following equivalent form of Eq.~\eqref{eq:intraband}:
\begin{eqnarray}\label{eq:intraband}
	({\rm intraband})= -\frac{g\mu_B}{2\hbar } {\rm Re}\left[i^{\ell-1} \sum_{m;l} \int [d\mathbf k]\sideset{}{'}\sum_{j_{x,y,z}} C_{j_{x,y,z}} s^i_{ml} \left(P_{m\mathbf k}^{(n_x-j_x,n_y-j_y,n_z-j_z)}\right)_{lm\mathbf k} G_{\mathbf k'}^{(j_x,j_y,j_z)'}\Big |_{\mathbf k' = \mathbf k}\right]
\end{eqnarray}

Due to the primed sum $(xy,) = 0$. For the other three terms we need to calculate
\begin{eqnarray}
	\partial_{x'} \left(\frac{(f_{m\mathbf k'} - f_{m \mathbf k})(\epsilon_{m\mathbf k} + \epsilon_{m \mathbf k'} -2\mu)}{\epsilon_{m \mathbf k'} - \epsilon_{m \mathbf k} }\right)\Bigg |_{\mathbf k' = \mathbf k} &=&(\partial_x \epsilon) f^{(1)} +(\epsilon - \mu)(\partial_{x}\epsilon) f^{(2)}\\\nonumber
	\partial_{x'} \partial_{y'}  \left(\frac{(f_{m\mathbf k'} - f_{m \mathbf k})(\epsilon_{m\mathbf k} + \epsilon_{m \mathbf k'} -2\mu)}{\epsilon_{m \mathbf k'} - \epsilon_{m \mathbf k} }\right)\Bigg |_{\mathbf k' = \mathbf k} &=& (\partial_x\partial_y\epsilon) \left[f^{(1) } + (\epsilon - \mu)f^{(2)} \right]+ \frac{1}{3} (\partial_x\epsilon)(\partial_y \epsilon) \left[3f^{(2)} + 2(\epsilon-\mu)f^{(3)}\right]
\end{eqnarray}
where the $k$-derivatives of $\epsilon_{m\mathbf k}$ can be calculated using FFT.

The derivatives of $P_{m }$ are calculated in the same way as for the interband contribution but are simpler:
\begin{eqnarray}
	(\partial_x \partial_y P_{m})_{lm} &=& -i r^{x;y}_{lm} - (r^y r^x)_{lm} - (r^x r^y)_{mm}\delta_{lm} \\\nonumber
	&=&-\frac{1}{2} \left[i r^{x;y}_{lm} + (r^x r^y)_{lm} +  (r^x r^y)_{mm}\delta_{lm} +(x\leftrightarrow y) \right] \\\nonumber
		(\partial_x P_{m})_{lm} 	&=& -ir^x_{ lm}
\end{eqnarray}

Finally,
\begin{eqnarray}\label{eq:octupoleintra}
	(x,y) &=&\frac{g\mu_B}{2\hbar} {\rm Re}\int[d\mathbf k] \sum_{m} \left[ f^{(1)}_{m } +(\epsilon_{m } - \mu) f^{(2)}_{m }\right] (\partial_y \epsilon_{m }) \sum_{l}s^z_{ml }(\partial_x P_{m })_{lm}\\\nonumber
	(y,x) &=& \frac{g\mu_B}{2\hbar} {\rm Re}\int[d\mathbf k] \sum_{m} \left[ f^{(1)}_{m } +(\epsilon_{m } - \mu)f^{(2)}_{m }\right] (\partial_x \epsilon_{m })\sum_{l}s^z_{ml }(\partial_y P_{m })_{lm}\\\nonumber
	(,xy) &=& \frac{g\mu_B}{2\hbar} {\rm Re}\int[d\mathbf k] \sum_{m} \left\{(\partial_x\partial_y\epsilon) \left[f^{(1) } + (\epsilon - \mu)f^{(2)} \right]+ \frac{1}{3} (\partial_x\epsilon)(\partial_y \epsilon) \left[3f^{(2)} + 2(\epsilon-\mu)f^{(3)}\right] \right\}_{m } s^z_{mm} 
\end{eqnarray}
which can be further simplified at zero temperature as discussed in the next section and we quote the results below
\begin{eqnarray}\label{eq:octupoleintraT0}
	(x,y) &=& (y,x) = 0\\\nonumber
(,xy) &=& \frac{g\mu_B}{2\hbar} {\rm Re}\int[d\mathbf k] \sum_{m} \left\{(\partial_x\partial_y\epsilon) \left[f^{(1) } + (\epsilon - \mu)f^{(2)} \right]+ \frac{1}{3} (\partial_x\epsilon)(\partial_y \epsilon) \left[3f^{(2)} + 2(\epsilon-\mu)f^{(3)}\right] \right\}_{m } s^z_{mm} 
\end{eqnarray}

\subsection*{Summary of all contributions}
For convenience we summarize all nonzero contributions at $T=0$ K below:
\begin{eqnarray}\label{eq:octupoleallT0}
\noindent{(\rm Fermi\, sea)}\\\nonumber
	(xy,) &=& \frac{g\mu_B}{2\hbar}{\rm Re}\int[d\mathbf k] \sum_{m\neq n} \frac{(f_{m}-f_n)(\epsilon_{m} + \epsilon_n - 2\mu) }{\epsilon_{m } -\epsilon_{n }}\sum_{l}s^z_{ml }(\partial_x \partial_y P_{n })_{lm} \\\nonumber
(x,y) &=& \frac{g\mu_B}{\hbar}{\rm Re}\int[d\mathbf k]\sum_{m\neq n} \frac{(f_{m}-f_n) (\epsilon_m - \mu) \partial_y \epsilon_{n }}{(\epsilon_{m } - \epsilon_{n  })^2} \sum_{l} s^z_{ml } (\partial_x P_{n })_{lm} \\\nonumber
(y,x) &=& \frac{g\mu_B}{\hbar}{\rm Re}\int[d\mathbf k]\sum_{m\neq n} \frac{(f_{m}-f_n) (\epsilon_m - \mu) \partial_x \epsilon_{n }}{(\epsilon_{m } - \epsilon_{n  })^2} \sum_{l} s^z_{ml } (\partial_y P_{n })_{lm} \\\nonumber
\noindent{(\rm Fermi\, surface)}\\\nonumber	
(x,y) &\approx& \frac{g\mu_B}{2\hbar}{\rm Re} \sum_{m\neq n} \int[d\mathbf k] g_n (\partial_y \epsilon_{n }) \sum_{l} s^z_{ml } (\partial_x P_{n })_{lm} \\\nonumber
(y,x) &\approx& \frac{g\mu_B}{2\hbar}{\rm Re} \sum_{m\neq n} \int[d\mathbf k] g_n (\partial_x \epsilon_{n }) \sum_{l} s^z_{ml } (\partial_y P_{n })_{lm}  \\\nonumber
(,xy) &\approx&  -\frac{g\mu_B}{12\hbar} \sum_m  \int[d\mathbf k] g_m \left[2(\partial_x\partial_y\epsilon_m)s_m^z +  (\partial_x\epsilon_m) (\partial_y s_m^z) +  (\partial_y\epsilon_m) (\partial_x s_m^z) \right]
\end{eqnarray}
where $g_n$ and $g_m$ in the Fermi surface terms are Gaussian functions serving as an approximation of the $\delta$ function.
    \end{widetext}

\bibliography{eqm3_ref}

\begin{thebibliography}{123}%
\makeatletter
\providecommand \@ifxundefined [1]{%
 \@ifx{#1\undefined}
}%
\providecommand \@ifnum [1]{%
 \ifnum #1\expandafter \@firstoftwo
 \else \expandafter \@secondoftwo
 \fi
}%
\providecommand \@ifx [1]{%
 \ifx #1\expandafter \@firstoftwo
 \else \expandafter \@secondoftwo
 \fi
}%
\providecommand \natexlab [1]{#1}%
\providecommand \enquote  [1]{``#1''}%
\providecommand \bibnamefont  [1]{#1}%
\providecommand \bibfnamefont [1]{#1}%
\providecommand \citenamefont [1]{#1}%
\providecommand \href@noop [0]{\@secondoftwo}%
\providecommand \href [0]{\begingroup \@sanitize@url \@href}%
\providecommand \@href[1]{\@@startlink{#1}\@@href}%
\providecommand \@@href[1]{\endgroup#1\@@endlink}%
\providecommand \@sanitize@url [0]{\catcode `\\12\catcode `\$12\catcode
  `\&12\catcode `\#12\catcode `\^12\catcode `\_12\catcode `\%12\relax}%
\providecommand \@@startlink[1]{}%
\providecommand \@@endlink[0]{}%
\providecommand \url  [0]{\begingroup\@sanitize@url \@url }%
\providecommand \@url [1]{\endgroup\@href {#1}{\urlprefix }}%
\providecommand \urlprefix  [0]{URL }%
\providecommand \Eprint [0]{\href }%
\providecommand \doibase [0]{https://doi.org/}%
\providecommand \selectlanguage [0]{\@gobble}%
\providecommand \bibinfo  [0]{\@secondoftwo}%
\providecommand \bibfield  [0]{\@secondoftwo}%
\providecommand \translation [1]{[#1]}%
\providecommand \BibitemOpen [0]{}%
\providecommand \bibitemStop [0]{}%
\providecommand \bibitemNoStop [0]{.\EOS\space}%
\providecommand \EOS [0]{\spacefactor3000\relax}%
\providecommand \BibitemShut  [1]{\csname bibitem#1\endcsname}%
\let\auto@bib@innerbib\@empty
\bibitem [{\citenamefont {Solovyev}(1997)}]{Solovyev1997}%
  \BibitemOpen
  \bibfield  {author} {\bibinfo {author} {\bibfnamefont {I.~V.}\ \bibnamefont
  {Solovyev}},\ }\bibfield  {title} {\bibinfo {title} {{Magneto-optical effect
  in the weak ferromagnets ${\mathrm{LaMO}}_{3}$ (M= Cr, Mn, and Fe)}},\ }\href
  {https://doi.org/10.1103/PhysRevB.55.8060} {\bibfield  {journal} {\bibinfo
  {journal} {Phys. Rev. B}\ }\textbf {\bibinfo {volume} {55}},\ \bibinfo
  {pages} {8060} (\bibinfo {year} {1997})}\BibitemShut {NoStop}%
\bibitem [{\citenamefont {Tomizawa}\ and\ \citenamefont
  {Kontani}(2009)}]{Tomizawa2009}%
  \BibitemOpen
  \bibfield  {author} {\bibinfo {author} {\bibfnamefont {T.}~\bibnamefont
  {Tomizawa}}\ and\ \bibinfo {author} {\bibfnamefont {H.}~\bibnamefont
  {Kontani}},\ }\bibfield  {title} {\bibinfo {title} {{Anomalous Hall effect in
  the ${t}_{2g}$ orbital kagome lattice due to noncollinearity: Significance of
  the orbital Aharonov-Bohm effect}},\ }\href
  {https://doi.org/10.1103/PhysRevB.80.100401} {\bibfield  {journal} {\bibinfo
  {journal} {Phys. Rev. B}\ }\textbf {\bibinfo {volume} {80}},\ \bibinfo
  {pages} {100401} (\bibinfo {year} {2009})}\BibitemShut {NoStop}%
\bibitem [{\citenamefont {Ohgushi}\ \emph {et~al.}(2000)\citenamefont
  {Ohgushi}, \citenamefont {Murakami},\ and\ \citenamefont
  {Nagaosa}}]{Ohgushi2000}%
  \BibitemOpen
  \bibfield  {author} {\bibinfo {author} {\bibfnamefont {K.}~\bibnamefont
  {Ohgushi}}, \bibinfo {author} {\bibfnamefont {S.}~\bibnamefont {Murakami}},\
  and\ \bibinfo {author} {\bibfnamefont {N.}~\bibnamefont {Nagaosa}},\
  }\bibfield  {title} {\bibinfo {title} {{Spin anisotropy and quantum Hall
  effect in the kagom\'e lattice: Chiral spin state based on a ferromagnet}},\
  }\href {https://doi.org/10.1103/PhysRevB.62.R6065} {\bibfield  {journal}
  {\bibinfo  {journal} {Phys. Rev. B}\ }\textbf {\bibinfo {volume} {62}},\
  \bibinfo {pages} {R6065} (\bibinfo {year} {2000})}\BibitemShut {NoStop}%
\bibitem [{\citenamefont {Shindou}\ and\ \citenamefont
  {Nagaosa}(2001)}]{Shindou2001}%
  \BibitemOpen
  \bibfield  {author} {\bibinfo {author} {\bibfnamefont {R.}~\bibnamefont
  {Shindou}}\ and\ \bibinfo {author} {\bibfnamefont {N.}~\bibnamefont
  {Nagaosa}},\ }\bibfield  {title} {\bibinfo {title} {{Orbital Ferromagnetism
  and Anomalous Hall Effect in Antiferromagnets on the Distorted fcc
  Lattice}},\ }\href {https://doi.org/10.1103/PhysRevLett.87.116801} {\bibfield
   {journal} {\bibinfo  {journal} {Phys. Rev. Lett.}\ }\textbf {\bibinfo
  {volume} {87}},\ \bibinfo {pages} {116801} (\bibinfo {year}
  {2001})}\BibitemShut {NoStop}%
\bibitem [{\citenamefont {Chen}\ \emph {et~al.}(2014)\citenamefont {Chen},
  \citenamefont {Niu},\ and\ \citenamefont {MacDonald}}]{chen_2014}%
  \BibitemOpen
  \bibfield  {author} {\bibinfo {author} {\bibfnamefont {H.}~\bibnamefont
  {Chen}}, \bibinfo {author} {\bibfnamefont {Q.}~\bibnamefont {Niu}},\ and\
  \bibinfo {author} {\bibfnamefont {A.~H.}\ \bibnamefont {MacDonald}},\
  }\bibfield  {title} {\bibinfo {title} {{Anomalous Hall Effect Arising from
  Noncollinear Antiferromagnetism}},\ }\href
  {https://doi.org/10.1103/PhysRevLett.112.017205} {\bibfield  {journal}
  {\bibinfo  {journal} {Phys. Rev. Lett.}\ }\textbf {\bibinfo {volume} {112}},\
  \bibinfo {pages} {017205} (\bibinfo {year} {2014})}\BibitemShut {NoStop}%
\bibitem [{\citenamefont {K\"{u}bler}\ and\ \citenamefont
  {Felser}(2014)}]{Kubler_2014}%
  \BibitemOpen
  \bibfield  {author} {\bibinfo {author} {\bibfnamefont {J.}~\bibnamefont
  {K\"{u}bler}}\ and\ \bibinfo {author} {\bibfnamefont {C.}~\bibnamefont
  {Felser}},\ }\bibfield  {title} {\bibinfo {title} {{Non-collinear
  antiferromagnets and the anomalous Hall effect}},\ }\href
  {https://doi.org/10.1209/0295-5075/108/67001} {\bibfield  {journal} {\bibinfo
   {journal} {{EPL} (Europhysics Letters)}\ }\textbf {\bibinfo {volume}
  {108}},\ \bibinfo {pages} {67001} (\bibinfo {year} {2014})}\BibitemShut
  {NoStop}%
\bibitem [{\citenamefont {Nakatsuji}\ \emph {et~al.}(2015)\citenamefont
  {Nakatsuji}, \citenamefont {Kiyohara},\ and\ \citenamefont
  {Higo}}]{Nakatsuji_2015}%
  \BibitemOpen
  \bibfield  {author} {\bibinfo {author} {\bibfnamefont {S.}~\bibnamefont
  {Nakatsuji}}, \bibinfo {author} {\bibfnamefont {N.}~\bibnamefont
  {Kiyohara}},\ and\ \bibinfo {author} {\bibfnamefont {T.}~\bibnamefont
  {Higo}},\ }\bibfield  {title} {\bibinfo {title} {{Large anomalous Hall effect
  in a non-collinear antiferromagnet at room temperature}},\ }\href@noop {}
  {\bibfield  {journal} {\bibinfo  {journal} {Nature}\ }\textbf {\bibinfo
  {volume} {527}},\ \bibinfo {pages} {212} (\bibinfo {year}
  {2015})}\BibitemShut {NoStop}%
\bibitem [{\citenamefont {Nayak}\ \emph {et~al.}(2016)\citenamefont {Nayak},
  \citenamefont {Fischer}, \citenamefont {Sun}, \citenamefont {Yan},
  \citenamefont {Karel}, \citenamefont {Komarek}, \citenamefont {Shekhar},
  \citenamefont {Kumar}, \citenamefont {Schnelle}, \citenamefont {K{\"u}bler},
  \citenamefont {Felser},\ and\ \citenamefont {Parkin}}]{Nayak2016}%
  \BibitemOpen
  \bibfield  {author} {\bibinfo {author} {\bibfnamefont {A.~K.}\ \bibnamefont
  {Nayak}}, \bibinfo {author} {\bibfnamefont {J.~E.}\ \bibnamefont {Fischer}},
  \bibinfo {author} {\bibfnamefont {Y.}~\bibnamefont {Sun}}, \bibinfo {author}
  {\bibfnamefont {B.}~\bibnamefont {Yan}}, \bibinfo {author} {\bibfnamefont
  {J.}~\bibnamefont {Karel}}, \bibinfo {author} {\bibfnamefont {A.~C.}\
  \bibnamefont {Komarek}}, \bibinfo {author} {\bibfnamefont {C.}~\bibnamefont
  {Shekhar}}, \bibinfo {author} {\bibfnamefont {N.}~\bibnamefont {Kumar}},
  \bibinfo {author} {\bibfnamefont {W.}~\bibnamefont {Schnelle}}, \bibinfo
  {author} {\bibfnamefont {J.}~\bibnamefont {K{\"u}bler}}, \bibinfo {author}
  {\bibfnamefont {C.}~\bibnamefont {Felser}},\ and\ \bibinfo {author}
  {\bibfnamefont {S.~S.~P.}\ \bibnamefont {Parkin}},\ }\bibfield  {title}
  {\bibinfo {title} {{Large anomalous Hall effect driven by a nonvanishing
  Berry curvature in the noncolinear antiferromagnet Mn$_3$Ge}},\ }\href
  {https://doi.org/10.1126/sciadv.1501870} {\bibfield  {journal} {\bibinfo
  {journal} {Science Advances}\ }\textbf {\bibinfo {volume} {2}},\ \bibinfo
  {pages} {e1501870} (\bibinfo {year} {2016})}\BibitemShut {NoStop}%
\bibitem [{\citenamefont {Zhou}\ \emph {et~al.}(2019)\citenamefont {Zhou},
  \citenamefont {Hanke}, \citenamefont {Feng}, \citenamefont {Li},
  \citenamefont {Guo}, \citenamefont {Yao}, \citenamefont {Bl\"ugel},\ and\
  \citenamefont {Mokrousov}}]{Zhou2019}%
  \BibitemOpen
  \bibfield  {author} {\bibinfo {author} {\bibfnamefont {X.}~\bibnamefont
  {Zhou}}, \bibinfo {author} {\bibfnamefont {J.-P.}\ \bibnamefont {Hanke}},
  \bibinfo {author} {\bibfnamefont {W.}~\bibnamefont {Feng}}, \bibinfo {author}
  {\bibfnamefont {F.}~\bibnamefont {Li}}, \bibinfo {author} {\bibfnamefont
  {G.-Y.}\ \bibnamefont {Guo}}, \bibinfo {author} {\bibfnamefont
  {Y.}~\bibnamefont {Yao}}, \bibinfo {author} {\bibfnamefont {S.}~\bibnamefont
  {Bl\"ugel}},\ and\ \bibinfo {author} {\bibfnamefont {Y.}~\bibnamefont
  {Mokrousov}},\ }\bibfield  {title} {\bibinfo {title} {{Spin-order dependent
  anomalous Hall effect and magneto-optical effect in the noncollinear
  antiferromagnets ${\mathrm{Mn}}_{3}X\mathrm{N}$ with $X=\mathrm{Ga}$, Zn, Ag,
  or Ni}},\ }\href {https://doi.org/10.1103/PhysRevB.99.104428} {\bibfield
  {journal} {\bibinfo  {journal} {Phys. Rev. B}\ }\textbf {\bibinfo {volume}
  {99}},\ \bibinfo {pages} {104428} (\bibinfo {year} {2019})}\BibitemShut
  {NoStop}%
\bibitem [{\citenamefont {Gurung}\ \emph {et~al.}(2019)\citenamefont {Gurung},
  \citenamefont {Shao}, \citenamefont {Paudel},\ and\ \citenamefont
  {Tsymbal}}]{Gurung2019}%
  \BibitemOpen
  \bibfield  {author} {\bibinfo {author} {\bibfnamefont {G.}~\bibnamefont
  {Gurung}}, \bibinfo {author} {\bibfnamefont {D.-F.}\ \bibnamefont {Shao}},
  \bibinfo {author} {\bibfnamefont {T.~R.}\ \bibnamefont {Paudel}},\ and\
  \bibinfo {author} {\bibfnamefont {E.~Y.}\ \bibnamefont {Tsymbal}},\
  }\bibfield  {title} {\bibinfo {title} {{Anomalous Hall conductivity of
  noncollinear magnetic antiperovskites}},\ }\href
  {https://doi.org/10.1103/PhysRevMaterials.3.044409} {\bibfield  {journal}
  {\bibinfo  {journal} {Phys. Rev. Materials}\ }\textbf {\bibinfo {volume}
  {3}},\ \bibinfo {pages} {044409} (\bibinfo {year} {2019})}\BibitemShut
  {NoStop}%
\bibitem [{\citenamefont {Boldrin}\ \emph {et~al.}(2019)\citenamefont
  {Boldrin}, \citenamefont {Samathrakis}, \citenamefont {Zemen}, \citenamefont
  {Mihai}, \citenamefont {Zou}, \citenamefont {Johnson}, \citenamefont {Esser},
  \citenamefont {McComb}, \citenamefont {Petrov}, \citenamefont {Zhang},\ and\
  \citenamefont {Cohen}}]{Boldrin2019}%
  \BibitemOpen
  \bibfield  {author} {\bibinfo {author} {\bibfnamefont {D.}~\bibnamefont
  {Boldrin}}, \bibinfo {author} {\bibfnamefont {I.}~\bibnamefont
  {Samathrakis}}, \bibinfo {author} {\bibfnamefont {J.}~\bibnamefont {Zemen}},
  \bibinfo {author} {\bibfnamefont {A.}~\bibnamefont {Mihai}}, \bibinfo
  {author} {\bibfnamefont {B.}~\bibnamefont {Zou}}, \bibinfo {author}
  {\bibfnamefont {F.}~\bibnamefont {Johnson}}, \bibinfo {author} {\bibfnamefont
  {B.~D.}\ \bibnamefont {Esser}}, \bibinfo {author} {\bibfnamefont {D.~W.}\
  \bibnamefont {McComb}}, \bibinfo {author} {\bibfnamefont {P.~K.}\
  \bibnamefont {Petrov}}, \bibinfo {author} {\bibfnamefont {H.}~\bibnamefont
  {Zhang}},\ and\ \bibinfo {author} {\bibfnamefont {L.~F.}\ \bibnamefont
  {Cohen}},\ }\bibfield  {title} {\bibinfo {title} {{Anomalous Hall effect in
  noncollinear antiferromagnetic ${\mathrm{Mn}}_{3}\mathrm{NiN}$ thin films}},\
  }\href {https://doi.org/10.1103/PhysRevMaterials.3.094409} {\bibfield
  {journal} {\bibinfo  {journal} {Phys. Rev. Materials}\ }\textbf {\bibinfo
  {volume} {3}},\ \bibinfo {pages} {094409} (\bibinfo {year}
  {2019})}\BibitemShut {NoStop}%
\bibitem [{\citenamefont {Zhao}\ \emph {et~al.}(2019)\citenamefont {Zhao},
  \citenamefont {Hajiri}, \citenamefont {Chen}, \citenamefont {Miki},
  \citenamefont {Asano},\ and\ \citenamefont {Gegenwart}}]{Zhao2019}%
  \BibitemOpen
  \bibfield  {author} {\bibinfo {author} {\bibfnamefont {K.}~\bibnamefont
  {Zhao}}, \bibinfo {author} {\bibfnamefont {T.}~\bibnamefont {Hajiri}},
  \bibinfo {author} {\bibfnamefont {H.}~\bibnamefont {Chen}}, \bibinfo {author}
  {\bibfnamefont {R.}~\bibnamefont {Miki}}, \bibinfo {author} {\bibfnamefont
  {H.}~\bibnamefont {Asano}},\ and\ \bibinfo {author} {\bibfnamefont
  {P.}~\bibnamefont {Gegenwart}},\ }\bibfield  {title} {\bibinfo {title}
  {{Anomalous Hall effect in the noncollinear antiferromagnetic antiperovskite
  ${\mathrm{Mn}}_{3}{\mathrm{Ni}}_{1\ensuremath{-}x}{\mathrm{Cu}}_{x}\mathrm{N}$}},\
  }\href {https://doi.org/10.1103/PhysRevB.100.045109} {\bibfield  {journal}
  {\bibinfo  {journal} {Phys. Rev. B}\ }\textbf {\bibinfo {volume} {100}},\
  \bibinfo {pages} {045109} (\bibinfo {year} {2019})}\BibitemShut {NoStop}%
\bibitem [{\citenamefont {Liu}\ \emph {et~al.}(2018)\citenamefont {Liu},
  \citenamefont {Chen}, \citenamefont {Wang}, \citenamefont {Liu},
  \citenamefont {Wang}, \citenamefont {Feng}, \citenamefont {Yan},
  \citenamefont {Wang}, \citenamefont {Jiang}, \citenamefont {Coey},\ and\
  \citenamefont {MacDonald}}]{Liu2018}%
  \BibitemOpen
  \bibfield  {author} {\bibinfo {author} {\bibfnamefont {Z.~Q.}\ \bibnamefont
  {Liu}}, \bibinfo {author} {\bibfnamefont {H.}~\bibnamefont {Chen}}, \bibinfo
  {author} {\bibfnamefont {J.~M.}\ \bibnamefont {Wang}}, \bibinfo {author}
  {\bibfnamefont {J.~H.}\ \bibnamefont {Liu}}, \bibinfo {author} {\bibfnamefont
  {K.}~\bibnamefont {Wang}}, \bibinfo {author} {\bibfnamefont {Z.~X.}\
  \bibnamefont {Feng}}, \bibinfo {author} {\bibfnamefont {H.}~\bibnamefont
  {Yan}}, \bibinfo {author} {\bibfnamefont {X.~R.}\ \bibnamefont {Wang}},
  \bibinfo {author} {\bibfnamefont {C.~B.}\ \bibnamefont {Jiang}}, \bibinfo
  {author} {\bibfnamefont {J.~M.~D.}\ \bibnamefont {Coey}},\ and\ \bibinfo
  {author} {\bibfnamefont {A.~H.}\ \bibnamefont {MacDonald}},\ }\bibfield
  {title} {\bibinfo {title} {{Electrical switching of the topological anomalous
  Hall effect in a non-collinear antiferromagnet above room temperature}},\
  }\href {https://doi.org/10.1038/s41928-018-0040-1} {\bibfield  {journal}
  {\bibinfo  {journal} {Nature Electronics}\ }\textbf {\bibinfo {volume} {1}},\
  \bibinfo {pages} {172} (\bibinfo {year} {2018})}\BibitemShut {NoStop}%
\bibitem [{\citenamefont {{\v S}mejkal}\ \emph {et~al.}(2020)\citenamefont {{\v
  S}mejkal}, \citenamefont {Gonz{\'a}lez-Hern{\'a}ndez}, \citenamefont
  {Jungwirth},\ and\ \citenamefont {Sinova}}]{Smejkal2020}%
  \BibitemOpen
  \bibfield  {author} {\bibinfo {author} {\bibfnamefont {L.}~\bibnamefont {{\v
  S}mejkal}}, \bibinfo {author} {\bibfnamefont {R.}~\bibnamefont
  {Gonz{\'a}lez-Hern{\'a}ndez}}, \bibinfo {author} {\bibfnamefont
  {T.}~\bibnamefont {Jungwirth}},\ and\ \bibinfo {author} {\bibfnamefont
  {J.}~\bibnamefont {Sinova}},\ }\bibfield  {title} {\bibinfo {title} {{Crystal
  time-reversal symmetry breaking and spontaneous Hall effect in collinear
  antiferromagnets}},\ }\href {https://doi.org/10.1126/sciadv.aaz8809}
  {\bibfield  {journal} {\bibinfo  {journal} {Science Advances}\ }\textbf
  {\bibinfo {volume} {6}},\ \bibinfo {pages} {eaaz8809} (\bibinfo {year}
  {2020})}\BibitemShut {NoStop}%
\bibitem [{\citenamefont {Chen}\ \emph {et~al.}(2020)\citenamefont {Chen},
  \citenamefont {Wang}, \citenamefont {Xiao}, \citenamefont {Guo},
  \citenamefont {Niu},\ and\ \citenamefont {MacDonald}}]{chen_2020}%
  \BibitemOpen
  \bibfield  {author} {\bibinfo {author} {\bibfnamefont {H.}~\bibnamefont
  {Chen}}, \bibinfo {author} {\bibfnamefont {T.-C.}\ \bibnamefont {Wang}},
  \bibinfo {author} {\bibfnamefont {D.}~\bibnamefont {Xiao}}, \bibinfo {author}
  {\bibfnamefont {G.-Y.}\ \bibnamefont {Guo}}, \bibinfo {author} {\bibfnamefont
  {Q.}~\bibnamefont {Niu}},\ and\ \bibinfo {author} {\bibfnamefont {A.~H.}\
  \bibnamefont {MacDonald}},\ }\bibfield  {title} {\bibinfo {title}
  {{Manipulating anomalous Hall antiferromagnets with magnetic fields}},\
  }\href {https://doi.org/10.1103/PhysRevB.101.104418} {\bibfield  {journal}
  {\bibinfo  {journal} {Phys. Rev. B}\ }\textbf {\bibinfo {volume} {101}},\
  \bibinfo {pages} {104418} (\bibinfo {year} {2020})}\BibitemShut {NoStop}%
\bibitem [{\citenamefont {Chen}(2022)}]{chen_2022}%
  \BibitemOpen
  \bibfield  {author} {\bibinfo {author} {\bibfnamefont {H.}~\bibnamefont
  {Chen}},\ }\bibfield  {title} {\bibinfo {title} {{Electronic chiralization as
  an indicator of the anomalous Hall effect in unconventional magnetic
  systems}},\ }\href {https://doi.org/10.1103/PhysRevB.106.024421} {\bibfield
  {journal} {\bibinfo  {journal} {Phys. Rev. B}\ }\textbf {\bibinfo {volume}
  {106}},\ \bibinfo {pages} {024421} (\bibinfo {year} {2022})}\BibitemShut
  {NoStop}%
\bibitem [{\citenamefont {Jackson}(1999)}]{Jackson:490457}%
  \BibitemOpen
  \bibfield  {author} {\bibinfo {author} {\bibfnamefont {J.~D.}\ \bibnamefont
  {Jackson}},\ }\href {https://cds.cern.ch/record/490457} {\emph {\bibinfo
  {title} {{Classical electrodynamics; 3rd ed.}}}}\ (\bibinfo  {publisher}
  {Wiley},\ \bibinfo {address} {New York, NY},\ \bibinfo {year}
  {1999})\BibitemShut {NoStop}%
\bibitem [{\citenamefont {Andreev}(1978)}]{Andreev1978}%
  \BibitemOpen
  \bibfield  {author} {\bibinfo {author} {\bibfnamefont {A.~F.}\ \bibnamefont
  {Andreev}},\ }\bibfield  {title} {\bibinfo {title} {{Magnetic properties of
  disordered media}},\ }\href {https://doi.org/10.1070/pu1978v021n06abeh005565}
  {\bibfield  {journal} {\bibinfo  {journal} {Soviet Physics Uspekhi}\ }\textbf
  {\bibinfo {volume} {21}},\ \bibinfo {pages} {541} (\bibinfo {year}
  {1978})}\BibitemShut {NoStop}%
\bibitem [{\citenamefont {Andreev}\ and\ \citenamefont
  {Marchenko}(1980)}]{Andreev_1980}%
  \BibitemOpen
  \bibfield  {author} {\bibinfo {author} {\bibfnamefont {A.~F.}\ \bibnamefont
  {Andreev}}\ and\ \bibinfo {author} {\bibfnamefont {V.~I.}\ \bibnamefont
  {Marchenko}},\ }\bibfield  {title} {\bibinfo {title} {{Symmetry and the
  macroscopic dynamics of magnetic materials}},\ }\href
  {https://doi.org/10.1070/pu1980v023n01abeh004859} {\bibfield  {journal}
  {\bibinfo  {journal} {Soviet Physics Uspekhi}\ }\textbf {\bibinfo {volume}
  {23}},\ \bibinfo {pages} {21} (\bibinfo {year} {1980})}\BibitemShut {NoStop}%
\bibitem [{\citenamefont {Dzyaloshinskii}(1992)}]{DZYALOSHINSKII1992579}%
  \BibitemOpen
  \bibfield  {author} {\bibinfo {author} {\bibfnamefont {I.}~\bibnamefont
  {Dzyaloshinskii}},\ }\bibfield  {title} {\bibinfo {title} {External magnetic
  fields of antiferromagnets},\ }\href
  {https://doi.org/https://doi.org/10.1016/0038-1098(92)90236-3} {\bibfield
  {journal} {\bibinfo  {journal} {Solid State Communications}\ }\textbf
  {\bibinfo {volume} {82}},\ \bibinfo {pages} {579 } (\bibinfo {year}
  {1992})}\BibitemShut {NoStop}%
\bibitem [{\citenamefont {Andreev}(1996)}]{Andreev1996}%
  \BibitemOpen
  \bibfield  {author} {\bibinfo {author} {\bibfnamefont {A.~F.}\ \bibnamefont
  {Andreev}},\ }\bibfield  {title} {\bibinfo {title} {{Macroscopic magnetic
  fields of antiferromagnets}},\ }\href {https://doi.org/10.1134/1.566978}
  {\bibfield  {journal} {\bibinfo  {journal} {Journal of Experimental and
  Theoretical Physics Letters}\ }\textbf {\bibinfo {volume} {63}},\ \bibinfo
  {pages} {758} (\bibinfo {year} {1996})}\BibitemShut {NoStop}%
\bibitem [{\citenamefont {Astrov}\ \emph {et~al.}(1996)\citenamefont {Astrov},
  \citenamefont {Ermakov}, \citenamefont {Borovik-Romanov}, \citenamefont
  {Kolevatov},\ and\ \citenamefont {Nizhankovskii}}]{Astrov1996}%
  \BibitemOpen
  \bibfield  {author} {\bibinfo {author} {\bibfnamefont {D.~N.}\ \bibnamefont
  {Astrov}}, \bibinfo {author} {\bibfnamefont {N.~B.}\ \bibnamefont {Ermakov}},
  \bibinfo {author} {\bibfnamefont {A.~S.}\ \bibnamefont {Borovik-Romanov}},
  \bibinfo {author} {\bibfnamefont {E.~G.}\ \bibnamefont {Kolevatov}},\ and\
  \bibinfo {author} {\bibfnamefont {V.~I.}\ \bibnamefont {Nizhankovskii}},\
  }\bibfield  {title} {\bibinfo {title} {{External quadrupole magnetic field of
  antiferromagnetic Cr2O3}},\ }\href {https://doi.org/10.1134/1.566976}
  {\bibfield  {journal} {\bibinfo  {journal} {Journal of Experimental and
  Theoretical Physics Letters}\ }\textbf {\bibinfo {volume} {63}},\ \bibinfo
  {pages} {745} (\bibinfo {year} {1996})}\BibitemShut {NoStop}%
\bibitem [{\citenamefont {Kuramoto}\ \emph {et~al.}(2009)\citenamefont
  {Kuramoto}, \citenamefont {Kusunose},\ and\ \citenamefont
  {Kiss}}]{kuramoto_2009}%
  \BibitemOpen
  \bibfield  {author} {\bibinfo {author} {\bibfnamefont {Y.}~\bibnamefont
  {Kuramoto}}, \bibinfo {author} {\bibfnamefont {H.}~\bibnamefont {Kusunose}},\
  and\ \bibinfo {author} {\bibfnamefont {A.}~\bibnamefont {Kiss}},\ }\bibfield
  {title} {\bibinfo {title} {{Multipole Orders and Fluctuations in Strongly
  Correlated Electron Systems}},\ }\href
  {https://doi.org/10.1143/JPSJ.78.072001} {\bibfield  {journal} {\bibinfo
  {journal} {Journal of the Physical Society of Japan}\ }\textbf {\bibinfo
  {volume} {78}},\ \bibinfo {pages} {072001} (\bibinfo {year} {2009})},\
  \Eprint {https://arxiv.org/abs/https://doi.org/10.1143/JPSJ.78.072001}
  {https://doi.org/10.1143/JPSJ.78.072001} \BibitemShut {NoStop}%
\bibitem [{\citenamefont {Santini}\ \emph {et~al.}(2009)\citenamefont
  {Santini}, \citenamefont {Carretta}, \citenamefont {Amoretti}, \citenamefont
  {Caciuffo}, \citenamefont {Magnani},\ and\ \citenamefont
  {Lander}}]{santini_2009}%
  \BibitemOpen
  \bibfield  {author} {\bibinfo {author} {\bibfnamefont {P.}~\bibnamefont
  {Santini}}, \bibinfo {author} {\bibfnamefont {S.}~\bibnamefont {Carretta}},
  \bibinfo {author} {\bibfnamefont {G.}~\bibnamefont {Amoretti}}, \bibinfo
  {author} {\bibfnamefont {R.}~\bibnamefont {Caciuffo}}, \bibinfo {author}
  {\bibfnamefont {N.}~\bibnamefont {Magnani}},\ and\ \bibinfo {author}
  {\bibfnamefont {G.~H.}\ \bibnamefont {Lander}},\ }\bibfield  {title}
  {\bibinfo {title} {{Multipolar interactions in $f$-electron systems: The
  paradigm of actinide dioxides}},\ }\href
  {https://doi.org/10.1103/RevModPhys.81.807} {\bibfield  {journal} {\bibinfo
  {journal} {Rev. Mod. Phys.}\ }\textbf {\bibinfo {volume} {81}},\ \bibinfo
  {pages} {807} (\bibinfo {year} {2009})}\BibitemShut {NoStop}%
\bibitem [{\citenamefont {Sakai}\ and\ \citenamefont
  {Nakatsuji}(2011)}]{sakai_2011}%
  \BibitemOpen
  \bibfield  {author} {\bibinfo {author} {\bibfnamefont {A.}~\bibnamefont
  {Sakai}}\ and\ \bibinfo {author} {\bibfnamefont {S.}~\bibnamefont
  {Nakatsuji}},\ }\bibfield  {title} {\bibinfo {title} {{Kondo Effects and
  Multipolar Order in the Cubic PrTr2Al20 (Tr=Ti, V)}},\ }\href
  {https://doi.org/10.1143/JPSJ.80.063701} {\bibfield  {journal} {\bibinfo
  {journal} {Journal of the Physical Society of Japan}\ }\textbf {\bibinfo
  {volume} {80}},\ \bibinfo {pages} {063701} (\bibinfo {year} {2011})},\
  \Eprint {https://arxiv.org/abs/https://doi.org/10.1143/JPSJ.80.063701}
  {https://doi.org/10.1143/JPSJ.80.063701} \BibitemShut {NoStop}%
\bibitem [{\citenamefont {Onimaru}\ \emph {et~al.}(2011)\citenamefont
  {Onimaru}, \citenamefont {Matsumoto}, \citenamefont {Inoue}, \citenamefont
  {Umeo}, \citenamefont {Sakakibara}, \citenamefont {Karaki}, \citenamefont
  {Kubota},\ and\ \citenamefont {Takabatake}}]{onimaru_2011}%
  \BibitemOpen
  \bibfield  {author} {\bibinfo {author} {\bibfnamefont {T.}~\bibnamefont
  {Onimaru}}, \bibinfo {author} {\bibfnamefont {K.~T.}\ \bibnamefont
  {Matsumoto}}, \bibinfo {author} {\bibfnamefont {Y.~F.}\ \bibnamefont
  {Inoue}}, \bibinfo {author} {\bibfnamefont {K.}~\bibnamefont {Umeo}},
  \bibinfo {author} {\bibfnamefont {T.}~\bibnamefont {Sakakibara}}, \bibinfo
  {author} {\bibfnamefont {Y.}~\bibnamefont {Karaki}}, \bibinfo {author}
  {\bibfnamefont {M.}~\bibnamefont {Kubota}},\ and\ \bibinfo {author}
  {\bibfnamefont {T.}~\bibnamefont {Takabatake}},\ }\bibfield  {title}
  {\bibinfo {title} {{Antiferroquadrupolar Ordering in a Pr-Based
  Superconductor ${\mathrm{PrIr}}_{2}{\mathrm{Zn}}_{20}$}},\ }\href
  {https://doi.org/10.1103/PhysRevLett.106.177001} {\bibfield  {journal}
  {\bibinfo  {journal} {Phys. Rev. Lett.}\ }\textbf {\bibinfo {volume} {106}},\
  \bibinfo {pages} {177001} (\bibinfo {year} {2011})}\BibitemShut {NoStop}%
\bibitem [{\citenamefont {Onimaru}\ and\ \citenamefont
  {Kusunose}(2016)}]{onimaru_2016}%
  \BibitemOpen
  \bibfield  {author} {\bibinfo {author} {\bibfnamefont {T.}~\bibnamefont
  {Onimaru}}\ and\ \bibinfo {author} {\bibfnamefont {H.}~\bibnamefont
  {Kusunose}},\ }\bibfield  {title} {\bibinfo {title} {{Exotic Quadrupolar
  Phenomena in Non-Kramers Doublet Systems — The Cases of PrT2Zn20 (T = Ir,
  Rh) and PrT2Al20 (T = V, Ti) —}},\ }\href
  {https://doi.org/10.7566/JPSJ.85.082002} {\bibfield  {journal} {\bibinfo
  {journal} {Journal of the Physical Society of Japan}\ }\textbf {\bibinfo
  {volume} {85}},\ \bibinfo {pages} {082002} (\bibinfo {year} {2016})},\
  \Eprint {https://arxiv.org/abs/https://doi.org/10.7566/JPSJ.85.082002}
  {https://doi.org/10.7566/JPSJ.85.082002} \BibitemShut {NoStop}%
\bibitem [{\citenamefont {Kubo}\ and\ \citenamefont
  {Kuramoto}(2004)}]{kubo_2004}%
  \BibitemOpen
  \bibfield  {author} {\bibinfo {author} {\bibfnamefont {K.}~\bibnamefont
  {Kubo}}\ and\ \bibinfo {author} {\bibfnamefont {Y.}~\bibnamefont
  {Kuramoto}},\ }\bibfield  {title} {\bibinfo {title} {{Octupole Ordering Model
  for the Phase IV of CexLa1-xB6}},\ }\href
  {https://doi.org/10.1143/JPSJ.73.216} {\bibfield  {journal} {\bibinfo
  {journal} {Journal of the Physical Society of Japan}\ }\textbf {\bibinfo
  {volume} {73}},\ \bibinfo {pages} {216} (\bibinfo {year} {2004})},\ \Eprint
  {https://arxiv.org/abs/https://doi.org/10.1143/JPSJ.73.216}
  {https://doi.org/10.1143/JPSJ.73.216} \BibitemShut {NoStop}%
\bibitem [{\citenamefont {Mannix}\ \emph {et~al.}(2005)\citenamefont {Mannix},
  \citenamefont {Tanaka}, \citenamefont {Carbone}, \citenamefont {Bernhoeft},\
  and\ \citenamefont {Kunii}}]{mannix_2005}%
  \BibitemOpen
  \bibfield  {author} {\bibinfo {author} {\bibfnamefont {D.}~\bibnamefont
  {Mannix}}, \bibinfo {author} {\bibfnamefont {Y.}~\bibnamefont {Tanaka}},
  \bibinfo {author} {\bibfnamefont {D.}~\bibnamefont {Carbone}}, \bibinfo
  {author} {\bibfnamefont {N.}~\bibnamefont {Bernhoeft}},\ and\ \bibinfo
  {author} {\bibfnamefont {S.}~\bibnamefont {Kunii}},\ }\bibfield  {title}
  {\bibinfo {title} {{Order Parameter Segregation in
  ${\mathrm{Ce}}_{0.7}{\mathrm{La}}_{0.3}{\mathrm{B}}_{6}$: $4f$ Octopole and
  $5d$ Dipole Magnetic Order}},\ }\href
  {https://doi.org/10.1103/PhysRevLett.95.117206} {\bibfield  {journal}
  {\bibinfo  {journal} {Phys. Rev. Lett.}\ }\textbf {\bibinfo {volume} {95}},\
  \bibinfo {pages} {117206} (\bibinfo {year} {2005})}\BibitemShut {NoStop}%
\bibitem [{\citenamefont {Kuwahara}\ \emph {et~al.}(2007)\citenamefont
  {Kuwahara}, \citenamefont {Iwasa}, \citenamefont {Kohgi}, \citenamefont
  {Aso}, \citenamefont {Sera},\ and\ \citenamefont {Iga}}]{kuwahara_2007}%
  \BibitemOpen
  \bibfield  {author} {\bibinfo {author} {\bibfnamefont {K.}~\bibnamefont
  {Kuwahara}}, \bibinfo {author} {\bibfnamefont {K.}~\bibnamefont {Iwasa}},
  \bibinfo {author} {\bibfnamefont {M.}~\bibnamefont {Kohgi}}, \bibinfo
  {author} {\bibfnamefont {N.}~\bibnamefont {Aso}}, \bibinfo {author}
  {\bibfnamefont {M.}~\bibnamefont {Sera}},\ and\ \bibinfo {author}
  {\bibfnamefont {F.}~\bibnamefont {Iga}},\ }\bibfield  {title} {\bibinfo
  {title} {{Detection of Neutron Scattering from Phase IV of Ce0.7La0.3B6: A
  Confirmation of the Octupole Order}},\ }\href
  {https://doi.org/10.1143/JPSJ.76.093702} {\bibfield  {journal} {\bibinfo
  {journal} {Journal of the Physical Society of Japan}\ }\textbf {\bibinfo
  {volume} {76}},\ \bibinfo {pages} {093702} (\bibinfo {year} {2007})},\
  \Eprint {https://arxiv.org/abs/https://doi.org/10.1143/JPSJ.76.093702}
  {https://doi.org/10.1143/JPSJ.76.093702} \BibitemShut {NoStop}%
\bibitem [{\citenamefont {Matsumura}\ \emph {et~al.}(2009)\citenamefont
  {Matsumura}, \citenamefont {Yonemura}, \citenamefont {Kunimori},
  \citenamefont {Sera},\ and\ \citenamefont {Iga}}]{matsumura_2009}%
  \BibitemOpen
  \bibfield  {author} {\bibinfo {author} {\bibfnamefont {T.}~\bibnamefont
  {Matsumura}}, \bibinfo {author} {\bibfnamefont {T.}~\bibnamefont {Yonemura}},
  \bibinfo {author} {\bibfnamefont {K.}~\bibnamefont {Kunimori}}, \bibinfo
  {author} {\bibfnamefont {M.}~\bibnamefont {Sera}},\ and\ \bibinfo {author}
  {\bibfnamefont {F.}~\bibnamefont {Iga}},\ }\bibfield  {title} {\bibinfo
  {title} {{Magnetic Field Induced $4f$ Octupole in ${\mathrm{CeB}}_{6}$ Probed
  by Resonant X-Ray Diffraction}},\ }\href
  {https://doi.org/10.1103/PhysRevLett.103.017203} {\bibfield  {journal}
  {\bibinfo  {journal} {Phys. Rev. Lett.}\ }\textbf {\bibinfo {volume} {103}},\
  \bibinfo {pages} {017203} (\bibinfo {year} {2009})}\BibitemShut {NoStop}%
\bibitem [{\citenamefont {Watanabe}\ and\ \citenamefont
  {Yanase}(2018)}]{watanabe_2018}%
  \BibitemOpen
  \bibfield  {author} {\bibinfo {author} {\bibfnamefont {H.}~\bibnamefont
  {Watanabe}}\ and\ \bibinfo {author} {\bibfnamefont {Y.}~\bibnamefont
  {Yanase}},\ }\bibfield  {title} {\bibinfo {title} {{Group-theoretical
  classification of multipole order: Emergent responses and candidate
  materials}},\ }\href {https://doi.org/10.1103/PhysRevB.98.245129} {\bibfield
  {journal} {\bibinfo  {journal} {Phys. Rev. B}\ }\textbf {\bibinfo {volume}
  {98}},\ \bibinfo {pages} {245129} (\bibinfo {year} {2018})}\BibitemShut
  {NoStop}%
\bibitem [{\citenamefont {Hayami}\ \emph {et~al.}(2018)\citenamefont {Hayami},
  \citenamefont {Yatsushiro}, \citenamefont {Yanagi},\ and\ \citenamefont
  {Kusunose}}]{hayami_2018}%
  \BibitemOpen
  \bibfield  {author} {\bibinfo {author} {\bibfnamefont {S.}~\bibnamefont
  {Hayami}}, \bibinfo {author} {\bibfnamefont {M.}~\bibnamefont {Yatsushiro}},
  \bibinfo {author} {\bibfnamefont {Y.}~\bibnamefont {Yanagi}},\ and\ \bibinfo
  {author} {\bibfnamefont {H.}~\bibnamefont {Kusunose}},\ }\bibfield  {title}
  {\bibinfo {title} {{Classification of atomic-scale multipoles under
  crystallographic point groups and application to linear response tensors}},\
  }\href {https://doi.org/10.1103/PhysRevB.98.165110} {\bibfield  {journal}
  {\bibinfo  {journal} {Phys. Rev. B}\ }\textbf {\bibinfo {volume} {98}},\
  \bibinfo {pages} {165110} (\bibinfo {year} {2018})}\BibitemShut {NoStop}%
\bibitem [{\citenamefont {Gao}\ \emph {et~al.}(2018)\citenamefont {Gao},
  \citenamefont {Vanderbilt},\ and\ \citenamefont {Xiao}}]{gao_2018}%
  \BibitemOpen
  \bibfield  {author} {\bibinfo {author} {\bibfnamefont {Y.}~\bibnamefont
  {Gao}}, \bibinfo {author} {\bibfnamefont {D.}~\bibnamefont {Vanderbilt}},\
  and\ \bibinfo {author} {\bibfnamefont {D.}~\bibnamefont {Xiao}},\ }\bibfield
  {title} {\bibinfo {title} {{Microscopic theory of spin toroidization in
  periodic crystals}},\ }\href {https://doi.org/10.1103/PhysRevB.97.134423}
  {\bibfield  {journal} {\bibinfo  {journal} {Phys. Rev. B}\ }\textbf {\bibinfo
  {volume} {97}},\ \bibinfo {pages} {134423} (\bibinfo {year}
  {2018})}\BibitemShut {NoStop}%
\bibitem [{\citenamefont {Gao}\ and\ \citenamefont {Xiao}(2018)}]{gao_2018_2}%
  \BibitemOpen
  \bibfield  {author} {\bibinfo {author} {\bibfnamefont {Y.}~\bibnamefont
  {Gao}}\ and\ \bibinfo {author} {\bibfnamefont {D.}~\bibnamefont {Xiao}},\
  }\bibfield  {title} {\bibinfo {title} {{Orbital magnetic quadrupole moment
  and nonlinear anomalous thermoelectric transport}},\ }\href
  {https://doi.org/10.1103/PhysRevB.98.060402} {\bibfield  {journal} {\bibinfo
  {journal} {Phys. Rev. B}\ }\textbf {\bibinfo {volume} {98}},\ \bibinfo
  {pages} {060402} (\bibinfo {year} {2018})}\BibitemShut {NoStop}%
\bibitem [{\citenamefont {Shitade}\ \emph {et~al.}(2018)\citenamefont
  {Shitade}, \citenamefont {Watanabe},\ and\ \citenamefont
  {Yanase}}]{shitade_2018}%
  \BibitemOpen
  \bibfield  {author} {\bibinfo {author} {\bibfnamefont {A.}~\bibnamefont
  {Shitade}}, \bibinfo {author} {\bibfnamefont {H.}~\bibnamefont {Watanabe}},\
  and\ \bibinfo {author} {\bibfnamefont {Y.}~\bibnamefont {Yanase}},\
  }\bibfield  {title} {\bibinfo {title} {{Theory of orbital magnetic quadrupole
  moment and magnetoelectric susceptibility}},\ }\href
  {https://doi.org/10.1103/PhysRevB.98.020407} {\bibfield  {journal} {\bibinfo
  {journal} {Phys. Rev. B}\ }\textbf {\bibinfo {volume} {98}},\ \bibinfo
  {pages} {020407} (\bibinfo {year} {2018})}\BibitemShut {NoStop}%
\bibitem [{\citenamefont {Shitade}\ \emph
  {et~al.}(2019{\natexlab{a}})\citenamefont {Shitade}, \citenamefont {Daido},\
  and\ \citenamefont {Yanase}}]{shitade_2019}%
  \BibitemOpen
  \bibfield  {author} {\bibinfo {author} {\bibfnamefont {A.}~\bibnamefont
  {Shitade}}, \bibinfo {author} {\bibfnamefont {A.}~\bibnamefont {Daido}},\
  and\ \bibinfo {author} {\bibfnamefont {Y.}~\bibnamefont {Yanase}},\
  }\bibfield  {title} {\bibinfo {title} {{Theory of spin magnetic quadrupole
  moment and temperature-gradient-induced magnetization}},\ }\href
  {https://doi.org/10.1103/PhysRevB.99.024404} {\bibfield  {journal} {\bibinfo
  {journal} {Phys. Rev. B}\ }\textbf {\bibinfo {volume} {99}},\ \bibinfo
  {pages} {024404} (\bibinfo {year} {2019}{\natexlab{a}})}\BibitemShut
  {NoStop}%
\bibitem [{\citenamefont {Dubovik}\ and\ \citenamefont
  {Tugushev}(1990)}]{Dubovik_1990}%
  \BibitemOpen
  \bibfield  {author} {\bibinfo {author} {\bibfnamefont {V.}~\bibnamefont
  {Dubovik}}\ and\ \bibinfo {author} {\bibfnamefont {V.}~\bibnamefont
  {Tugushev}},\ }\bibfield  {title} {\bibinfo {title} {{Toroid moments in
  electrodynamics and solid-state physics}},\ }\href
  {https://doi.org/https://doi.org/10.1016/0370-1573(90)90042-Z} {\bibfield
  {journal} {\bibinfo  {journal} {Physics Reports}\ }\textbf {\bibinfo {volume}
  {187}},\ \bibinfo {pages} {145} (\bibinfo {year} {1990})}\BibitemShut
  {NoStop}%
\bibitem [{\citenamefont {Gorbatsevich}\ and\ \citenamefont
  {Kopaev}(1994)}]{Gorbatsevich_1994}%
  \BibitemOpen
  \bibfield  {author} {\bibinfo {author} {\bibfnamefont {A.~A.}\ \bibnamefont
  {Gorbatsevich}}\ and\ \bibinfo {author} {\bibfnamefont {Y.~V.}\ \bibnamefont
  {Kopaev}},\ }\bibfield  {title} {\bibinfo {title} {{Toroidal order in
  crystals}},\ }\href {https://doi.org/10.1080/00150199408213381} {\bibfield
  {journal} {\bibinfo  {journal} {Ferroelectrics}\ }\textbf {\bibinfo {volume}
  {161}},\ \bibinfo {pages} {321} (\bibinfo {year} {1994})},\ \Eprint
  {https://arxiv.org/abs/https://doi.org/10.1080/00150199408213381}
  {https://doi.org/10.1080/00150199408213381} \BibitemShut {NoStop}%
\bibitem [{\citenamefont {Ederer}\ and\ \citenamefont
  {Spaldin}(2007)}]{ederer_2007}%
  \BibitemOpen
  \bibfield  {author} {\bibinfo {author} {\bibfnamefont {C.}~\bibnamefont
  {Ederer}}\ and\ \bibinfo {author} {\bibfnamefont {N.~A.}\ \bibnamefont
  {Spaldin}},\ }\bibfield  {title} {\bibinfo {title} {{Towards a microscopic
  theory of toroidal moments in bulk periodic crystals}},\ }\href
  {https://doi.org/10.1103/PhysRevB.76.214404} {\bibfield  {journal} {\bibinfo
  {journal} {Phys. Rev. B}\ }\textbf {\bibinfo {volume} {76}},\ \bibinfo
  {pages} {214404} (\bibinfo {year} {2007})}\BibitemShut {NoStop}%
\bibitem [{\citenamefont {Spaldin}\ \emph {et~al.}(2008)\citenamefont
  {Spaldin}, \citenamefont {Fiebig},\ and\ \citenamefont
  {Mostovoy}}]{Spaldin_2008}%
  \BibitemOpen
  \bibfield  {author} {\bibinfo {author} {\bibfnamefont {N.~A.}\ \bibnamefont
  {Spaldin}}, \bibinfo {author} {\bibfnamefont {M.}~\bibnamefont {Fiebig}},\
  and\ \bibinfo {author} {\bibfnamefont {M.}~\bibnamefont {Mostovoy}},\
  }\bibfield  {title} {\bibinfo {title} {{The toroidal moment in
  condensed-matter physics and its relation to the magnetoelectric effect}},\
  }\href {https://doi.org/10.1088/0953-8984/20/43/434203} {\bibfield  {journal}
  {\bibinfo  {journal} {Journal of Physics: Condensed Matter}\ }\textbf
  {\bibinfo {volume} {20}},\ \bibinfo {pages} {434203} (\bibinfo {year}
  {2008})}\BibitemShut {NoStop}%
\bibitem [{\citenamefont {Arima}\ \emph {et~al.}(2005)\citenamefont {Arima},
  \citenamefont {Jung}, \citenamefont {Matsubara}, \citenamefont {Kubota},
  \citenamefont {He}, \citenamefont {Kaneko},\ and\ \citenamefont
  {Tokura}}]{arima_2005}%
  \BibitemOpen
  \bibfield  {author} {\bibinfo {author} {\bibfnamefont {T.-h.}\ \bibnamefont
  {Arima}}, \bibinfo {author} {\bibfnamefont {J.-H.}\ \bibnamefont {Jung}},
  \bibinfo {author} {\bibfnamefont {M.}~\bibnamefont {Matsubara}}, \bibinfo
  {author} {\bibfnamefont {M.}~\bibnamefont {Kubota}}, \bibinfo {author}
  {\bibfnamefont {J.-P.}\ \bibnamefont {He}}, \bibinfo {author} {\bibfnamefont
  {Y.}~\bibnamefont {Kaneko}},\ and\ \bibinfo {author} {\bibfnamefont
  {Y.}~\bibnamefont {Tokura}},\ }\bibfield  {title} {\bibinfo {title}
  {{Resonant Magnetoelectric X-ray Scattering in GaFeO3: Observation of
  Ordering of Toroidal Moments}},\ }\href
  {https://doi.org/10.1143/JPSJ.74.1419} {\bibfield  {journal} {\bibinfo
  {journal} {Journal of the Physical Society of Japan}\ }\textbf {\bibinfo
  {volume} {74}},\ \bibinfo {pages} {1419} (\bibinfo {year} {2005})},\ \Eprint
  {https://arxiv.org/abs/https://doi.org/10.1143/JPSJ.74.1419}
  {https://doi.org/10.1143/JPSJ.74.1419} \BibitemShut {NoStop}%
\bibitem [{\citenamefont {Van~Aken}\ \emph {et~al.}(2007)\citenamefont
  {Van~Aken}, \citenamefont {Rivera}, \citenamefont {Schmid},\ and\
  \citenamefont {Fiebig}}]{VanAken_2007}%
  \BibitemOpen
  \bibfield  {author} {\bibinfo {author} {\bibfnamefont {B.~B.}\ \bibnamefont
  {Van~Aken}}, \bibinfo {author} {\bibfnamefont {J.-P.}\ \bibnamefont
  {Rivera}}, \bibinfo {author} {\bibfnamefont {H.}~\bibnamefont {Schmid}},\
  and\ \bibinfo {author} {\bibfnamefont {M.}~\bibnamefont {Fiebig}},\
  }\bibfield  {title} {\bibinfo {title} {{Observation of ferrotoroidic
  domains}},\ }\href {https://doi.org/10.1038/nature06139} {\bibfield
  {journal} {\bibinfo  {journal} {Nature}\ }\textbf {\bibinfo {volume} {449}},\
  \bibinfo {pages} {702} (\bibinfo {year} {2007})}\BibitemShut {NoStop}%
\bibitem [{\citenamefont {Hayami}\ \emph {et~al.}(2014)\citenamefont {Hayami},
  \citenamefont {Kusunose},\ and\ \citenamefont {Motome}}]{Hayami_2014}%
  \BibitemOpen
  \bibfield  {author} {\bibinfo {author} {\bibfnamefont {S.}~\bibnamefont
  {Hayami}}, \bibinfo {author} {\bibfnamefont {H.}~\bibnamefont {Kusunose}},\
  and\ \bibinfo {author} {\bibfnamefont {Y.}~\bibnamefont {Motome}},\
  }\bibfield  {title} {\bibinfo {title} {{Toroidal order in metals without
  local inversion symmetry}},\ }\href
  {https://doi.org/10.1103/PhysRevB.90.024432} {\bibfield  {journal} {\bibinfo
  {journal} {Phys. Rev. B}\ }\textbf {\bibinfo {volume} {90}},\ \bibinfo
  {pages} {024432} (\bibinfo {year} {2014})}\BibitemShut {NoStop}%
\bibitem [{\citenamefont {Batista}\ \emph {et~al.}(2008)\citenamefont
  {Batista}, \citenamefont {Ortiz},\ and\ \citenamefont
  {Aligia}}]{Batista_2008}%
  \BibitemOpen
  \bibfield  {author} {\bibinfo {author} {\bibfnamefont {C.~D.}\ \bibnamefont
  {Batista}}, \bibinfo {author} {\bibfnamefont {G.}~\bibnamefont {Ortiz}},\
  and\ \bibinfo {author} {\bibfnamefont {A.~A.}\ \bibnamefont {Aligia}},\
  }\bibfield  {title} {\bibinfo {title} {{Ferrotoroidic Moment as a Quantum
  Geometric Phase}},\ }\href {https://doi.org/10.1103/PhysRevLett.101.077203}
  {\bibfield  {journal} {\bibinfo  {journal} {Phys. Rev. Lett.}\ }\textbf
  {\bibinfo {volume} {101}},\ \bibinfo {pages} {077203} (\bibinfo {year}
  {2008})}\BibitemShut {NoStop}%
\bibitem [{\citenamefont {Th\"ole}\ \emph {et~al.}(2016)\citenamefont
  {Th\"ole}, \citenamefont {Fechner},\ and\ \citenamefont
  {Spaldin}}]{thole_2016}%
  \BibitemOpen
  \bibfield  {author} {\bibinfo {author} {\bibfnamefont {F.}~\bibnamefont
  {Th\"ole}}, \bibinfo {author} {\bibfnamefont {M.}~\bibnamefont {Fechner}},\
  and\ \bibinfo {author} {\bibfnamefont {N.~A.}\ \bibnamefont {Spaldin}},\
  }\bibfield  {title} {\bibinfo {title} {{First-principles calculation of the
  bulk magnetoelectric monopole density: Berry phase and Wannier function
  approaches}},\ }\href {https://doi.org/10.1103/PhysRevB.93.195167} {\bibfield
   {journal} {\bibinfo  {journal} {Phys. Rev. B}\ }\textbf {\bibinfo {volume}
  {93}},\ \bibinfo {pages} {195167} (\bibinfo {year} {2016})}\BibitemShut
  {NoStop}%
\bibitem [{\citenamefont {Watanabe}\ and\ \citenamefont
  {Yanase}(2017)}]{watanabe_2017_1}%
  \BibitemOpen
  \bibfield  {author} {\bibinfo {author} {\bibfnamefont {H.}~\bibnamefont
  {Watanabe}}\ and\ \bibinfo {author} {\bibfnamefont {Y.}~\bibnamefont
  {Yanase}},\ }\bibfield  {title} {\bibinfo {title} {{Magnetic hexadecapole
  order and magnetopiezoelectric metal state in
  ${\mathbf{Ba}}_{1\ensuremath{-}x}{\mathbf{K}}_{x}{\mathbf{Mn}}_{2}{\mathbf{As}}_{2}$}},\
  }\href {https://doi.org/10.1103/PhysRevB.96.064432} {\bibfield  {journal}
  {\bibinfo  {journal} {Phys. Rev. B}\ }\textbf {\bibinfo {volume} {96}},\
  \bibinfo {pages} {064432} (\bibinfo {year} {2017})}\BibitemShut {NoStop}%
\bibitem [{\citenamefont {Suzuki}\ \emph {et~al.}(2017)\citenamefont {Suzuki},
  \citenamefont {Koretsune}, \citenamefont {Ochi},\ and\ \citenamefont
  {Arita}}]{Suzuki2017}%
  \BibitemOpen
  \bibfield  {author} {\bibinfo {author} {\bibfnamefont {M.-T.}\ \bibnamefont
  {Suzuki}}, \bibinfo {author} {\bibfnamefont {T.}~\bibnamefont {Koretsune}},
  \bibinfo {author} {\bibfnamefont {M.}~\bibnamefont {Ochi}},\ and\ \bibinfo
  {author} {\bibfnamefont {R.}~\bibnamefont {Arita}},\ }\bibfield  {title}
  {\bibinfo {title} {{Cluster multipole theory for anomalous Hall effect in
  antiferromagnets}},\ }\href {https://doi.org/10.1103/PhysRevB.95.094406}
  {\bibfield  {journal} {\bibinfo  {journal} {Phys. Rev. B}\ }\textbf {\bibinfo
  {volume} {95}},\ \bibinfo {pages} {094406} (\bibinfo {year}
  {2017})}\BibitemShut {NoStop}%
\bibitem [{\citenamefont {Tahir}\ and\ \citenamefont {Chen}(2023)}]{Tahir2023}%
  \BibitemOpen
  \bibfield  {author} {\bibinfo {author} {\bibfnamefont {M.}~\bibnamefont
  {Tahir}}\ and\ \bibinfo {author} {\bibfnamefont {H.}~\bibnamefont {Chen}},\
  }\bibfield  {title} {\bibinfo {title} {{Transport of Spin Magnetic Multipole
  Moments Carried by Bloch Quasiparticles}},\ }\href
  {https://doi.org/10.1103/physrevlett.131.106701} {\bibfield  {journal}
  {\bibinfo  {journal} {Physical Review Letters}\ }\textbf {\bibinfo {volume}
  {131}},\ \bibinfo {pages} {106701} (\bibinfo {year} {2023})}\BibitemShut
  {NoStop}%
\bibitem [{\citenamefont {Resta}(1994)}]{Resta_1994}%
  \BibitemOpen
  \bibfield  {author} {\bibinfo {author} {\bibfnamefont {R.}~\bibnamefont
  {Resta}},\ }\bibfield  {title} {\bibinfo {title} {{Macroscopic polarization
  in crystalline dielectrics: the geometric phase approach}},\ }\href
  {https://doi.org/10.1103/RevModPhys.66.899} {\bibfield  {journal} {\bibinfo
  {journal} {Rev. Mod. Phys.}\ }\textbf {\bibinfo {volume} {66}},\ \bibinfo
  {pages} {899} (\bibinfo {year} {1994})}\BibitemShut {NoStop}%
\bibitem [{\citenamefont {King-Smith}\ and\ \citenamefont
  {Vanderbilt}(1993)}]{king-smith_1993}%
  \BibitemOpen
  \bibfield  {author} {\bibinfo {author} {\bibfnamefont {R.~D.}\ \bibnamefont
  {King-Smith}}\ and\ \bibinfo {author} {\bibfnamefont {D.}~\bibnamefont
  {Vanderbilt}},\ }\bibfield  {title} {\bibinfo {title} {{Theory of
  polarization of crystalline solids}},\ }\href
  {https://doi.org/10.1103/PhysRevB.47.1651} {\bibfield  {journal} {\bibinfo
  {journal} {Phys. Rev. B}\ }\textbf {\bibinfo {volume} {47}},\ \bibinfo
  {pages} {1651} (\bibinfo {year} {1993})}\BibitemShut {NoStop}%
\bibitem [{\citenamefont {Xiao}\ \emph {et~al.}(2005)\citenamefont {Xiao},
  \citenamefont {Shi},\ and\ \citenamefont {Niu}}]{xiao_2005}%
  \BibitemOpen
  \bibfield  {author} {\bibinfo {author} {\bibfnamefont {D.}~\bibnamefont
  {Xiao}}, \bibinfo {author} {\bibfnamefont {J.}~\bibnamefont {Shi}},\ and\
  \bibinfo {author} {\bibfnamefont {Q.}~\bibnamefont {Niu}},\ }\bibfield
  {title} {\bibinfo {title} {{Berry Phase Correction to Electron Density of
  States in Solids}},\ }\href {https://doi.org/10.1103/PhysRevLett.95.137204}
  {\bibfield  {journal} {\bibinfo  {journal} {Phys. Rev. Lett.}\ }\textbf
  {\bibinfo {volume} {95}},\ \bibinfo {pages} {137204} (\bibinfo {year}
  {2005})}\BibitemShut {NoStop}%
\bibitem [{\citenamefont {Thonhauser}\ \emph {et~al.}(2005)\citenamefont
  {Thonhauser}, \citenamefont {Ceresoli}, \citenamefont {Vanderbilt},\ and\
  \citenamefont {Resta}}]{thonhauser_2005}%
  \BibitemOpen
  \bibfield  {author} {\bibinfo {author} {\bibfnamefont {T.}~\bibnamefont
  {Thonhauser}}, \bibinfo {author} {\bibfnamefont {D.}~\bibnamefont
  {Ceresoli}}, \bibinfo {author} {\bibfnamefont {D.}~\bibnamefont
  {Vanderbilt}},\ and\ \bibinfo {author} {\bibfnamefont {R.}~\bibnamefont
  {Resta}},\ }\bibfield  {title} {\bibinfo {title} {{Orbital Magnetization in
  Periodic Insulators}},\ }\href
  {https://doi.org/10.1103/PhysRevLett.95.137205} {\bibfield  {journal}
  {\bibinfo  {journal} {Phys. Rev. Lett.}\ }\textbf {\bibinfo {volume} {95}},\
  \bibinfo {pages} {137205} (\bibinfo {year} {2005})}\BibitemShut {NoStop}%
\bibitem [{\citenamefont {Shi}\ \emph {et~al.}(2007)\citenamefont {Shi},
  \citenamefont {Vignale}, \citenamefont {Xiao},\ and\ \citenamefont
  {Niu}}]{shi_2007}%
  \BibitemOpen
  \bibfield  {author} {\bibinfo {author} {\bibfnamefont {J.}~\bibnamefont
  {Shi}}, \bibinfo {author} {\bibfnamefont {G.}~\bibnamefont {Vignale}},
  \bibinfo {author} {\bibfnamefont {D.}~\bibnamefont {Xiao}},\ and\ \bibinfo
  {author} {\bibfnamefont {Q.}~\bibnamefont {Niu}},\ }\bibfield  {title}
  {\bibinfo {title} {{Quantum Theory of Orbital Magnetization and Its
  Generalization to Interacting Systems}},\ }\href
  {https://doi.org/10.1103/PhysRevLett.99.197202} {\bibfield  {journal}
  {\bibinfo  {journal} {Phys. Rev. Lett.}\ }\textbf {\bibinfo {volume} {99}},\
  \bibinfo {pages} {197202} (\bibinfo {year} {2007})}\BibitemShut {NoStop}%
\bibitem [{\citenamefont {Suzuki}\ \emph {et~al.}(2019)\citenamefont {Suzuki},
  \citenamefont {Nomoto}, \citenamefont {Arita}, \citenamefont {Yanagi},
  \citenamefont {Hayami},\ and\ \citenamefont {Kusunose}}]{PhysRevB.99.174407}%
  \BibitemOpen
  \bibfield  {author} {\bibinfo {author} {\bibfnamefont {M.-T.}\ \bibnamefont
  {Suzuki}}, \bibinfo {author} {\bibfnamefont {T.}~\bibnamefont {Nomoto}},
  \bibinfo {author} {\bibfnamefont {R.}~\bibnamefont {Arita}}, \bibinfo
  {author} {\bibfnamefont {Y.}~\bibnamefont {Yanagi}}, \bibinfo {author}
  {\bibfnamefont {S.}~\bibnamefont {Hayami}},\ and\ \bibinfo {author}
  {\bibfnamefont {H.}~\bibnamefont {Kusunose}},\ }\bibfield  {title} {\bibinfo
  {title} {{Multipole expansion for magnetic structures: A generation scheme
  for a symmetry-adapted orthonormal basis set in the crystallographic point
  group}},\ }\href {https://doi.org/10.1103/PhysRevB.99.174407} {\bibfield
  {journal} {\bibinfo  {journal} {Phys. Rev. B}\ }\textbf {\bibinfo {volume}
  {99}},\ \bibinfo {pages} {174407} (\bibinfo {year} {2019})}\BibitemShut
  {NoStop}%
\bibitem [{\citenamefont {Nomoto}\ and\ \citenamefont
  {Arita}(2020)}]{PhysRevResearch.2.012045}%
  \BibitemOpen
  \bibfield  {author} {\bibinfo {author} {\bibfnamefont {T.}~\bibnamefont
  {Nomoto}}\ and\ \bibinfo {author} {\bibfnamefont {R.}~\bibnamefont {Arita}},\
  }\bibfield  {title} {\bibinfo {title} {{Cluster multipole dynamics in
  noncollinear antiferromagnets}},\ }\href
  {https://doi.org/10.1103/PhysRevResearch.2.012045} {\bibfield  {journal}
  {\bibinfo  {journal} {Phys. Rev. Research}\ }\textbf {\bibinfo {volume}
  {2}},\ \bibinfo {pages} {012045} (\bibinfo {year} {2020})}\BibitemShut
  {NoStop}%
\bibitem [{\citenamefont {Lapa}\ and\ \citenamefont
  {Hughes}(2019)}]{lapa_2019}%
  \BibitemOpen
  \bibfield  {author} {\bibinfo {author} {\bibfnamefont {M.~F.}\ \bibnamefont
  {Lapa}}\ and\ \bibinfo {author} {\bibfnamefont {T.~L.}\ \bibnamefont
  {Hughes}},\ }\bibfield  {title} {\bibinfo {title} {{Semiclassical wave packet
  dynamics in nonuniform electric fields}},\ }\href
  {https://doi.org/10.1103/PhysRevB.99.121111} {\bibfield  {journal} {\bibinfo
  {journal} {Phys. Rev. B}\ }\textbf {\bibinfo {volume} {99}},\ \bibinfo
  {pages} {121111} (\bibinfo {year} {2019})}\BibitemShut {NoStop}%
\bibitem [{\citenamefont {Daido}\ \emph {et~al.}(2020)\citenamefont {Daido},
  \citenamefont {Shitade},\ and\ \citenamefont {Yanase}}]{daido_2020}%
  \BibitemOpen
  \bibfield  {author} {\bibinfo {author} {\bibfnamefont {A.}~\bibnamefont
  {Daido}}, \bibinfo {author} {\bibfnamefont {A.}~\bibnamefont {Shitade}},\
  and\ \bibinfo {author} {\bibfnamefont {Y.}~\bibnamefont {Yanase}},\
  }\bibfield  {title} {\bibinfo {title} {{Thermodynamic approach to electric
  quadrupole moments}},\ }\href {https://doi.org/10.1103/PhysRevB.102.235149}
  {\bibfield  {journal} {\bibinfo  {journal} {Phys. Rev. B}\ }\textbf {\bibinfo
  {volume} {102}},\ \bibinfo {pages} {235149} (\bibinfo {year}
  {2020})}\BibitemShut {NoStop}%
\bibitem [{\citenamefont {Resta}(1998)}]{Resta1998}%
  \BibitemOpen
  \bibfield  {author} {\bibinfo {author} {\bibfnamefont {R.}~\bibnamefont
  {Resta}},\ }\bibfield  {title} {\bibinfo {title} {{Quantum-Mechanical
  Position Operator in Extended Systems}},\ }\href
  {https://doi.org/10.1103/physrevlett.80.1800} {\bibfield  {journal} {\bibinfo
   {journal} {Physical Review Letters}\ }\textbf {\bibinfo {volume} {80}},\
  \bibinfo {pages} {1800} (\bibinfo {year} {1998})}\BibitemShut {NoStop}%
\bibitem [{\citenamefont {Benalcazar}\ \emph
  {et~al.}(2017{\natexlab{a}})\citenamefont {Benalcazar}, \citenamefont
  {Bernevig},\ and\ \citenamefont {Hughes}}]{Benalcazar61}%
  \BibitemOpen
  \bibfield  {author} {\bibinfo {author} {\bibfnamefont {W.~A.}\ \bibnamefont
  {Benalcazar}}, \bibinfo {author} {\bibfnamefont {B.~A.}\ \bibnamefont
  {Bernevig}},\ and\ \bibinfo {author} {\bibfnamefont {T.~L.}\ \bibnamefont
  {Hughes}},\ }\bibfield  {title} {\bibinfo {title} {{Quantized electric
  multipole insulators}},\ }\href {https://doi.org/10.1126/science.aah6442}
  {\bibfield  {journal} {\bibinfo  {journal} {Science}\ }\textbf {\bibinfo
  {volume} {357}},\ \bibinfo {pages} {61} (\bibinfo {year}
  {2017}{\natexlab{a}})},\ \Eprint
  {https://arxiv.org/abs/https://science.sciencemag.org/content/357/6346/61.full.pdf}
  {https://science.sciencemag.org/content/357/6346/61.full.pdf} \BibitemShut
  {NoStop}%
\bibitem [{\citenamefont {Benalcazar}\ \emph
  {et~al.}(2017{\natexlab{b}})\citenamefont {Benalcazar}, \citenamefont
  {Bernevig},\ and\ \citenamefont {Hughes}}]{Benalcazar2017}%
  \BibitemOpen
  \bibfield  {author} {\bibinfo {author} {\bibfnamefont {W.~A.}\ \bibnamefont
  {Benalcazar}}, \bibinfo {author} {\bibfnamefont {B.~A.}\ \bibnamefont
  {Bernevig}},\ and\ \bibinfo {author} {\bibfnamefont {T.~L.}\ \bibnamefont
  {Hughes}},\ }\bibfield  {title} {\bibinfo {title} {{Electric multipole
  moments, topological multipole moment pumping, and chiral hinge states in
  crystalline insulators}},\ }\href
  {https://doi.org/10.1103/physrevb.96.245115} {\bibfield  {journal} {\bibinfo
  {journal} {Physical Review B}\ }\textbf {\bibinfo {volume} {96}},\ \bibinfo
  {pages} {245115} (\bibinfo {year} {2017}{\natexlab{b}})}\BibitemShut
  {NoStop}%
\bibitem [{\citenamefont {Schindler}\ \emph {et~al.}(2018)\citenamefont
  {Schindler}, \citenamefont {Cook}, \citenamefont {Vergniory}, \citenamefont
  {Wang}, \citenamefont {Parkin}, \citenamefont {Bernevig},\ and\ \citenamefont
  {Neupert}}]{Schindler2018}%
  \BibitemOpen
  \bibfield  {author} {\bibinfo {author} {\bibfnamefont {F.}~\bibnamefont
  {Schindler}}, \bibinfo {author} {\bibfnamefont {A.~M.}\ \bibnamefont {Cook}},
  \bibinfo {author} {\bibfnamefont {M.~G.}\ \bibnamefont {Vergniory}}, \bibinfo
  {author} {\bibfnamefont {Z.}~\bibnamefont {Wang}}, \bibinfo {author}
  {\bibfnamefont {S.~S.~P.}\ \bibnamefont {Parkin}}, \bibinfo {author}
  {\bibfnamefont {B.~A.}\ \bibnamefont {Bernevig}},\ and\ \bibinfo {author}
  {\bibfnamefont {T.}~\bibnamefont {Neupert}},\ }\bibfield  {title} {\bibinfo
  {title} {{Higher-order topological insulators}},\ }\bibfield  {journal}
  {\bibinfo  {journal} {Science Advances}\ }\textbf {\bibinfo {volume} {4}},\
  \href {https://doi.org/10.1126/sciadv.aat0346} {10.1126/sciadv.aat0346}
  (\bibinfo {year} {2018})\BibitemShut {NoStop}%
\bibitem [{\citenamefont {Kang}\ \emph {et~al.}(2019)\citenamefont {Kang},
  \citenamefont {Shiozaki},\ and\ \citenamefont {Cho}}]{Kang2019}%
  \BibitemOpen
  \bibfield  {author} {\bibinfo {author} {\bibfnamefont {B.}~\bibnamefont
  {Kang}}, \bibinfo {author} {\bibfnamefont {K.}~\bibnamefont {Shiozaki}},\
  and\ \bibinfo {author} {\bibfnamefont {G.~Y.}\ \bibnamefont {Cho}},\
  }\bibfield  {title} {\bibinfo {title} {{Many-body order parameters for
  multipoles in solids}},\ }\href {https://doi.org/10.1103/physrevb.100.245134}
  {\bibfield  {journal} {\bibinfo  {journal} {Physical Review B}\ }\textbf
  {\bibinfo {volume} {100}},\ \bibinfo {pages} {245134} (\bibinfo {year}
  {2019})}\BibitemShut {NoStop}%
\bibitem [{\citenamefont {Watanabe}\ and\ \citenamefont
  {Ono}(2020)}]{Watanabe2020}%
  \BibitemOpen
  \bibfield  {author} {\bibinfo {author} {\bibfnamefont {H.}~\bibnamefont
  {Watanabe}}\ and\ \bibinfo {author} {\bibfnamefont {S.}~\bibnamefont {Ono}},\
  }\bibfield  {title} {\bibinfo {title} {{Corner charge and bulk multipole
  moment in periodic systems}},\ }\href
  {https://doi.org/10.1103/physrevb.102.165120} {\bibfield  {journal} {\bibinfo
   {journal} {Physical Review B}\ }\textbf {\bibinfo {volume} {102}},\ \bibinfo
  {pages} {165120} (\bibinfo {year} {2020})}\BibitemShut {NoStop}%
\bibitem [{\citenamefont {Trifunovic}(2020)}]{Trifunovic2020}%
  \BibitemOpen
  \bibfield  {author} {\bibinfo {author} {\bibfnamefont {L.}~\bibnamefont
  {Trifunovic}},\ }\bibfield  {title} {\bibinfo {title} {{Bulk-and-edge to
  corner correspondence}},\ }\href
  {https://doi.org/10.1103/physrevresearch.2.043012} {\bibfield  {journal}
  {\bibinfo  {journal} {Physical Review Research}\ }\textbf {\bibinfo {volume}
  {2}},\ \bibinfo {pages} {043012} (\bibinfo {year} {2020})}\BibitemShut
  {NoStop}%
\bibitem [{\citenamefont {Ren}\ \emph {et~al.}(2021)\citenamefont {Ren},
  \citenamefont {Souza},\ and\ \citenamefont {Vanderbilt}}]{Ren2021}%
  \BibitemOpen
  \bibfield  {author} {\bibinfo {author} {\bibfnamefont {S.}~\bibnamefont
  {Ren}}, \bibinfo {author} {\bibfnamefont {I.}~\bibnamefont {Souza}},\ and\
  \bibinfo {author} {\bibfnamefont {D.}~\bibnamefont {Vanderbilt}},\ }\bibfield
   {title} {\bibinfo {title} {{Quadrupole moments, edge polarizations, and
  corner charges in the Wannier representation}},\ }\href
  {https://doi.org/10.1103/physrevb.103.035147} {\bibfield  {journal} {\bibinfo
   {journal} {Physical Review B}\ }\textbf {\bibinfo {volume} {103}},\ \bibinfo
  {pages} {035147} (\bibinfo {year} {2021})}\BibitemShut {NoStop}%
\bibitem [{\citenamefont {Ōiké}\ \emph {et~al.}(2025)\citenamefont {Ōiké},
  \citenamefont {Peters},\ and\ \citenamefont {Shinada}}]{Oike2025a}%
  \BibitemOpen
  \bibfield  {author} {\bibinfo {author} {\bibfnamefont {J.}~\bibnamefont
  {Ōiké}}, \bibinfo {author} {\bibfnamefont {R.}~\bibnamefont {Peters}},\
  and\ \bibinfo {author} {\bibfnamefont {K.}~\bibnamefont {Shinada}},\
  }\bibfield  {title} {\bibinfo {title} {{Thermodynamic formulation of the spin
  magnetic octupole moment in bulk crystals}},\ }\bibfield  {journal} {\bibinfo
   {journal} {Physical Review B}\ }\textbf {\bibinfo {volume} {112}},\ \href
  {https://doi.org/10.1103/frq1-9xx7} {10.1103/frq1-9xx7} (\bibinfo {year}
  {2025})\BibitemShut {NoStop}%
\bibitem [{\citenamefont {Sato}\ and\ \citenamefont {Hayami}(2026)}]{Sato2026}%
  \BibitemOpen
  \bibfield  {author} {\bibinfo {author} {\bibfnamefont {T.}~\bibnamefont
  {Sato}}\ and\ \bibinfo {author} {\bibfnamefont {S.}~\bibnamefont {Hayami}},\
  }\bibfield  {title} {\bibinfo {title} {{Quantum theory of magnetic octupole
  in periodic crystals and application to d-wave altermagnets}},\ }\bibfield
  {journal} {\bibinfo  {journal} {npj Quantum Materials}\ }\href
  {https://doi.org/10.1038/s41535-026-00865-9} {10.1038/s41535-026-00865-9}
  (\bibinfo {year} {2026})\BibitemShut {NoStop}%
\bibitem [{\citenamefont {Chen}\ and\ \citenamefont {Lee}(2012)}]{Chen2012}%
  \BibitemOpen
  \bibfield  {author} {\bibinfo {author} {\bibfnamefont {K.-T.}\ \bibnamefont
  {Chen}}\ and\ \bibinfo {author} {\bibfnamefont {P.~A.}\ \bibnamefont {Lee}},\
  }\bibfield  {title} {\bibinfo {title} {{Effect of the boundary on
  thermodynamic quantities such as magnetization}},\ }\href
  {https://doi.org/10.1103/physrevb.86.195111} {\bibfield  {journal} {\bibinfo
  {journal} {Physical Review B}\ }\textbf {\bibinfo {volume} {86}},\ \bibinfo
  {pages} {195111} (\bibinfo {year} {2012})}\BibitemShut {NoStop}%
\bibitem [{\citenamefont {Moseni}\ and\ \citenamefont
  {Coh}(2024)}]{Moseni2024}%
  \BibitemOpen
  \bibfield  {author} {\bibinfo {author} {\bibfnamefont {K.}~\bibnamefont
  {Moseni}}\ and\ \bibinfo {author} {\bibfnamefont {S.}~\bibnamefont {Coh}},\
  }\bibfield  {title} {\bibinfo {title} {{Orbital magnetization of a metal is
  not a bulk property in the mesoscopic regime}},\ }\href
  {https://doi.org/10.1103/physrevb.109.174431} {\bibfield  {journal} {\bibinfo
   {journal} {Physical Review B}\ }\textbf {\bibinfo {volume} {109}},\ \bibinfo
  {pages} {174431} (\bibinfo {year} {2024})}\BibitemShut {NoStop}%
\bibitem [{\citenamefont {Bianco}\ and\ \citenamefont
  {Resta}(2013)}]{Bianco2013}%
  \BibitemOpen
  \bibfield  {author} {\bibinfo {author} {\bibfnamefont {R.}~\bibnamefont
  {Bianco}}\ and\ \bibinfo {author} {\bibfnamefont {R.}~\bibnamefont {Resta}},\
  }\bibfield  {title} {\bibinfo {title} {{Orbital Magnetization as a Local
  Property}},\ }\href {https://doi.org/10.1103/physrevlett.110.087202}
  {\bibfield  {journal} {\bibinfo  {journal} {Physical Review Letters}\
  }\textbf {\bibinfo {volume} {110}},\ \bibinfo {pages} {087202} (\bibinfo
  {year} {2013})}\BibitemShut {NoStop}%
\bibitem [{\citenamefont {Seleznev}\ and\ \citenamefont
  {Vanderbilt}(2023)}]{Seleznev2023}%
  \BibitemOpen
  \bibfield  {author} {\bibinfo {author} {\bibfnamefont {D.}~\bibnamefont
  {Seleznev}}\ and\ \bibinfo {author} {\bibfnamefont {D.}~\bibnamefont
  {Vanderbilt}},\ }\bibfield  {title} {\bibinfo {title} {{Towards a theory of
  surface orbital magnetization}},\ }\href
  {https://doi.org/10.1103/physrevb.107.115102} {\bibfield  {journal} {\bibinfo
   {journal} {Physical Review B}\ }\textbf {\bibinfo {volume} {107}},\ \bibinfo
  {pages} {115102} (\bibinfo {year} {2023})}\BibitemShut {NoStop}%
\bibitem [{\citenamefont {Šmejkal}\ \emph
  {et~al.}(2022{\natexlab{a}})\citenamefont {Šmejkal}, \citenamefont
  {Sinova},\ and\ \citenamefont {Jungwirth}}]{Smejkal2022a}%
  \BibitemOpen
  \bibfield  {author} {\bibinfo {author} {\bibfnamefont {L.}~\bibnamefont
  {Šmejkal}}, \bibinfo {author} {\bibfnamefont {J.}~\bibnamefont {Sinova}},\
  and\ \bibinfo {author} {\bibfnamefont {T.}~\bibnamefont {Jungwirth}},\
  }\bibfield  {title} {\bibinfo {title} {{Emerging Research Landscape of
  Altermagnetism}},\ }\href {https://doi.org/10.1103/physrevx.12.040501}
  {\bibfield  {journal} {\bibinfo  {journal} {Physical Review X}\ }\textbf
  {\bibinfo {volume} {12}},\ \bibinfo {pages} {040501} (\bibinfo {year}
  {2022}{\natexlab{a}})}\BibitemShut {NoStop}%
\bibitem [{\citenamefont {Šmejkal}\ \emph
  {et~al.}(2022{\natexlab{b}})\citenamefont {Šmejkal}, \citenamefont
  {Sinova},\ and\ \citenamefont {Jungwirth}}]{Smejkal2022}%
  \BibitemOpen
  \bibfield  {author} {\bibinfo {author} {\bibfnamefont {L.}~\bibnamefont
  {Šmejkal}}, \bibinfo {author} {\bibfnamefont {J.}~\bibnamefont {Sinova}},\
  and\ \bibinfo {author} {\bibfnamefont {T.}~\bibnamefont {Jungwirth}},\
  }\bibfield  {title} {\bibinfo {title} {{Beyond Conventional Ferromagnetism
  and Antiferromagnetism: A Phase with Nonrelativistic Spin and Crystal
  Rotation Symmetry}},\ }\href {https://doi.org/10.1103/physrevx.12.031042}
  {\bibfield  {journal} {\bibinfo  {journal} {Physical Review X}\ }\textbf
  {\bibinfo {volume} {12}},\ \bibinfo {pages} {031042} (\bibinfo {year}
  {2022}{\natexlab{b}})}\BibitemShut {NoStop}%
\bibitem [{\citenamefont {Liu}\ \emph {et~al.}(2022)\citenamefont {Liu},
  \citenamefont {Li}, \citenamefont {Han}, \citenamefont {Wan},\ and\
  \citenamefont {Liu}}]{Liu2022}%
  \BibitemOpen
  \bibfield  {author} {\bibinfo {author} {\bibfnamefont {P.}~\bibnamefont
  {Liu}}, \bibinfo {author} {\bibfnamefont {J.}~\bibnamefont {Li}}, \bibinfo
  {author} {\bibfnamefont {J.}~\bibnamefont {Han}}, \bibinfo {author}
  {\bibfnamefont {X.}~\bibnamefont {Wan}},\ and\ \bibinfo {author}
  {\bibfnamefont {Q.}~\bibnamefont {Liu}},\ }\bibfield  {title} {\bibinfo
  {title} {{Spin-Group Symmetry in Magnetic Materials with Negligible
  Spin-Orbit Coupling}},\ }\href {https://doi.org/10.1103/physrevx.12.021016}
  {\bibfield  {journal} {\bibinfo  {journal} {Physical Review X}\ }\textbf
  {\bibinfo {volume} {12}},\ \bibinfo {pages} {021016} (\bibinfo {year}
  {2022})}\BibitemShut {NoStop}%
\bibitem [{\citenamefont {Guo}\ \emph {et~al.}(2025)\citenamefont {Guo},
  \citenamefont {Wang}, \citenamefont {Wang}, \citenamefont {Zhang},
  \citenamefont {Zhou},\ and\ \citenamefont {Cheng}}]{Guo2025}%
  \BibitemOpen
  \bibfield  {author} {\bibinfo {author} {\bibfnamefont {Z.}~\bibnamefont
  {Guo}}, \bibinfo {author} {\bibfnamefont {X.}~\bibnamefont {Wang}}, \bibinfo
  {author} {\bibfnamefont {W.}~\bibnamefont {Wang}}, \bibinfo {author}
  {\bibfnamefont {G.}~\bibnamefont {Zhang}}, \bibinfo {author} {\bibfnamefont
  {X.}~\bibnamefont {Zhou}},\ and\ \bibinfo {author} {\bibfnamefont
  {Z.}~\bibnamefont {Cheng}},\ }\bibfield  {title} {\bibinfo {title}
  {{Spin‐Polarized Antiferromagnets for Spintronics}},\ }\bibfield  {journal}
  {\bibinfo  {journal} {Advanced Materials}\ }\textbf {\bibinfo {volume}
  {37}},\ \href {https://doi.org/10.1002/adma.202505779}
  {10.1002/adma.202505779} (\bibinfo {year} {2025})\BibitemShut {NoStop}%
\bibitem [{\citenamefont {Shim}\ \emph {et~al.}(2025)\citenamefont {Shim},
  \citenamefont {Mehraeen}, \citenamefont {Sklenar}, \citenamefont {Zhang},
  \citenamefont {Hoffmann},\ and\ \citenamefont {Mason}}]{Shim2025}%
  \BibitemOpen
  \bibfield  {author} {\bibinfo {author} {\bibfnamefont {S.}~\bibnamefont
  {Shim}}, \bibinfo {author} {\bibfnamefont {M.}~\bibnamefont {Mehraeen}},
  \bibinfo {author} {\bibfnamefont {J.}~\bibnamefont {Sklenar}}, \bibinfo
  {author} {\bibfnamefont {S.~S.-L.}\ \bibnamefont {Zhang}}, \bibinfo {author}
  {\bibfnamefont {A.}~\bibnamefont {Hoffmann}},\ and\ \bibinfo {author}
  {\bibfnamefont {N.}~\bibnamefont {Mason}},\ }\bibfield  {title} {\bibinfo
  {title} {{Spin-Polarized Antiferromagnetic Metals}},\ }\href
  {https://doi.org/10.1146/annurev-conmatphys-042924-123620} {\bibfield
  {journal} {\bibinfo  {journal} {Annual Review of Condensed Matter Physics}\
  }\textbf {\bibinfo {volume} {16}},\ \bibinfo {pages} {103} (\bibinfo {year}
  {2025})}\BibitemShut {NoStop}%
\bibitem [{\citenamefont {Tamang}\ \emph {et~al.}(2025)\citenamefont {Tamang},
  \citenamefont {Gurung}, \citenamefont {Rai}, \citenamefont {Brahimi},\ and\
  \citenamefont {Lounis}}]{Tamang2025}%
  \BibitemOpen
  \bibfield  {author} {\bibinfo {author} {\bibfnamefont {R.}~\bibnamefont
  {Tamang}}, \bibinfo {author} {\bibfnamefont {S.}~\bibnamefont {Gurung}},
  \bibinfo {author} {\bibfnamefont {D.~P.}\ \bibnamefont {Rai}}, \bibinfo
  {author} {\bibfnamefont {S.}~\bibnamefont {Brahimi}},\ and\ \bibinfo {author}
  {\bibfnamefont {S.}~\bibnamefont {Lounis}},\ }\bibfield  {title} {\bibinfo
  {title} {{Altermagnetism and Altermagnets: A Brief Review}},\ }\href
  {https://doi.org/10.3390/magnetism5030017} {\bibfield  {journal} {\bibinfo
  {journal} {Magnetism}\ }\textbf {\bibinfo {volume} {5}},\ \bibinfo {pages}
  {17} (\bibinfo {year} {2025})}\BibitemShut {NoStop}%
\bibitem [{\citenamefont {Song}\ \emph {et~al.}(2025)\citenamefont {Song},
  \citenamefont {Bai}, \citenamefont {Zhou}, \citenamefont {Han}, \citenamefont
  {Reichlova}, \citenamefont {Dil}, \citenamefont {Liu}, \citenamefont {Chen},\
  and\ \citenamefont {Pan}}]{Song2025}%
  \BibitemOpen
  \bibfield  {author} {\bibinfo {author} {\bibfnamefont {C.}~\bibnamefont
  {Song}}, \bibinfo {author} {\bibfnamefont {H.}~\bibnamefont {Bai}}, \bibinfo
  {author} {\bibfnamefont {Z.}~\bibnamefont {Zhou}}, \bibinfo {author}
  {\bibfnamefont {L.}~\bibnamefont {Han}}, \bibinfo {author} {\bibfnamefont
  {H.}~\bibnamefont {Reichlova}}, \bibinfo {author} {\bibfnamefont {J.~H.}\
  \bibnamefont {Dil}}, \bibinfo {author} {\bibfnamefont {J.}~\bibnamefont
  {Liu}}, \bibinfo {author} {\bibfnamefont {X.}~\bibnamefont {Chen}},\ and\
  \bibinfo {author} {\bibfnamefont {F.}~\bibnamefont {Pan}},\ }\bibfield
  {title} {\bibinfo {title} {{Altermagnets as a new class of functional
  materials}},\ }\href {https://doi.org/10.1038/s41578-025-00779-1} {\bibfield
  {journal} {\bibinfo  {journal} {Nature Reviews Materials}\ }\textbf {\bibinfo
  {volume} {10}},\ \bibinfo {pages} {473} (\bibinfo {year} {2025})}\BibitemShut
  {NoStop}%
\bibitem [{\citenamefont {Jungwirth}\ \emph {et~al.}(2026)\citenamefont
  {Jungwirth}, \citenamefont {Sinova}, \citenamefont {Fernandes}, \citenamefont
  {Liu}, \citenamefont {Watanabe}, \citenamefont {Murakami}, \citenamefont
  {Nakatsuji},\ and\ \citenamefont {Šmejkal}}]{Jungwirth2026}%
  \BibitemOpen
  \bibfield  {author} {\bibinfo {author} {\bibfnamefont {T.}~\bibnamefont
  {Jungwirth}}, \bibinfo {author} {\bibfnamefont {J.}~\bibnamefont {Sinova}},
  \bibinfo {author} {\bibfnamefont {R.~M.}\ \bibnamefont {Fernandes}}, \bibinfo
  {author} {\bibfnamefont {Q.}~\bibnamefont {Liu}}, \bibinfo {author}
  {\bibfnamefont {H.}~\bibnamefont {Watanabe}}, \bibinfo {author}
  {\bibfnamefont {S.}~\bibnamefont {Murakami}}, \bibinfo {author}
  {\bibfnamefont {S.}~\bibnamefont {Nakatsuji}},\ and\ \bibinfo {author}
  {\bibfnamefont {L.}~\bibnamefont {Šmejkal}},\ }\bibfield  {title} {\bibinfo
  {title} {{Symmetry, microscopy and spectroscopy signatures of
  altermagnetism}},\ }\href {https://doi.org/10.1038/s41586-025-09883-2}
  {\bibfield  {journal} {\bibinfo  {journal} {Nature}\ }\textbf {\bibinfo
  {volume} {649}},\ \bibinfo {pages} {837} (\bibinfo {year}
  {2026})}\BibitemShut {NoStop}%
\bibitem [{\citenamefont {Brinkman}\ and\ \citenamefont
  {Elliott}(1966)}]{Brinkman1966}%
  \BibitemOpen
  \bibfield  {author} {\bibinfo {author} {\bibfnamefont {W.~F.}\ \bibnamefont
  {Brinkman}}\ and\ \bibinfo {author} {\bibfnamefont {R.~J.}\ \bibnamefont
  {Elliott}},\ }\bibfield  {title} {\bibinfo {title} {{Theory of Spin-Space
  Groups}},\ }\href {http://www.jstor.org/stable/2415409} {\bibfield  {journal}
  {\bibinfo  {journal} {Proceedings of the Royal Society of London. Series A,
  Mathematical and Physical Sciences}\ }\textbf {\bibinfo {volume} {294}},\
  \bibinfo {pages} {343} (\bibinfo {year} {1966})}\BibitemShut {NoStop}%
\bibitem [{\citenamefont {Litvin}\ and\ \citenamefont
  {Opechowski}(1974)}]{Litvin1974}%
  \BibitemOpen
  \bibfield  {author} {\bibinfo {author} {\bibfnamefont {D.}~\bibnamefont
  {Litvin}}\ and\ \bibinfo {author} {\bibfnamefont {W.}~\bibnamefont
  {Opechowski}},\ }\bibfield  {title} {\bibinfo {title} {{Spin groups}},\
  }\href {https://doi.org/10.1016/0031-8914(74)90157-8} {\bibfield  {journal}
  {\bibinfo  {journal} {Physica}\ }\textbf {\bibinfo {volume} {76}},\ \bibinfo
  {pages} {538} (\bibinfo {year} {1974})}\BibitemShut {NoStop}%
\bibitem [{\citenamefont {Litvin}(1977)}]{Litvin1977}%
  \BibitemOpen
  \bibfield  {author} {\bibinfo {author} {\bibfnamefont {D.~B.}\ \bibnamefont
  {Litvin}},\ }\bibfield  {title} {\bibinfo {title} {{Spin point groups}},\
  }\href {https://doi.org/10.1107/s0567739477000709} {\bibfield  {journal}
  {\bibinfo  {journal} {Acta Crystallographica Section A}\ }\textbf {\bibinfo
  {volume} {33}},\ \bibinfo {pages} {279} (\bibinfo {year} {1977})}\BibitemShut
  {NoStop}%
\bibitem [{\citenamefont {Chen}\ \emph {et~al.}(2024)\citenamefont {Chen},
  \citenamefont {Ren}, \citenamefont {Zhu}, \citenamefont {Yu}, \citenamefont
  {Zhang}, \citenamefont {Liu}, \citenamefont {Li}, \citenamefont {Liu},
  \citenamefont {Li},\ and\ \citenamefont {Liu}}]{Chen2024}%
  \BibitemOpen
  \bibfield  {author} {\bibinfo {author} {\bibfnamefont {X.}~\bibnamefont
  {Chen}}, \bibinfo {author} {\bibfnamefont {J.}~\bibnamefont {Ren}}, \bibinfo
  {author} {\bibfnamefont {Y.}~\bibnamefont {Zhu}}, \bibinfo {author}
  {\bibfnamefont {Y.}~\bibnamefont {Yu}}, \bibinfo {author} {\bibfnamefont
  {A.}~\bibnamefont {Zhang}}, \bibinfo {author} {\bibfnamefont
  {P.}~\bibnamefont {Liu}}, \bibinfo {author} {\bibfnamefont {J.}~\bibnamefont
  {Li}}, \bibinfo {author} {\bibfnamefont {Y.}~\bibnamefont {Liu}}, \bibinfo
  {author} {\bibfnamefont {C.}~\bibnamefont {Li}},\ and\ \bibinfo {author}
  {\bibfnamefont {Q.}~\bibnamefont {Liu}},\ }\bibfield  {title} {\bibinfo
  {title} {{Enumeration and Representation Theory of Spin Space Groups}},\
  }\href {https://doi.org/10.1103/physrevx.14.031038} {\bibfield  {journal}
  {\bibinfo  {journal} {Physical Review X}\ }\textbf {\bibinfo {volume} {14}},\
  \bibinfo {pages} {031038} (\bibinfo {year} {2024})}\BibitemShut {NoStop}%
\bibitem [{\citenamefont {Jiang}\ \emph {et~al.}(2024)\citenamefont {Jiang},
  \citenamefont {Song}, \citenamefont {Zhu}, \citenamefont {Fang},
  \citenamefont {Weng}, \citenamefont {Liu}, \citenamefont {Yang},\ and\
  \citenamefont {Fang}}]{Jiang2024}%
  \BibitemOpen
  \bibfield  {author} {\bibinfo {author} {\bibfnamefont {Y.}~\bibnamefont
  {Jiang}}, \bibinfo {author} {\bibfnamefont {Z.}~\bibnamefont {Song}},
  \bibinfo {author} {\bibfnamefont {T.}~\bibnamefont {Zhu}}, \bibinfo {author}
  {\bibfnamefont {Z.}~\bibnamefont {Fang}}, \bibinfo {author} {\bibfnamefont
  {H.}~\bibnamefont {Weng}}, \bibinfo {author} {\bibfnamefont {Z.-X.}\
  \bibnamefont {Liu}}, \bibinfo {author} {\bibfnamefont {J.}~\bibnamefont
  {Yang}},\ and\ \bibinfo {author} {\bibfnamefont {C.}~\bibnamefont {Fang}},\
  }\bibfield  {title} {\bibinfo {title} {{Enumeration of Spin-Space Groups:
  Toward a Complete Description of Symmetries of Magnetic Orders}},\ }\href
  {https://doi.org/10.1103/physrevx.14.031039} {\bibfield  {journal} {\bibinfo
  {journal} {Physical Review X}\ }\textbf {\bibinfo {volume} {14}},\ \bibinfo
  {pages} {031039} (\bibinfo {year} {2024})}\BibitemShut {NoStop}%
\bibitem [{\citenamefont {Xiao}\ \emph {et~al.}(2024)\citenamefont {Xiao},
  \citenamefont {Zhao}, \citenamefont {Li}, \citenamefont {Shindou},\ and\
  \citenamefont {Song}}]{Xiao2024}%
  \BibitemOpen
  \bibfield  {author} {\bibinfo {author} {\bibfnamefont {Z.}~\bibnamefont
  {Xiao}}, \bibinfo {author} {\bibfnamefont {J.}~\bibnamefont {Zhao}}, \bibinfo
  {author} {\bibfnamefont {Y.}~\bibnamefont {Li}}, \bibinfo {author}
  {\bibfnamefont {R.}~\bibnamefont {Shindou}},\ and\ \bibinfo {author}
  {\bibfnamefont {Z.-D.}\ \bibnamefont {Song}},\ }\bibfield  {title} {\bibinfo
  {title} {{Spin Space Groups: Full Classification and Applications}},\ }\href
  {https://doi.org/10.1103/physrevx.14.031037} {\bibfield  {journal} {\bibinfo
  {journal} {Physical Review X}\ }\textbf {\bibinfo {volume} {14}},\ \bibinfo
  {pages} {031037} (\bibinfo {year} {2024})}\BibitemShut {NoStop}%
\bibitem [{\citenamefont {Watanabe}\ \emph {et~al.}(2024)\citenamefont
  {Watanabe}, \citenamefont {Shinohara}, \citenamefont {Nomoto}, \citenamefont
  {Togo},\ and\ \citenamefont {Arita}}]{Watanabe2024}%
  \BibitemOpen
  \bibfield  {author} {\bibinfo {author} {\bibfnamefont {H.}~\bibnamefont
  {Watanabe}}, \bibinfo {author} {\bibfnamefont {K.}~\bibnamefont {Shinohara}},
  \bibinfo {author} {\bibfnamefont {T.}~\bibnamefont {Nomoto}}, \bibinfo
  {author} {\bibfnamefont {A.}~\bibnamefont {Togo}},\ and\ \bibinfo {author}
  {\bibfnamefont {R.}~\bibnamefont {Arita}},\ }\bibfield  {title} {\bibinfo
  {title} {{Symmetry analysis with spin crystallographic groups: Disentangling
  effects free of spin-orbit coupling in emergent electromagnetism}},\ }\href
  {https://doi.org/10.1103/physrevb.109.094438} {\bibfield  {journal} {\bibinfo
   {journal} {Physical Review B}\ }\textbf {\bibinfo {volume} {109}},\ \bibinfo
  {pages} {094438} (\bibinfo {year} {2024})}\BibitemShut {NoStop}%
\bibitem [{\citenamefont {Etxebarria}\ \emph {et~al.}(2025)\citenamefont
  {Etxebarria}, \citenamefont {Perez-Mato}, \citenamefont {Tasci},\ and\
  \citenamefont {Elcoro}}]{Etxebarria2025}%
  \BibitemOpen
  \bibfield  {author} {\bibinfo {author} {\bibfnamefont {J.}~\bibnamefont
  {Etxebarria}}, \bibinfo {author} {\bibfnamefont {J.~M.}\ \bibnamefont
  {Perez-Mato}}, \bibinfo {author} {\bibfnamefont {E.~S.}\ \bibnamefont
  {Tasci}},\ and\ \bibinfo {author} {\bibfnamefont {L.}~\bibnamefont
  {Elcoro}},\ }\bibfield  {title} {\bibinfo {title} {{Crystal tensor properties
  of magnetic materials with and without spin–orbit coupling. Application of
  spin point groups as approximate symmetries}},\ }\href
  {https://doi.org/10.1107/s2053273325004127} {\bibfield  {journal} {\bibinfo
  {journal} {Acta Crystallographica Section A Foundations and Advances}\
  }\textbf {\bibinfo {volume} {81}},\ \bibinfo {pages} {317} (\bibinfo {year}
  {2025})}\BibitemShut {NoStop}%
\bibitem [{\citenamefont {Liu}\ \emph {et~al.}(2025)\citenamefont {Liu},
  \citenamefont {Wei}, \citenamefont {Peng}, \citenamefont {Hou}, \citenamefont
  {Gao},\ and\ \citenamefont {Niu}}]{Liu2025}%
  \BibitemOpen
  \bibfield  {author} {\bibinfo {author} {\bibfnamefont {Z.}~\bibnamefont
  {Liu}}, \bibinfo {author} {\bibfnamefont {M.}~\bibnamefont {Wei}}, \bibinfo
  {author} {\bibfnamefont {W.}~\bibnamefont {Peng}}, \bibinfo {author}
  {\bibfnamefont {D.}~\bibnamefont {Hou}}, \bibinfo {author} {\bibfnamefont
  {Y.}~\bibnamefont {Gao}},\ and\ \bibinfo {author} {\bibfnamefont
  {Q.}~\bibnamefont {Niu}},\ }\bibfield  {title} {\bibinfo {title} {{Multipolar
  Anisotropy in Anomalous Hall Effect from Spin-Group Symmetry Breaking}},\
  }\href {https://doi.org/10.1103/physrevx.15.031006} {\bibfield  {journal}
  {\bibinfo  {journal} {Physical Review X}\ }\textbf {\bibinfo {volume} {15}},\
  \bibinfo {pages} {031006} (\bibinfo {year} {2025})}\BibitemShut {NoStop}%
\bibitem [{\citenamefont {Liu}\ \emph {et~al.}(2026)\citenamefont {Liu},
  \citenamefont {Gao},\ and\ \citenamefont {Niu}}]{Liu2026}%
  \BibitemOpen
  \bibfield  {author} {\bibinfo {author} {\bibfnamefont {Z.}~\bibnamefont
  {Liu}}, \bibinfo {author} {\bibfnamefont {Y.}~\bibnamefont {Gao}},\ and\
  \bibinfo {author} {\bibfnamefont {Q.}~\bibnamefont {Niu}},\ }\bibfield
  {title} {\bibinfo {title} {{Rigid-Body Anisotropy in Noncollinear
  Antiferromagnets}},\ }\bibfield  {journal} {\bibinfo  {journal} {Physical
  Review Letters}\ }\textbf {\bibinfo {volume} {136}},\ \href
  {https://doi.org/10.1103/64wt-51gd} {10.1103/64wt-51gd} (\bibinfo {year}
  {2026})\BibitemShut {NoStop}%
\bibitem [{\citenamefont {Russakoff}(1970)}]{Russakoff1970}%
  \BibitemOpen
  \bibfield  {author} {\bibinfo {author} {\bibfnamefont {G.}~\bibnamefont
  {Russakoff}},\ }\bibfield  {title} {\bibinfo {title} {{A Derivation of the
  Macroscopic Maxwell Equations}},\ }\href {https://doi.org/10.1119/1.1976000}
  {\bibfield  {journal} {\bibinfo  {journal} {American Journal of Physics}\
  }\textbf {\bibinfo {volume} {38}},\ \bibinfo {pages} {1188} (\bibinfo {year}
  {1970})}\BibitemShut {NoStop}%
\bibitem [{\citenamefont {Robinson}(1973)}]{Robinson}%
  \BibitemOpen
  \bibfield  {author} {\bibinfo {author} {\bibfnamefont {F.~N.~H.}\
  \bibnamefont {Robinson}},\ }\href@noop {} {\emph {\bibinfo {title}
  {Macroscopic Electromagnetism}}}\ (\bibinfo  {publisher} {Pergamon Press},\
  \bibinfo {address} {Oxford},\ \bibinfo {year} {1973})\BibitemShut {NoStop}%
\bibitem [{\citenamefont {Robinson}(1971)}]{Robinson1971}%
  \BibitemOpen
  \bibfield  {author} {\bibinfo {author} {\bibfnamefont {F.}~\bibnamefont
  {Robinson}},\ }\bibfield  {title} {\bibinfo {title} {{The microscopic and
  macroscopic equations of the electromagnetic field}},\ }\href
  {https://doi.org/10.1016/0031-8914(71)90180-7} {\bibfield  {journal}
  {\bibinfo  {journal} {Physica}\ }\textbf {\bibinfo {volume} {54}},\ \bibinfo
  {pages} {329} (\bibinfo {year} {1971})}\BibitemShut {NoStop}%
\bibitem [{\citenamefont {Culcer}\ \emph {et~al.}(2004)\citenamefont {Culcer},
  \citenamefont {Sinova}, \citenamefont {Sinitsyn}, \citenamefont {Jungwirth},
  \citenamefont {MacDonald},\ and\ \citenamefont {Niu}}]{Culcer2004}%
  \BibitemOpen
  \bibfield  {author} {\bibinfo {author} {\bibfnamefont {D.}~\bibnamefont
  {Culcer}}, \bibinfo {author} {\bibfnamefont {J.}~\bibnamefont {Sinova}},
  \bibinfo {author} {\bibfnamefont {N.~A.}\ \bibnamefont {Sinitsyn}}, \bibinfo
  {author} {\bibfnamefont {T.}~\bibnamefont {Jungwirth}}, \bibinfo {author}
  {\bibfnamefont {A.~H.}\ \bibnamefont {MacDonald}},\ and\ \bibinfo {author}
  {\bibfnamefont {Q.}~\bibnamefont {Niu}},\ }\bibfield  {title} {\bibinfo
  {title} {{Semiclassical Spin Transport in Spin-Orbit-Coupled Bands}},\ }\href
  {https://doi.org/10.1103/physrevlett.93.046602} {\bibfield  {journal}
  {\bibinfo  {journal} {Physical Review Letters}\ }\textbf {\bibinfo {volume}
  {93}},\ \bibinfo {pages} {046602} (\bibinfo {year} {2004})}\BibitemShut
  {NoStop}%
\bibitem [{\citenamefont {Altland}\ and\ \citenamefont
  {Simons}(2023)}]{Altland2023}%
  \BibitemOpen
  \bibfield  {author} {\bibinfo {author} {\bibfnamefont {A.}~\bibnamefont
  {Altland}}\ and\ \bibinfo {author} {\bibfnamefont {B.~D.}\ \bibnamefont
  {Simons}},\ }\href {https://doi.org/10.1017/9781108781244} {\emph {\bibinfo
  {title} {{Condensed Matter Field Theory}}}},\ \bibinfo {edition} {3rd}\ ed.\
  (\bibinfo  {publisher} {Cambridge University Press},\ \bibinfo {year}
  {2023})\BibitemShut {NoStop}%
\bibitem [{\citenamefont {Belashchenko}(2010)}]{Belashchenko2010}%
  \BibitemOpen
  \bibfield  {author} {\bibinfo {author} {\bibfnamefont {K.~D.}\ \bibnamefont
  {Belashchenko}},\ }\bibfield  {title} {\bibinfo {title} {{Equilibrium
  Magnetization at the Boundary of a Magnetoelectric Antiferromagnet}},\ }\href
  {https://doi.org/10.1103/physrevlett.105.147204} {\bibfield  {journal}
  {\bibinfo  {journal} {Physical Review Letters}\ }\textbf {\bibinfo {volume}
  {105}},\ \bibinfo {pages} {147204} (\bibinfo {year} {2010})}\BibitemShut
  {NoStop}%
\bibitem [{\citenamefont {Chen}\ \emph {et~al.}(2019)\citenamefont {Chen},
  \citenamefont {Niu},\ and\ \citenamefont {MacDonald}}]{chen2019spin}%
  \BibitemOpen
  \bibfield  {author} {\bibinfo {author} {\bibfnamefont {H.}~\bibnamefont
  {Chen}}, \bibinfo {author} {\bibfnamefont {Q.}~\bibnamefont {Niu}},\ and\
  \bibinfo {author} {\bibfnamefont {A.~H.}\ \bibnamefont {MacDonald}},\
  }\href@noop {} {\bibinfo {title} {{Spin Hall effects without spin currents in
  magnetic insulators}}} (\bibinfo {year} {2019}),\ \Eprint
  {https://arxiv.org/abs/1803.01294} {arXiv:1803.01294 [cond-mat.mes-hall]}
  \BibitemShut {NoStop}%
\bibitem [{\citenamefont {Spaldin}(2021)}]{Spaldin2021a}%
  \BibitemOpen
  \bibfield  {author} {\bibinfo {author} {\bibfnamefont {N.~A.}\ \bibnamefont
  {Spaldin}},\ }\bibfield  {title} {\bibinfo {title} {{Analogy between the
  Magnetic Dipole Moment at the Surface of a Magnetoelectric and the Electric
  Charge at the Surface of a Ferroelectric}},\ }\href
  {https://doi.org/10.1134/s1063776121040208} {\bibfield  {journal} {\bibinfo
  {journal} {Journal of Experimental and Theoretical Physics}\ }\textbf
  {\bibinfo {volume} {132}},\ \bibinfo {pages} {493} (\bibinfo {year}
  {2021})}\BibitemShut {NoStop}%
\bibitem [{\citenamefont {Qin}\ \emph {et~al.}(2011)\citenamefont {Qin},
  \citenamefont {Niu},\ and\ \citenamefont {Shi}}]{Qin2011}%
  \BibitemOpen
  \bibfield  {author} {\bibinfo {author} {\bibfnamefont {T.}~\bibnamefont
  {Qin}}, \bibinfo {author} {\bibfnamefont {Q.}~\bibnamefont {Niu}},\ and\
  \bibinfo {author} {\bibfnamefont {J.}~\bibnamefont {Shi}},\ }\bibfield
  {title} {\bibinfo {title} {{Energy Magnetization and the Thermal Hall
  Effect}},\ }\href {https://doi.org/10.1103/physrevlett.107.236601} {\bibfield
   {journal} {\bibinfo  {journal} {Physical Review Letters}\ }\textbf {\bibinfo
  {volume} {107}},\ \bibinfo {pages} {236601} (\bibinfo {year}
  {2011})}\BibitemShut {NoStop}%
\bibitem [{\citenamefont {Luttinger}(1964)}]{Luttinger1964}%
  \BibitemOpen
  \bibfield  {author} {\bibinfo {author} {\bibfnamefont {J.~M.}\ \bibnamefont
  {Luttinger}},\ }\bibfield  {title} {\bibinfo {title} {{Theory of Thermal
  Transport Coefficients}},\ }\href {https://doi.org/10.1103/physrev.135.a1505}
  {\bibfield  {journal} {\bibinfo  {journal} {Physical Review}\ }\textbf
  {\bibinfo {volume} {135}},\ \bibinfo {pages} {A1505} (\bibinfo {year}
  {1964})}\BibitemShut {NoStop}%
\bibitem [{\citenamefont {Tolman}(1930)}]{Tolman1930}%
  \BibitemOpen
  \bibfield  {author} {\bibinfo {author} {\bibfnamefont {R.~C.}\ \bibnamefont
  {Tolman}},\ }\bibfield  {title} {\bibinfo {title} {{On the Weight of Heat and
  Thermal Equilibrium in General Relativity}},\ }\href
  {https://doi.org/10.1103/physrev.35.904} {\bibfield  {journal} {\bibinfo
  {journal} {Physical Review}\ }\textbf {\bibinfo {volume} {35}},\ \bibinfo
  {pages} {904} (\bibinfo {year} {1930})}\BibitemShut {NoStop}%
\bibitem [{\citenamefont {Tolman}\ and\ \citenamefont
  {Ehrenfest}(1930)}]{Tolman1930a}%
  \BibitemOpen
  \bibfield  {author} {\bibinfo {author} {\bibfnamefont {R.~C.}\ \bibnamefont
  {Tolman}}\ and\ \bibinfo {author} {\bibfnamefont {P.}~\bibnamefont
  {Ehrenfest}},\ }\bibfield  {title} {\bibinfo {title} {{Temperature
  Equilibrium in a Static Gravitational Field}},\ }\href
  {https://doi.org/10.1103/physrev.36.1791} {\bibfield  {journal} {\bibinfo
  {journal} {Physical Review}\ }\textbf {\bibinfo {volume} {36}},\ \bibinfo
  {pages} {1791} (\bibinfo {year} {1930})}\BibitemShut {NoStop}%
\bibitem [{\citenamefont {Klein}(1949)}]{Klein1949}%
  \BibitemOpen
  \bibfield  {author} {\bibinfo {author} {\bibfnamefont {O.}~\bibnamefont
  {Klein}},\ }\bibfield  {title} {\bibinfo {title} {{On the Thermodynamical
  Equilibrium of Fluids in Gravitational Fields}},\ }\href
  {https://doi.org/10.1103/revmodphys.21.531} {\bibfield  {journal} {\bibinfo
  {journal} {Reviews of Modern Physics}\ }\textbf {\bibinfo {volume} {21}},\
  \bibinfo {pages} {531} (\bibinfo {year} {1949})}\BibitemShut {NoStop}%
\bibitem [{\citenamefont {Shitade}\ \emph
  {et~al.}(2019{\natexlab{b}})\citenamefont {Shitade}, \citenamefont {Daido},\
  and\ \citenamefont {Yanase}}]{Shitade2019}%
  \BibitemOpen
  \bibfield  {author} {\bibinfo {author} {\bibfnamefont {A.}~\bibnamefont
  {Shitade}}, \bibinfo {author} {\bibfnamefont {A.}~\bibnamefont {Daido}},\
  and\ \bibinfo {author} {\bibfnamefont {Y.}~\bibnamefont {Yanase}},\
  }\bibfield  {title} {\bibinfo {title} {{Theory of spin magnetic quadrupole
  moment and temperature-gradient-induced magnetization}},\ }\href
  {https://doi.org/10.1103/physrevb.99.024404} {\bibfield  {journal} {\bibinfo
  {journal} {Physical Review B}\ }\textbf {\bibinfo {volume} {99}},\ \bibinfo
  {pages} {024404} (\bibinfo {year} {2019}{\natexlab{b}})}\BibitemShut
  {NoStop}%
\bibitem [{\citenamefont {Moriya}(1960)}]{Moriya1960}%
  \BibitemOpen
  \bibfield  {author} {\bibinfo {author} {\bibfnamefont {T.}~\bibnamefont
  {Moriya}},\ }\bibfield  {title} {\bibinfo {title} {{Anisotropic Superexchange
  Interaction and Weak Ferromagnetism}},\ }\href
  {https://doi.org/10.1103/physrev.120.91} {\bibfield  {journal} {\bibinfo
  {journal} {Physical Review}\ }\textbf {\bibinfo {volume} {120}},\ \bibinfo
  {pages} {91} (\bibinfo {year} {1960})}\BibitemShut {NoStop}%
\bibitem [{\citenamefont {Zhu}\ \emph {et~al.}(2025)\citenamefont {Zhu},
  \citenamefont {Li}, \citenamefont {Chen}, \citenamefont {Yu},\ and\
  \citenamefont {Liu}}]{Zhu2025}%
  \BibitemOpen
  \bibfield  {author} {\bibinfo {author} {\bibfnamefont {H.}~\bibnamefont
  {Zhu}}, \bibinfo {author} {\bibfnamefont {J.}~\bibnamefont {Li}}, \bibinfo
  {author} {\bibfnamefont {X.}~\bibnamefont {Chen}}, \bibinfo {author}
  {\bibfnamefont {Y.}~\bibnamefont {Yu}},\ and\ \bibinfo {author}
  {\bibfnamefont {Q.}~\bibnamefont {Liu}},\ }\bibfield  {title} {\bibinfo
  {title} {{Magnetic geometry induced quantum geometry and nonlinear
  transports}},\ }\bibfield  {journal} {\bibinfo  {journal} {Nature
  Communications}\ }\textbf {\bibinfo {volume} {16}},\ \href
  {https://doi.org/10.1038/s41467-025-60128-2} {10.1038/s41467-025-60128-2}
  (\bibinfo {year} {2025})\BibitemShut {NoStop}%
\bibitem [{\citenamefont {Giannozzi}\ \emph {et~al.}(2009)\citenamefont
  {Giannozzi}, \citenamefont {Baroni}, \citenamefont {Bonini}, \citenamefont
  {Calandra}, \citenamefont {Car}, \citenamefont {Cavazzoni}, \citenamefont
  {Ceresoli}, \citenamefont {Chiarotti}, \citenamefont {Cococcioni},
  \citenamefont {Dabo}, \citenamefont {Dal~Corso}, \citenamefont
  {de~Gironcoli}, \citenamefont {Fabris}, \citenamefont {Fratesi},
  \citenamefont {Gebauer}, \citenamefont {Gerstmann}, \citenamefont
  {Gougoussis}, \citenamefont {Kokalj}, \citenamefont {Lazzeri}, \citenamefont
  {Martin-Samos}, \citenamefont {Marzari}, \citenamefont {Mauri}, \citenamefont
  {Mazzarello}, \citenamefont {Paolini}, \citenamefont {Pasquarello},
  \citenamefont {Paulatto}, \citenamefont {Sbraccia}, \citenamefont {Scandolo},
  \citenamefont {Sclauzero}, \citenamefont {Seitsonen}, \citenamefont
  {Smogunov}, \citenamefont {Umari},\ and\ \citenamefont
  {Wentzcovitch}}]{Giannozzi2009}%
  \BibitemOpen
  \bibfield  {author} {\bibinfo {author} {\bibfnamefont {P.}~\bibnamefont
  {Giannozzi}}, \bibinfo {author} {\bibfnamefont {S.}~\bibnamefont {Baroni}},
  \bibinfo {author} {\bibfnamefont {N.}~\bibnamefont {Bonini}}, \bibinfo
  {author} {\bibfnamefont {M.}~\bibnamefont {Calandra}}, \bibinfo {author}
  {\bibfnamefont {R.}~\bibnamefont {Car}}, \bibinfo {author} {\bibfnamefont
  {C.}~\bibnamefont {Cavazzoni}}, \bibinfo {author} {\bibfnamefont
  {D.}~\bibnamefont {Ceresoli}}, \bibinfo {author} {\bibfnamefont {G.~L.}\
  \bibnamefont {Chiarotti}}, \bibinfo {author} {\bibfnamefont {M.}~\bibnamefont
  {Cococcioni}}, \bibinfo {author} {\bibfnamefont {I.}~\bibnamefont {Dabo}},
  \bibinfo {author} {\bibfnamefont {A.}~\bibnamefont {Dal~Corso}}, \bibinfo
  {author} {\bibfnamefont {S.}~\bibnamefont {de~Gironcoli}}, \bibinfo {author}
  {\bibfnamefont {S.}~\bibnamefont {Fabris}}, \bibinfo {author} {\bibfnamefont
  {G.}~\bibnamefont {Fratesi}}, \bibinfo {author} {\bibfnamefont
  {R.}~\bibnamefont {Gebauer}}, \bibinfo {author} {\bibfnamefont
  {U.}~\bibnamefont {Gerstmann}}, \bibinfo {author} {\bibfnamefont
  {C.}~\bibnamefont {Gougoussis}}, \bibinfo {author} {\bibfnamefont
  {A.}~\bibnamefont {Kokalj}}, \bibinfo {author} {\bibfnamefont
  {M.}~\bibnamefont {Lazzeri}}, \bibinfo {author} {\bibfnamefont
  {L.}~\bibnamefont {Martin-Samos}}, \bibinfo {author} {\bibfnamefont
  {N.}~\bibnamefont {Marzari}}, \bibinfo {author} {\bibfnamefont
  {F.}~\bibnamefont {Mauri}}, \bibinfo {author} {\bibfnamefont
  {R.}~\bibnamefont {Mazzarello}}, \bibinfo {author} {\bibfnamefont
  {S.}~\bibnamefont {Paolini}}, \bibinfo {author} {\bibfnamefont
  {A.}~\bibnamefont {Pasquarello}}, \bibinfo {author} {\bibfnamefont
  {L.}~\bibnamefont {Paulatto}}, \bibinfo {author} {\bibfnamefont
  {C.}~\bibnamefont {Sbraccia}}, \bibinfo {author} {\bibfnamefont
  {S.}~\bibnamefont {Scandolo}}, \bibinfo {author} {\bibfnamefont
  {G.}~\bibnamefont {Sclauzero}}, \bibinfo {author} {\bibfnamefont {A.~P.}\
  \bibnamefont {Seitsonen}}, \bibinfo {author} {\bibfnamefont {A.}~\bibnamefont
  {Smogunov}}, \bibinfo {author} {\bibfnamefont {P.}~\bibnamefont {Umari}},\
  and\ \bibinfo {author} {\bibfnamefont {R.~M.}\ \bibnamefont {Wentzcovitch}},\
  }\bibfield  {title} {\bibinfo {title} {{QUANTUM ESPRESSO: a modular and
  open-source software project for quantum simulations of materials}},\ }\href
  {https://doi.org/10.1088/0953-8984/21/39/395502} {\bibfield  {journal}
  {\bibinfo  {journal} {Journal of Physics: Condensed Matter}\ }\textbf
  {\bibinfo {volume} {21}},\ \bibinfo {pages} {395502} (\bibinfo {year}
  {2009})}\BibitemShut {NoStop}%
\bibitem [{\citenamefont {Giannozzi}\ \emph {et~al.}(2017)\citenamefont
  {Giannozzi}, \citenamefont {Andreussi}, \citenamefont {Brumme}, \citenamefont
  {Bunau}, \citenamefont {Buongiorno~Nardelli}, \citenamefont {Calandra},
  \citenamefont {Car}, \citenamefont {Cavazzoni}, \citenamefont {Ceresoli},
  \citenamefont {Cococcioni}, \citenamefont {Colonna}, \citenamefont
  {Carnimeo}, \citenamefont {Dal~Corso}, \citenamefont {de~Gironcoli},
  \citenamefont {Delugas}, \citenamefont {DiStasio}, \citenamefont {Ferretti},
  \citenamefont {Floris}, \citenamefont {Fratesi}, \citenamefont {Fugallo},
  \citenamefont {Gebauer}, \citenamefont {Gerstmann}, \citenamefont {Giustino},
  \citenamefont {Gorni}, \citenamefont {Jia}, \citenamefont {Kawamura},
  \citenamefont {Ko}, \citenamefont {Kokalj}, \citenamefont {Küçükbenli},
  \citenamefont {Lazzeri}, \citenamefont {Marsili}, \citenamefont {Marzari},
  \citenamefont {Mauri}, \citenamefont {Nguyen}, \citenamefont {Nguyen},
  \citenamefont {Otero-de-la Roza}, \citenamefont {Paulatto}, \citenamefont
  {Poncé}, \citenamefont {Rocca}, \citenamefont {Sabatini}, \citenamefont
  {Santra}, \citenamefont {Schlipf}, \citenamefont {Seitsonen}, \citenamefont
  {Smogunov}, \citenamefont {Timrov}, \citenamefont {Thonhauser}, \citenamefont
  {Umari}, \citenamefont {Vast}, \citenamefont {Wu},\ and\ \citenamefont
  {Baroni}}]{Giannozzi2017}%
  \BibitemOpen
  \bibfield  {author} {\bibinfo {author} {\bibfnamefont {P.}~\bibnamefont
  {Giannozzi}}, \bibinfo {author} {\bibfnamefont {O.}~\bibnamefont
  {Andreussi}}, \bibinfo {author} {\bibfnamefont {T.}~\bibnamefont {Brumme}},
  \bibinfo {author} {\bibfnamefont {O.}~\bibnamefont {Bunau}}, \bibinfo
  {author} {\bibfnamefont {M.}~\bibnamefont {Buongiorno~Nardelli}}, \bibinfo
  {author} {\bibfnamefont {M.}~\bibnamefont {Calandra}}, \bibinfo {author}
  {\bibfnamefont {R.}~\bibnamefont {Car}}, \bibinfo {author} {\bibfnamefont
  {C.}~\bibnamefont {Cavazzoni}}, \bibinfo {author} {\bibfnamefont
  {D.}~\bibnamefont {Ceresoli}}, \bibinfo {author} {\bibfnamefont
  {M.}~\bibnamefont {Cococcioni}}, \bibinfo {author} {\bibfnamefont
  {N.}~\bibnamefont {Colonna}}, \bibinfo {author} {\bibfnamefont
  {I.}~\bibnamefont {Carnimeo}}, \bibinfo {author} {\bibfnamefont
  {A.}~\bibnamefont {Dal~Corso}}, \bibinfo {author} {\bibfnamefont
  {S.}~\bibnamefont {de~Gironcoli}}, \bibinfo {author} {\bibfnamefont
  {P.}~\bibnamefont {Delugas}}, \bibinfo {author} {\bibfnamefont {R.~A.}\
  \bibnamefont {DiStasio}}, \bibinfo {author} {\bibfnamefont {A.}~\bibnamefont
  {Ferretti}}, \bibinfo {author} {\bibfnamefont {A.}~\bibnamefont {Floris}},
  \bibinfo {author} {\bibfnamefont {G.}~\bibnamefont {Fratesi}}, \bibinfo
  {author} {\bibfnamefont {G.}~\bibnamefont {Fugallo}}, \bibinfo {author}
  {\bibfnamefont {R.}~\bibnamefont {Gebauer}}, \bibinfo {author} {\bibfnamefont
  {U.}~\bibnamefont {Gerstmann}}, \bibinfo {author} {\bibfnamefont
  {F.}~\bibnamefont {Giustino}}, \bibinfo {author} {\bibfnamefont
  {T.}~\bibnamefont {Gorni}}, \bibinfo {author} {\bibfnamefont
  {J.}~\bibnamefont {Jia}}, \bibinfo {author} {\bibfnamefont {M.}~\bibnamefont
  {Kawamura}}, \bibinfo {author} {\bibfnamefont {H.-Y.}\ \bibnamefont {Ko}},
  \bibinfo {author} {\bibfnamefont {A.}~\bibnamefont {Kokalj}}, \bibinfo
  {author} {\bibfnamefont {E.}~\bibnamefont {Küçükbenli}}, \bibinfo {author}
  {\bibfnamefont {M.}~\bibnamefont {Lazzeri}}, \bibinfo {author} {\bibfnamefont
  {M.}~\bibnamefont {Marsili}}, \bibinfo {author} {\bibfnamefont
  {N.}~\bibnamefont {Marzari}}, \bibinfo {author} {\bibfnamefont
  {F.}~\bibnamefont {Mauri}}, \bibinfo {author} {\bibfnamefont {N.~L.}\
  \bibnamefont {Nguyen}}, \bibinfo {author} {\bibfnamefont {H.-V.}\
  \bibnamefont {Nguyen}}, \bibinfo {author} {\bibfnamefont {A.}~\bibnamefont
  {Otero-de-la Roza}}, \bibinfo {author} {\bibfnamefont {L.}~\bibnamefont
  {Paulatto}}, \bibinfo {author} {\bibfnamefont {S.}~\bibnamefont {Poncé}},
  \bibinfo {author} {\bibfnamefont {D.}~\bibnamefont {Rocca}}, \bibinfo
  {author} {\bibfnamefont {R.}~\bibnamefont {Sabatini}}, \bibinfo {author}
  {\bibfnamefont {B.}~\bibnamefont {Santra}}, \bibinfo {author} {\bibfnamefont
  {M.}~\bibnamefont {Schlipf}}, \bibinfo {author} {\bibfnamefont {A.~P.}\
  \bibnamefont {Seitsonen}}, \bibinfo {author} {\bibfnamefont {A.}~\bibnamefont
  {Smogunov}}, \bibinfo {author} {\bibfnamefont {I.}~\bibnamefont {Timrov}},
  \bibinfo {author} {\bibfnamefont {T.}~\bibnamefont {Thonhauser}}, \bibinfo
  {author} {\bibfnamefont {P.}~\bibnamefont {Umari}}, \bibinfo {author}
  {\bibfnamefont {N.}~\bibnamefont {Vast}}, \bibinfo {author} {\bibfnamefont
  {X.}~\bibnamefont {Wu}},\ and\ \bibinfo {author} {\bibfnamefont
  {S.}~\bibnamefont {Baroni}},\ }\bibfield  {title} {\bibinfo {title}
  {{Advanced capabilities for materials modelling with Quantum ESPRESSO}},\
  }\href {https://doi.org/10.1088/1361-648x/aa8f79} {\bibfield  {journal}
  {\bibinfo  {journal} {Journal of Physics: Condensed Matter}\ }\textbf
  {\bibinfo {volume} {29}},\ \bibinfo {pages} {465901} (\bibinfo {year}
  {2017})}\BibitemShut {NoStop}%
\bibitem [{\citenamefont {Hamann}(2013)}]{Hamann2013}%
  \BibitemOpen
  \bibfield  {author} {\bibinfo {author} {\bibfnamefont {D.~R.}\ \bibnamefont
  {Hamann}},\ }\bibfield  {title} {\bibinfo {title} {{Optimized norm-conserving
  Vanderbilt pseudopotentials}},\ }\href
  {https://doi.org/10.1103/physrevb.88.085117} {\bibfield  {journal} {\bibinfo
  {journal} {Physical Review B}\ }\textbf {\bibinfo {volume} {88}},\ \bibinfo
  {pages} {085117} (\bibinfo {year} {2013})}\BibitemShut {NoStop}%
\bibitem [{\citenamefont {Tsukamoto}\ \emph {et~al.}(2025)\citenamefont
  {Tsukamoto}, \citenamefont {Xu}, \citenamefont {Higo}, \citenamefont
  {Kondou}, \citenamefont {Sasaki}, \citenamefont {Asakura}, \citenamefont
  {Sakamoto}, \citenamefont {Gambardella}, \citenamefont {Miwa}, \citenamefont
  {Otani}, \citenamefont {Nakatsuji}, \citenamefont {Degen},\ and\
  \citenamefont {Kobayashi}}]{Tsukamoto2025}%
  \BibitemOpen
  \bibfield  {author} {\bibinfo {author} {\bibfnamefont {M.}~\bibnamefont
  {Tsukamoto}}, \bibinfo {author} {\bibfnamefont {Z.}~\bibnamefont {Xu}},
  \bibinfo {author} {\bibfnamefont {T.}~\bibnamefont {Higo}}, \bibinfo {author}
  {\bibfnamefont {K.}~\bibnamefont {Kondou}}, \bibinfo {author} {\bibfnamefont
  {K.}~\bibnamefont {Sasaki}}, \bibinfo {author} {\bibfnamefont
  {M.}~\bibnamefont {Asakura}}, \bibinfo {author} {\bibfnamefont
  {S.}~\bibnamefont {Sakamoto}}, \bibinfo {author} {\bibfnamefont
  {P.}~\bibnamefont {Gambardella}}, \bibinfo {author} {\bibfnamefont
  {S.}~\bibnamefont {Miwa}}, \bibinfo {author} {\bibfnamefont {Y.}~\bibnamefont
  {Otani}}, \bibinfo {author} {\bibfnamefont {S.}~\bibnamefont {Nakatsuji}},
  \bibinfo {author} {\bibfnamefont {C.~L.}\ \bibnamefont {Degen}},\ and\
  \bibinfo {author} {\bibfnamefont {K.}~\bibnamefont {Kobayashi}},\ }\bibfield
  {title} {\bibinfo {title} {{Observation of chiral domain walls in an
  octupole-ordered antiferromagnet}},\ }\bibfield  {journal} {\bibinfo
  {journal} {Physical Review B}\ }\textbf {\bibinfo {volume} {112}},\ \href
  {https://doi.org/10.1103/njnm-nl6n} {10.1103/njnm-nl6n} (\bibinfo {year}
  {2025})\BibitemShut {NoStop}%
\bibitem [{\citenamefont {Li}\ \emph {et~al.}(2023)\citenamefont {Li},
  \citenamefont {Huang}, \citenamefont {Lu}, \citenamefont {McLaughlin},
  \citenamefont {Xiao}, \citenamefont {Zhou}, \citenamefont {Fullerton},
  \citenamefont {Chen}, \citenamefont {Wang},\ and\ \citenamefont
  {Du}}]{Li2023}%
  \BibitemOpen
  \bibfield  {author} {\bibinfo {author} {\bibfnamefont {S.}~\bibnamefont
  {Li}}, \bibinfo {author} {\bibfnamefont {M.}~\bibnamefont {Huang}}, \bibinfo
  {author} {\bibfnamefont {H.}~\bibnamefont {Lu}}, \bibinfo {author}
  {\bibfnamefont {N.~J.}\ \bibnamefont {McLaughlin}}, \bibinfo {author}
  {\bibfnamefont {Y.}~\bibnamefont {Xiao}}, \bibinfo {author} {\bibfnamefont
  {J.}~\bibnamefont {Zhou}}, \bibinfo {author} {\bibfnamefont {E.~E.}\
  \bibnamefont {Fullerton}}, \bibinfo {author} {\bibfnamefont {H.}~\bibnamefont
  {Chen}}, \bibinfo {author} {\bibfnamefont {H.}~\bibnamefont {Wang}},\ and\
  \bibinfo {author} {\bibfnamefont {C.~R.}\ \bibnamefont {Du}},\ }\bibfield
  {title} {\bibinfo {title} {{Nanoscale Magnetic Domains in Polycrystalline
  Mn3Sn Films Imaged by a Scanning Single-Spin Magnetometer}},\ }\href
  {https://doi.org/10.1021/acs.nanolett.3c01523} {\bibfield  {journal}
  {\bibinfo  {journal} {Nano Letters}\ }\textbf {\bibinfo {volume} {23}},\
  \bibinfo {pages} {5326} (\bibinfo {year} {2023})}\BibitemShut {NoStop}%
\bibitem [{\citenamefont {Huxter}\ \emph {et~al.}(2022)\citenamefont {Huxter},
  \citenamefont {Palm}, \citenamefont {Davis}, \citenamefont {Welter},
  \citenamefont {Lambert}, \citenamefont {Trassin},\ and\ \citenamefont
  {Degen}}]{Huxter2022}%
  \BibitemOpen
  \bibfield  {author} {\bibinfo {author} {\bibfnamefont {W.~S.}\ \bibnamefont
  {Huxter}}, \bibinfo {author} {\bibfnamefont {M.~L.}\ \bibnamefont {Palm}},
  \bibinfo {author} {\bibfnamefont {M.~L.}\ \bibnamefont {Davis}}, \bibinfo
  {author} {\bibfnamefont {P.}~\bibnamefont {Welter}}, \bibinfo {author}
  {\bibfnamefont {C.-H.}\ \bibnamefont {Lambert}}, \bibinfo {author}
  {\bibfnamefont {M.}~\bibnamefont {Trassin}},\ and\ \bibinfo {author}
  {\bibfnamefont {C.~L.}\ \bibnamefont {Degen}},\ }\bibfield  {title} {\bibinfo
  {title} {{Scanning gradiometry with a single spin quantum magnetometer}},\
  }\bibfield  {journal} {\bibinfo  {journal} {Nature Communications}\ }\textbf
  {\bibinfo {volume} {13}},\ \href {https://doi.org/10.1038/s41467-022-31454-6}
  {10.1038/s41467-022-31454-6} (\bibinfo {year} {2022})\BibitemShut {NoStop}%
\bibitem [{\citenamefont {Takenaka}\ and\ \citenamefont
  {Takagi}(2005)}]{Takenaka2005}%
  \BibitemOpen
  \bibfield  {author} {\bibinfo {author} {\bibfnamefont {K.}~\bibnamefont
  {Takenaka}}\ and\ \bibinfo {author} {\bibfnamefont {H.}~\bibnamefont
  {Takagi}},\ }\bibfield  {title} {\bibinfo {title} {{Giant negative thermal
  expansion in Ge-doped anti-perovskite manganese nitrides}},\ }\bibfield
  {journal} {\bibinfo  {journal} {Applied Physics Letters}\ }\textbf {\bibinfo
  {volume} {87}},\ \href {https://doi.org/10.1063/1.2147726}
  {10.1063/1.2147726} (\bibinfo {year} {2005})\BibitemShut {NoStop}%
\bibitem [{\citenamefont {Takenaka}\ \emph {et~al.}(2014)\citenamefont
  {Takenaka}, \citenamefont {Ichigo}, \citenamefont {Hamada}, \citenamefont
  {Ozawa}, \citenamefont {Shibayama}, \citenamefont {Inagaki},\ and\
  \citenamefont {Asano}}]{Takenaka2014}%
  \BibitemOpen
  \bibfield  {author} {\bibinfo {author} {\bibfnamefont {K.}~\bibnamefont
  {Takenaka}}, \bibinfo {author} {\bibfnamefont {M.}~\bibnamefont {Ichigo}},
  \bibinfo {author} {\bibfnamefont {T.}~\bibnamefont {Hamada}}, \bibinfo
  {author} {\bibfnamefont {A.}~\bibnamefont {Ozawa}}, \bibinfo {author}
  {\bibfnamefont {T.}~\bibnamefont {Shibayama}}, \bibinfo {author}
  {\bibfnamefont {T.}~\bibnamefont {Inagaki}},\ and\ \bibinfo {author}
  {\bibfnamefont {K.}~\bibnamefont {Asano}},\ }\bibfield  {title} {\bibinfo
  {title} {{Magnetovolume effects in manganese nitrides with antiperovskite
  structure}},\ }\href {https://doi.org/10.1088/1468-6996/15/1/015009}
  {\bibfield  {journal} {\bibinfo  {journal} {Science and Technology of
  Advanced Materials}\ }\textbf {\bibinfo {volume} {15}},\ \bibinfo {pages}
  {015009} (\bibinfo {year} {2014})}\BibitemShut {NoStop}%
\bibitem [{\citenamefont {Ding}\ \emph {et~al.}(2011)\citenamefont {Ding},
  \citenamefont {Wang}, \citenamefont {Chu}, \citenamefont {Yan}, \citenamefont
  {Na}, \citenamefont {Huang},\ and\ \citenamefont {Chen}}]{Ding2011}%
  \BibitemOpen
  \bibfield  {author} {\bibinfo {author} {\bibfnamefont {L.}~\bibnamefont
  {Ding}}, \bibinfo {author} {\bibfnamefont {C.}~\bibnamefont {Wang}}, \bibinfo
  {author} {\bibfnamefont {L.}~\bibnamefont {Chu}}, \bibinfo {author}
  {\bibfnamefont {J.}~\bibnamefont {Yan}}, \bibinfo {author} {\bibfnamefont
  {Y.}~\bibnamefont {Na}}, \bibinfo {author} {\bibfnamefont {Q.}~\bibnamefont
  {Huang}},\ and\ \bibinfo {author} {\bibfnamefont {X.}~\bibnamefont {Chen}},\
  }\bibfield  {title} {\bibinfo {title} {{Near zero temperature coefficient of
  resistivity in antiperovskite Mn3Ni1-xCuxN}},\ }\bibfield  {journal}
  {\bibinfo  {journal} {Applied Physics Letters}\ }\textbf {\bibinfo {volume}
  {99}},\ \href {https://doi.org/10.1063/1.3671183} {10.1063/1.3671183}
  (\bibinfo {year} {2011})\BibitemShut {NoStop}%
\bibitem [{\citenamefont {Matsunami}\ \emph {et~al.}(2014)\citenamefont
  {Matsunami}, \citenamefont {Fujita}, \citenamefont {Takenaka},\ and\
  \citenamefont {Kano}}]{Matsunami2014}%
  \BibitemOpen
  \bibfield  {author} {\bibinfo {author} {\bibfnamefont {D.}~\bibnamefont
  {Matsunami}}, \bibinfo {author} {\bibfnamefont {A.}~\bibnamefont {Fujita}},
  \bibinfo {author} {\bibfnamefont {K.}~\bibnamefont {Takenaka}},\ and\
  \bibinfo {author} {\bibfnamefont {M.}~\bibnamefont {Kano}},\ }\bibfield
  {title} {\bibinfo {title} {{Giant barocaloric effect enhanced by the
  frustration of the antiferromagnetic phase in Mn3GaN}},\ }\href
  {https://doi.org/10.1038/nmat4117} {\bibfield  {journal} {\bibinfo  {journal}
  {Nature Materials}\ }\textbf {\bibinfo {volume} {14}},\ \bibinfo {pages} {73}
  (\bibinfo {year} {2014})}\BibitemShut {NoStop}%
\bibitem [{\citenamefont {Boldrin}\ \emph {et~al.}(2018)\citenamefont
  {Boldrin}, \citenamefont {Mendive-Tapia}, \citenamefont {Zemen},
  \citenamefont {Staunton}, \citenamefont {Hansen}, \citenamefont {Aznar},
  \citenamefont {Tamarit}, \citenamefont {Barrio}, \citenamefont {Lloveras},
  \citenamefont {Kim}, \citenamefont {Moya},\ and\ \citenamefont
  {Cohen}}]{Boldrin2018}%
  \BibitemOpen
  \bibfield  {author} {\bibinfo {author} {\bibfnamefont {D.}~\bibnamefont
  {Boldrin}}, \bibinfo {author} {\bibfnamefont {E.}~\bibnamefont
  {Mendive-Tapia}}, \bibinfo {author} {\bibfnamefont {J.}~\bibnamefont
  {Zemen}}, \bibinfo {author} {\bibfnamefont {J.~B.}\ \bibnamefont {Staunton}},
  \bibinfo {author} {\bibfnamefont {T.}~\bibnamefont {Hansen}}, \bibinfo
  {author} {\bibfnamefont {A.}~\bibnamefont {Aznar}}, \bibinfo {author}
  {\bibfnamefont {J.-L.}\ \bibnamefont {Tamarit}}, \bibinfo {author}
  {\bibfnamefont {M.}~\bibnamefont {Barrio}}, \bibinfo {author} {\bibfnamefont
  {P.}~\bibnamefont {Lloveras}}, \bibinfo {author} {\bibfnamefont
  {J.}~\bibnamefont {Kim}}, \bibinfo {author} {\bibfnamefont {X.}~\bibnamefont
  {Moya}},\ and\ \bibinfo {author} {\bibfnamefont {L.~F.}\ \bibnamefont
  {Cohen}},\ }\bibfield  {title} {\bibinfo {title} {{Multisite
  Exchange-Enhanced Barocaloric Response in Mn3NiN}},\ }\href
  {https://doi.org/10.1103/physrevx.8.041035} {\bibfield  {journal} {\bibinfo
  {journal} {Physical Review X}\ }\textbf {\bibinfo {volume} {8}},\ \bibinfo
  {pages} {041035} (\bibinfo {year} {2018})}\BibitemShut {NoStop}%
\bibitem [{\citenamefont {Martin}\ and\ \citenamefont
  {Batista}(2008)}]{Martin2008}%
  \BibitemOpen
  \bibfield  {author} {\bibinfo {author} {\bibfnamefont {I.}~\bibnamefont
  {Martin}}\ and\ \bibinfo {author} {\bibfnamefont {C.~D.}\ \bibnamefont
  {Batista}},\ }\bibfield  {title} {\bibinfo {title} {{Itinerant
  Electron-Driven Chiral Magnetic Ordering and Spontaneous Quantum Hall Effect
  in Triangular Lattice Models}},\ }\href
  {https://doi.org/10.1103/physrevlett.101.156402} {\bibfield  {journal}
  {\bibinfo  {journal} {Physical Review Letters}\ }\textbf {\bibinfo {volume}
  {101}},\ \bibinfo {pages} {156402} (\bibinfo {year} {2008})}\BibitemShut
  {NoStop}%
\bibitem [{\citenamefont {Zhou}\ \emph {et~al.}(2016)\citenamefont {Zhou},
  \citenamefont {Liang}, \citenamefont {Weng}, \citenamefont {Chen},
  \citenamefont {Yao}, \citenamefont {Chen}, \citenamefont {Dong},\ and\
  \citenamefont {Guo}}]{Zhou2016}%
  \BibitemOpen
  \bibfield  {author} {\bibinfo {author} {\bibfnamefont {J.}~\bibnamefont
  {Zhou}}, \bibinfo {author} {\bibfnamefont {Q.-F.}\ \bibnamefont {Liang}},
  \bibinfo {author} {\bibfnamefont {H.}~\bibnamefont {Weng}}, \bibinfo {author}
  {\bibfnamefont {Y.}~\bibnamefont {Chen}}, \bibinfo {author} {\bibfnamefont
  {S.-H.}\ \bibnamefont {Yao}}, \bibinfo {author} {\bibfnamefont {Y.-F.}\
  \bibnamefont {Chen}}, \bibinfo {author} {\bibfnamefont {J.}~\bibnamefont
  {Dong}},\ and\ \bibinfo {author} {\bibfnamefont {G.-Y.}\ \bibnamefont
  {Guo}},\ }\bibfield  {title} {\bibinfo {title} {{Predicted Quantum
  Topological Hall Effect and Noncoplanar Antiferromagnetism inK0.5RhO2}},\
  }\href {https://doi.org/10.1103/physrevlett.116.256601} {\bibfield  {journal}
  {\bibinfo  {journal} {Physical Review Letters}\ }\textbf {\bibinfo {volume}
  {116}},\ \bibinfo {pages} {256601} (\bibinfo {year} {2016})}\BibitemShut
  {NoStop}%
\bibitem [{\citenamefont {Boerner}\ \emph {et~al.}(2023)\citenamefont
  {Boerner}, \citenamefont {Deems}, \citenamefont {Furlani}, \citenamefont
  {Knuth},\ and\ \citenamefont {Towns}}]{Boerner2023}%
  \BibitemOpen
  \bibfield  {author} {\bibinfo {author} {\bibfnamefont {T.~J.}\ \bibnamefont
  {Boerner}}, \bibinfo {author} {\bibfnamefont {S.}~\bibnamefont {Deems}},
  \bibinfo {author} {\bibfnamefont {T.~R.}\ \bibnamefont {Furlani}}, \bibinfo
  {author} {\bibfnamefont {S.~L.}\ \bibnamefont {Knuth}},\ and\ \bibinfo
  {author} {\bibfnamefont {J.}~\bibnamefont {Towns}},\ }\bibfield  {title}
  {\bibinfo {title} {{ACCESS: Advancing Innovation: NSF’s Advanced
  Cyberinfrastructure Coordination Ecosystem: Services \& Support}},\ }in\
  \href {https://doi.org/10.1145/3569951.3597559} {\emph {\bibinfo {booktitle}
  {Practice and Experience in Advanced Research Computing}}},\ \bibinfo {series
  and number} {PEARC ’23}\ (\bibinfo  {publisher} {ACM},\ \bibinfo {year}
  {2023})\ pp.\ \bibinfo {pages} {173--176}\BibitemShut {NoStop}%
\bibitem [{\citenamefont {Ibañez-Azpiroz}\ \emph {et~al.}(2018)\citenamefont
  {Ibañez-Azpiroz}, \citenamefont {Tsirkin},\ and\ \citenamefont
  {Souza}}]{IbanezAzpiroz2018}%
  \BibitemOpen
  \bibfield  {author} {\bibinfo {author} {\bibfnamefont {J.}~\bibnamefont
  {Ibañez-Azpiroz}}, \bibinfo {author} {\bibfnamefont {S.~S.}\ \bibnamefont
  {Tsirkin}},\ and\ \bibinfo {author} {\bibfnamefont {I.}~\bibnamefont
  {Souza}},\ }\bibfield  {title} {\bibinfo {title} {{Ab initio calculation of
  the shift photocurrent by Wannier interpolation}},\ }\href
  {https://doi.org/10.1103/physrevb.97.245143} {\bibfield  {journal} {\bibinfo
  {journal} {Physical Review B}\ }\textbf {\bibinfo {volume} {97}},\ \bibinfo
  {pages} {245143} (\bibinfo {year} {2018})}\BibitemShut {NoStop}%
\bibitem [{\citenamefont {Lihm}(2021)}]{Lihm2021}%
  \BibitemOpen
  \bibfield  {author} {\bibinfo {author} {\bibfnamefont {J.-M.}\ \bibnamefont
  {Lihm}},\ }\bibfield  {title} {\bibinfo {title} {{Comment on “ Ab initio
  calculation of the shift photocurrent by Wannier interpolation”}},\ }\href
  {https://doi.org/10.1103/physrevb.103.247101} {\bibfield  {journal} {\bibinfo
   {journal} {Physical Review B}\ }\textbf {\bibinfo {volume} {103}},\ \bibinfo
  {pages} {247101} (\bibinfo {year} {2021})}\BibitemShut {NoStop}%
\bibitem [{\citenamefont {Mostofi}\ \emph {et~al.}(2014)\citenamefont
  {Mostofi}, \citenamefont {Yates}, \citenamefont {Pizzi}, \citenamefont {Lee},
  \citenamefont {Souza}, \citenamefont {Vanderbilt},\ and\ \citenamefont
  {Marzari}}]{Mostofi2014}%
  \BibitemOpen
  \bibfield  {author} {\bibinfo {author} {\bibfnamefont {A.~A.}\ \bibnamefont
  {Mostofi}}, \bibinfo {author} {\bibfnamefont {J.~R.}\ \bibnamefont {Yates}},
  \bibinfo {author} {\bibfnamefont {G.}~\bibnamefont {Pizzi}}, \bibinfo
  {author} {\bibfnamefont {Y.-S.}\ \bibnamefont {Lee}}, \bibinfo {author}
  {\bibfnamefont {I.}~\bibnamefont {Souza}}, \bibinfo {author} {\bibfnamefont
  {D.}~\bibnamefont {Vanderbilt}},\ and\ \bibinfo {author} {\bibfnamefont
  {N.}~\bibnamefont {Marzari}},\ }\bibfield  {title} {\bibinfo {title} {{An
  updated version of wannier90: A tool for obtaining maximally-localised
  Wannier functions}},\ }\href {https://doi.org/10.1016/j.cpc.2014.05.003}
  {\bibfield  {journal} {\bibinfo  {journal} {Computer Physics Communications}\
  }\textbf {\bibinfo {volume} {185}},\ \bibinfo {pages} {2309} (\bibinfo {year}
  {2014})}\BibitemShut {NoStop}%
\end{thebibliography}%

\end{document}